\documentclass[a4paper,11pt]{article}
\pdfoutput=1 

\usepackage{jcappub} 
\usepackage{etoolbox}
\notoc
\makeatletter
\renewcommand{\@fpheader}{}
\patchcmd{\maketitle}{\Huge \flushleft\sffamily\bfseries}{\LARGE \centering\normalfont\bfseries}{}%
  {\PackageError{main}{Unable to center the title}{Check the jcappub title definition.}}
\patchcmd{\maketitle}
  {\Large \bfseries\raggedright\sffamily\the\auth@toks\par}
  {\large \normalfont\centering\the\auth@toks\par}{}%
  {\PackageError{main}{Unable to center the authors}{Check the jcappub author definition.}}
\patchcmd{\maketitle}
  {\noindent\hspace{0.28cm}\begin{minipage}[l]{.9\textwidth}
   \begin{flushleft}
   E-mail: \the\email@toks
   \end{flushleft}
   \end{minipage}}{}{}%
  {\PackageError{main}{Unable to move the email}{Check the jcappub email definition.}}
\newcommand{\titlepageEmail}{\the\email@toks}
\patchcmd{\maketitle}{\noindent{\bfseries\large\sffamily{Abstract.}}\ \@abstract}{\noindent\@abstract}{}%
  {\PackageError{main}{Unable to remove the abstract label}{Check the jcappub title definition.}}
\patchcmd{\maketitle}{\the\affil@toks}{\centering\itshape\the\affil@toks}{}%
  {\PackageError{main}{Unable to center the affiliations}{Check the jcappub title definition.}}
\makeatother
\renewcommand{\afterProceedingsSpace}{\vskip6pt}
\renewcommand{\afterTitleSpace}{\vskip12pt}
\renewcommand{\afterAuthorSpace}{\vskip0pt}
\renewcommand{\afterAffiliationSpace}{\vskip6pt}
\renewcommand{\afterEmailSpace}{\vskip0pt}
\renewcommand{\afterAbstractSpace}{\vskip12pt}

\usepackage[T1]{fontenc} 

\usepackage{amsmath}
\usepackage{booktabs}
\usepackage{caption} 
\usepackage{array}
\usepackage{siunitx}
\usepackage{amssymb}
\usepackage{graphicx}
\usepackage{subcaption}  
\usepackage{hyperref}
\usepackage{mathtools}
\usepackage{placeins}
\usepackage{cleveref}
\usepackage{physics}
\usepackage{xcolor,ulem}
\newcommand{\rev}[1]{#1}

\DeclareRobustCommand{\numrev}[1]{#1}
\newcommand{\D}{\mathrm{d}}
\title{\boldmath\texorpdfstring{\rev{Probing parity-breaking black-hole geometry with extreme-mass-ratio inspirals}}{Probing parity-breaking black-hole geometry with extreme-mass-ratio inspirals}}

\author{Zhen-Bo Wu,}
\author{Qing Ou,}
\author[1]{and Peng-Cheng Li\note{Corresponding author. E-mail: \titlepageEmail.}}
\affiliation{School of Physics and Optoelectronics, South China University of Technology, Guangzhou 510641, China}

\emailAdd{pchli2021@scut.edu.cn}

\abstract{
We investigate the gravitational-wave signatures of equatorial reflection symmetry breaking in extreme-mass-ratio inspirals (EMRIs) around a parity-violating conformal Kerr black hole. For the circular-orbit branch connected to equatorial Kerr orbits, we derive the dynamics through second order in the dimensionless deformation parameter $\varepsilon$ and construct adiabatic inspirals and gravitational-wave signals using a mass-quadrupole prescription. Equatorial reflection maps $\varepsilon$ to $-\varepsilon$, making the polar displacement odd and the orbital frequency and energy even in the deformation. The induced first harmonic has a small direct contribution for the configurations studied, while second-order dynamical corrections accumulate over many cycles to produce appreciable dephasing. We find that rapid black-hole rotation enhances these phase effects and the associated noise-weighted waveform mismatch. For a one-year inspiral of a $10M_\odot$ object into a $10^6M_\odot$ black hole with $a/M=0.9$ and viewing inclination $\iota=2\pi/3$, the fixed-parameter mismatch reaches the comparison benchmark $\mathcal M=0.01$ at approximately $\varepsilon=10^{-3}$, without optimization over time, phase, or Kerr parameters. A single-parameter Fisher analysis also indicates improved local sensitivity to $\varepsilon$ at high spin. These results connect the reflection properties of the circular branch to accumulated source-phase effects and complement tests of propagation birefringence. }

\begin{document}

\maketitle
\flushbottom

\section{Introduction}
\label{sec1}

Gravitational-wave (GW) observations of binary black-hole coalescences enable tests of general relativity (GR) in the dynamical strong-field regime \cite{LIGOScientific:2016aoc,LIGOScientific:2021grtests}. Under the standard uniqueness assumptions, regular, stationary, asymptotically flat, uncharged black holes in four-dimensional vacuum GR are described by Kerr \cite{Carter:1971zc,Chrusciel:2012uniqueness}. Testing this description probes strong-field gravity \cite{Berti:2015grtests}.

An extreme-mass-ratio inspiral (EMRI) consists of a stellar-mass compact object of mass $m$ inspiraling into a massive black hole of mass $M$, with $\mu=m/M\ll1$. Its slowly evolving signal can contain $\sim10^4$--$10^5$ cycles, allowing small orbital and inspiral-rate corrections to accumulate in the phase \cite{Babak:2017tow}. EMRIs are targets for LISA \cite{LISA:2017pwj}, TianQin \cite{TianQin:2015yph}, and Taiji \cite{Ruan:2018tsw}. They enable tests of Kerr multipole relations \cite{Ryan:1995wh,Glampedakis:2005cf,Barack:2006pq,Moore:2017lxy,Zi:2021pdp,Cardenas-Avendano:2024mqp} and horizon absorption \cite{Cardoso:2019rvt,Zi:2023geb,Datta:2024vll,Xia:2026aty}, and probes of quantum-corrected or regular geometries \cite{Zi:2023qfk,Fu:2024cfk,Zi:2024jla,Zare:2025aek,Zhao:2025sck} and other deformed backgrounds and extensions of GR \cite{Tan:2024utr,Babichev:2024hjf,Kumar:2025jsi,LaHaye:2025ley,Lu:2025xlp,Long:2026dcb,Du:2026xre}. EMRIs also probe additional fields \cite{Maselli:2020zgv,Barsanti:2022ana,Zi:2022hcc,Zi:2023omh,Hu:2026way}, accretion disks \cite{Kocsis:2011dr,Speri:2022upm,HegadeKR:2025dur}, and dark-matter halos or spikes \cite{Dai:2023cft,Zhao:2024bpp,Mitra:2025tag,Vicente:2025gsg,Wang:2025rrj,Haroon:2025rzx,Rahman:2025mip,Das:2025vja,Das:2025eiv,Azreg-Ainou:2026xcc,Zhao:2026yis,Fu:2026gcu}. Related studies address environmental waveform modeling \cite{Barausse:2014tra,Cardoso:2022whc,Jiang:2024lwg,Dyson:2025dlj,Polcar:2025yto}, and distinguishing environmental from modified-gravity effects \cite{Kejriwal:2025jao,Copparoni:2025vty,Zi:2026zpw}; broader reviews are given in Refs.~\cite{LISA:2022kgy,Luo:2025ewp}.

A fundamental question is whether gravity preserves spatial parity. Interactions involving the Pontryagin density, as in Chern-Simons gravity, provide a setting for studying parity violation \cite{Alexander:2009cs,Qiao:2022mln} and its effects on wave generation and propagation \cite{Alexander:2017jmt}. Propagation birefringence causes the two circular tensor polarizations to undergo different amplification or phase evolution \cite{Zhao:2019xmm}, motivating LISA forecasts \cite{Yunes:2008bu} and catalog tests \cite{Wang:2021gqm,Ng:2023jjt,Gong:2021jgg,Zhu:2023rrx}. Modified source dynamics can also produce accumulated orbital dephasing without requiring propagation birefringence.

Source-phase tests have already been developed for dynamical Chern-Simons (dCS) EMRIs. Sopuerta and Yunes constructed semirelativistic waveforms from modified geodesics \cite{Sopuerta:2009cs}; Canizares, Gair, and Sopuerta modeled eccentric, inclined adiabatic inspirals using GR post-Newtonian (PN) fluxes calibrated to Teukolsky results and mass-quadrupole kludge waveforms \cite{Canizares:2012is}. Their slow-rotation background modifies frame dragging while preserving equatorial reflection symmetry. Complementary PN binary studies retain general mass ratios and include leading 2PN corrections to conservative dynamics, radiation, and source phasing, together with cosmological amplitude birefringence \cite{Li:2022dcswaveform}. No leading velocity birefringence appears in that propagation approximation. The source calculation has also been extended to eccentric binaries \cite{Li:2023dcseccentric}.

Multipolar models have also been used to study equatorial reflection symmetry breaking. Fransen and Mayerson included the current quadrupole and mass octupole, $S_2$ and $M_3$, which vanish in axisymmetric spacetimes with equatorial reflection symmetry. In an expansion about circular, equatorial orbits, they found linear polar displacements, quadratic binding-energy corrections, and a change in the radiation's harmonic content at linear order in the deformation \cite{Fransen:2022jtw}. Muguruza and Sopuerta used PN analytic kludge models with axisymmetric and nonaxisymmetric mass quadrupole and octupole deformations to forecast LISA sensitivity to equatorial and axial symmetry breaking \cite{Muguruza:2026hqn}.

Here we study the exact conformal Kerr family of Tahara et al.~\cite{Tahara:2023pyg}. Its invertible, curvature-dependent conformal transformation retains the two tensor degrees of freedom of GR in the complete pure-gravity theory. The conformal factor contains a nonzero odd part in the Pontryagin density, breaking equatorial reflection symmetry and displacing the circular branch connected to equatorial Kerr orbits. Unparametrized null geodesics follow the same paths as in Kerr, whereas timelike trajectories change, providing a source-side probe of the deformation. We extend the geodesic analysis of this background to radiation-driven EMRI signals and their spin dependence. The exact geometry requires no slow-rotation expansion, while our orbital calculation assumes weak deformation.

Reflection maps the deformation $\varepsilon$ to $-\varepsilon$, making the polar displacement odd and the orbital frequency and energy even on the locally unique analytic circular branch. In the mass-quadrupole prescription, this produces a linear first harmonic and quadratic corrections to the second harmonic and energy balance. Changes in both the frequency and inspiral rate accumulate in the phase, requiring a consistent second-order treatment of the orbit and energy balance. We quantify how this symmetry structure shapes the harmonics, dephasing, and spin dependence; their relative importance depends on the orbital coefficients and observation, beyond the powers of $\varepsilon$ alone.

We treat the smaller body as nonspinning and minimally coupled to the conformal metric, following the circular branch adiabatically using the mass-quadrupole energy flux. The equilibrium polar angle drifts during inspiral; independent polar oscillations and eccentricity are excluded. Kludge waveforms \cite{Glampedakis:2002cb,Babak:2006uv} provide the polarizations and a LISA-like detector strain. We calculate the dephasing relative to Kerr, the noise-weighted mismatch, and conditional single-parameter Fisher uncertainties. The mismatch is evaluated with aligned initial phases and fixed Kerr parameters, without optimization over relative time or phase. The Fisher widths describe local precision at nonzero fiducial deformations. These estimates isolate orbital effects within the adopted radiation model; observational bounds require the full theory's radiation dynamics and multiparameter inference.

Section~\ref{sec2} introduces the spacetime, Sec.~\ref{sec3} develops the orbital dynamics, waveforms, and sensitivity measures, and Sec.~\ref{sec4} presents the numerical results. Section~\ref{sec5} gives our conclusions. We use geometric units with $G=c=1$.

\section{Conformal Kerr spacetime in parity-violating gravity}
\label{sec2}

We adopt the conformal Kerr family of rotating black-hole solutions constructed in Ref.~\cite{Tahara:2023pyg}. The construction uses an invertible, curvature-dependent conformal transformation of the Einstein-Hilbert action: $S[\bar g]\equiv S_{\rm EH}[g[\bar g]]$, where $g[\bar g]$ is obtained from the inverse conformal transformation and $\bar g_{\mu\nu}$ is regarded as the physical metric. The complete transformed pure-gravity action retains the two propagating degrees of freedom of GR and is free from Ostrogradsky ghosts. This statement concerns the untruncated pure-gravity theory; matter couplings can spoil the degeneracy responsible for this property, as discussed in Ref.~\cite{Tahara:2023pyg}. Here, we use its vacuum black-hole background and treat the smaller compact object as a test body minimally coupled to $\bar g_{\mu\nu}$.

An invertible transformation maps a vacuum solution of GR into a solution of the transformed theory. Taking the Kerr metric $g_{\mu\nu}$ as the seed and restricting the conformal factor to depend on the gravitational Pontryagin density gives
\begin{equation}
    \bar g_{\mu\nu}=\Omega(\mathcal{P}) g_{\mu\nu}.
 \label{conformal kerr metric}
\end{equation}
In Boyer-Lindquist coordinates $(t,r,\theta,\varphi)$, the seed Kerr metric with signature $(-,+,+,+)$ has the line element
\begin{equation}
    \begin{aligned}
        \D s^2=g_{\mu\nu}\D x^\mu\D x^\nu
        &=-\left(1-\frac{2Mr}{\rho^2}\right)\D t^2
          -\frac{4aMr\sin^2\theta}{\rho^2}\D t\D\varphi\\
        &\quad+\frac{\rho^2}{\Delta}\D r^2+\rho^2\D\theta^2
          +\left(r^2+a^2+\frac{2Ma^2r\sin^2\theta}{\rho^2}\right)
           \sin^2\theta\D\varphi^2,
    \end{aligned}
    \label{eq:kerr-seed-metric}
\end{equation}

where $\Delta=r^2-2Mr+a^2$, $\rho^2=r^2+a^2\cos^2\theta$, $M$ is the black-hole mass, and $a=J/M$ is the spin parameter.

\rev{The Pontryagin density, also called the Chern-Simons scalar, is defined by}
\begin{equation}
    \mathcal{P} \equiv \frac{1}{2}\varepsilon^{\alpha\beta\gamma\delta}R^{\mu\nu}{}_{\alpha\beta}R_{\mu\nu\gamma\delta} = \frac{1}{2}\varepsilon^{\alpha\beta\gamma\delta} \mathcal{W}^{\mu\nu}{}_{\alpha\beta}\mathcal{W}_{\mu\nu\gamma\delta},
\end{equation}
where the Riemann tensor $R_{\mu\nu\alpha\beta}$, Weyl tensor $\mathcal W_{\mu\nu\alpha\beta}$, and Levi-Civita tensor $\varepsilon^{\alpha\beta\gamma\delta}$ are constructed from the seed metric. For the Kerr metric in Eq.~\eqref{eq:kerr-seed-metric}, the Pontryagin density is
\begin{equation}
    \mathcal{P}=-\frac{96aM^2r\cos\theta(r^2-3a^2\cos^2\theta)(3r^2-a^2\cos^2\theta)}{\rho^{12}}.
    \label{chen-simons term}
\end{equation}
The density is odd under equatorial reflection, $\mathcal P(r,\pi-\theta)=-\mathcal P(r,\theta)$, and vanishes in the nonrotating limit. A conformal factor with a nonzero odd part in $\mathcal P$ therefore breaks equatorial reflection symmetry; an even function of $\mathcal P$ would preserve it.

The conformal factor must be positive and the transformation locally invertible. For $\Omega=\Omega(\mathcal P)$, the invertibility condition is \cite{Takahashi:2022mew,Tahara:2023pyg}
\begin{equation}
    1-\frac{2\mathcal{P}}{\Omega}\frac{\D\Omega}{\D\mathcal{P}} \neq 0.
    \label{eq:invertibility_condition}
\end{equation}
This condition follows from the conformal covariance of the Pontryagin density, $\bar{\mathcal P}=\Omega^{-2}\mathcal P$, where $\bar{\mathcal P}$ is computed from $\bar g_{\mu\nu}$. Differentiating this relation gives
\begin{equation}
    \frac{\D\bar{\mathcal P}}{\D\mathcal P}
    =\frac{1}{\Omega^2}\left(1-\frac{2\mathcal P}{\Omega}
      \frac{\D\Omega}{\D\mathcal P}\right).
\end{equation}
Since $\Omega>0$, Eq.~\eqref{eq:invertibility_condition} ensures that this derivative is nonzero. The inverse function theorem then permits a local reconstruction $\mathcal P=\mathcal P(\bar{\mathcal P})$, and hence recovery of the seed metric through $g_{\mu\nu}=\bar g_{\mu\nu}/\Omega[\mathcal P(\bar{\mathcal P})]$.

Following the concrete model of Ref.~\cite{Tahara:2023pyg}, we take
\begin{equation}
    \Omega(\mathcal P)=1+\tanh x,\qquad
    x\equiv\frac{\alpha\mathcal P}{\Lambda^4}
      =\varepsilon M^4\mathcal P,\qquad
    \varepsilon\equiv\frac{\alpha}{M^4\Lambda^4},
    \label{eq:exact-conformal-factor}
\end{equation}
where $\alpha$ is dimensionless and $\Lambda$ is a mass scale, expressed in inverse-length units in the metric formulas. The parameter $M$ has dimensions of length, so $M\Lambda$ is dimensionless. Although $\alpha$ is already dimensionless, it does not by itself specify the deformation of a black hole of a given mass. Equation~\eqref{chen-simons term} can be written as
\begin{equation}
    \mathcal P(r,\theta)=\frac{1}{M^4}
    \widehat{\mathcal P}\!\left(\frac rM,\frac aM,\theta\right),
    \qquad x=\varepsilon\widehat{\mathcal P},
\end{equation}
where $\widehat{\mathcal P}=M^4\mathcal P$ is dimensionless. Thus, $\varepsilon$ combines the coupling with the black-hole curvature scale and provides a convenient expansion parameter. It introduces no additional independent coupling and coincides with the parameter $\hat\alpha$ used in Ref.~\cite{Tahara:2023pyg}.

The construction in Ref.~\cite{Tahara:2023pyg} does not fix a numerical value of $\Lambda$. In this model, $\alpha$ and $\Lambda$ enter only through the combination \numrev{$\kappa_{\mathcal P}\equiv\alpha/\Lambda^4=\varepsilon M^4$}; waveform information therefore cannot determine them separately. For fixed theoretical couplings, $\varepsilon\propto M^{-4}$, and comparisons between black holes of different masses should use the common coupling \numrev{$\kappa_{\mathcal P}$}. Assigning a value to $\varepsilon$ at a specified $M$ amounts to choosing a fiducial coupling, rather than inferring a physical value of $\Lambda$. For illustration, after restoring units and denoting the energy equivalent of $\Lambda$ by $\Lambda_E$, one has
\begin{equation}
    \Lambda_E=\frac{\hbar c\,|\alpha|^{1/4}}
    {r_g\,|\varepsilon|^{1/4}},\qquad
    r_g\equiv\frac{G M_{\rm BH}}{c^2}.
\end{equation}
A black hole of physical mass $M_{\rm BH}=10^6M_\odot$, with the illustrative choices $\alpha=1$ and $\varepsilon=10^{-3}$, corresponds to $\Lambda_E\simeq7.5\times10^{-16}\,\mathrm{eV}$, or a length $\hbar c/\Lambda_E\simeq2.6\times10^5\,\mathrm{km}$. These are conditional conversions of the chosen parameters, not theoretical predictions or observational bounds on the coupling scale.

The conformal factor in Eq.~\eqref{eq:exact-conformal-factor} is positive and satisfies Eq.~\eqref{eq:invertibility_condition} for any finite $\mathcal P$. Since $\mathcal P=\mathcal O(r^{-7})$ at large radius and $\Omega(0)=1$, the spacetime is asymptotically flat with the same asymptotic mass and angular momentum as the seed Kerr solution. Stationarity and axisymmetry are preserved because $\Omega$ depends only on $r$ and $\theta$. Where the conformal factor is smooth and positive, the outer horizon remains at $r_+=M+\sqrt{M^2-a^2}$.

Motivated by the consistency of existing gravitational-wave tests with GR \cite{LIGOScientific:2021grtests}, we focus on weak deformations of Kerr and develop a small-$\varepsilon$ expansion of the exact background. Asymptotic flatness holds for any fixed finite $\varepsilon$ and does not impose this small-deformation condition. Our waveform predictions and sensitivity estimates are conditional on this perturbative regime; we do not infer $\varepsilon$ or $\Lambda$ from observational data. For $|x|\ll1$,
\begin{equation}
    \Omega=1+x-\frac{x^3}{3}+\mathcal O(x^5)
    \simeq 1+\varepsilon M^4\mathcal P.
    \label{parity-break coefficient}
\end{equation}
Since the quadratic term vanishes, $\Omega_{\rm lin}=1+\varepsilon\widehat{\mathcal P}$ agrees with the exact conformal factor through second order in $\varepsilon$ at fixed $(r,\theta)$, with an error of $\mathcal O(\varepsilon^3)$. The nonlinear circular-geodesic conditions still generate second-order orbital corrections. For the linear approximation $\Omega_{\rm lin}=1+x$, the invertibility expression is exactly
\begin{equation}
    1-\frac{2\mathcal P}{\Omega_{\rm lin}}
       \frac{\D\Omega_{\rm lin}}{\D\mathcal P}
    =\frac{1-x}{1+x}=1-2x+\mathcal O(x^2).
    \label{eq:linear-conformal-invertibility}
\end{equation}

Positivity and local invertibility of this approximation hold in the weak-deformation region. Applying it to the circular orbit also requires $|\varepsilon\theta_1|\ll1$, $|\varepsilon^2\omega_2/\omega_0|\ll1$, and $|\varepsilon^2E_2/E_0|\ll1$, in addition to $|x|=|\varepsilon\widehat{\mathcal P}|\ll1$, throughout the modeled inspiral. The orbital coefficients are defined in Sec.~\ref{sec3}. Their growth at small radii makes these conditions dependent on spin and orbital radius as well as on $\varepsilon$. These conditions test internal consistency; controlling the accumulated phase additionally requires estimating the effect of omitted orders. Expanding the background metric does not replace the complete transformed action by a finite higher-derivative truncation.

A positive conformal transformation preserves unparametrized null-geodesic trajectories, although their affine parametrization changes. Timelike geodesics respond to gradients of $\Omega$. In particular, Eq.~\eqref{chen-simons term} gives

\begin{equation}
    \mathcal P(r,\pi/2)=0,\qquad
    \left.\partial_r\mathcal P\right|_{\theta=\pi/2}=0,\qquad
    \left.\partial_\theta\mathcal P\right|_{\theta=\pi/2}
    =\frac{288aM^2}{r^7}.
    \label{eq:equatorial-pontryagin-gradient}
\end{equation}

Thus, the metric components on the equatorial plane coincide with those of Kerr, but their polar derivatives differ for $a\varepsilon\ne0$. This shifts the equilibrium position of a timelike circular orbit away from $\theta=\pi/2$. Reference~\cite{Tahara:2023pyg} derives the first-order polar displacement and shows that neither the orbital frequency nor the ISCO radius receives a first-order correction for this model. On the branch continuously connected to equatorial Kerr circular orbits, $\theta_a-\pi/2=\mathcal O(\varepsilon)$ and $\omega-\omega_0=\mathcal O(\varepsilon^2)$.

The magnitude of $\widehat{\mathcal P}$ depends on position and spin. For example, at $r=6M$ and $a/M=0.5$, Eq.~\eqref{chen-simons term} gives
\begin{equation}
    \widehat{\mathcal P}
    =-\frac{288\cos\theta\,
      (36-0.75\cos^2\theta)(108-0.25\cos^2\theta)}
      {(36+0.25\cos^2\theta)^6}.
\end{equation}
It vanishes at $\theta=\pi/2$, while its maximum absolute value over $0\leq\theta\leq\pi$ is approximately $4.8208\times10^{-4}$, attained on the rotation axis. For a small displacement $\delta\theta\equiv\theta-\pi/2$, measured in radians,
\begin{equation}
    \widehat{\mathcal P}
    =5.1440\times10^{-4}\,\delta\theta
     +\mathcal O(\delta\theta^3).
\end{equation}
Thus, an illustrative value $|\varepsilon|=0.1$ gives $|x|\lesssim4.82\times10^{-5}$ at this radius and spin, with further suppression near the equatorial plane. This local estimate concerns the conformal-factor expansion; assessing the orbital approximation requires the corresponding corrections to be checked over the full inspiral domain.

The stability of these circular orbits involves coupled radial and polar perturbations. Reference~\cite{Tahara:2023pyg} also finds that multiple marginally stable circular orbits can occur at larger deformations, so the stable portion of the circular branch must be identified when specifying an inspiral domain. In the following section, we extend the circular-orbit expansion through second order and construct adiabatic inspirals, radiation reaction, and waveforms using a mass-quadrupole prescription. This calculation describes the influence of the deformed background within that approximation.

\section{Adiabatic evolution and waveform}
\label{sec3}

We consider an extreme-mass-ratio binary consisting of a nonspinning compact object of mass $m$ and a rotating black hole of mass $M$, with $\mu\equiv m/M\ll1$. 
We treat the black-hole mass $M$, spin parameter $a$, and dimensionless parity-violating parameter $\varepsilon$ as constants during the inspiral.
We model the conservative motion using constant-polar-angle circular geodesics in the parity-violating background and adopt the leading mass-quadrupole formula for radiation reaction and waveform generation.
It follows the adiabatic kludge strategy of Ref.~\cite{Glampedakis:2002cb}, while retaining the off-equatorial geometry characteristic of the conformal Kerr solution \cite{Tahara:2023pyg}.
The resulting waveform describes the influence of the parity-violating parameter $\varepsilon$ within this approximation.

\subsection{\texorpdfstring{Circular orbits and adiabatic evolution}{Circular orbits and adiabatic evolution}}

We first determine the circular geodesics that provide the conservative part of our inspiral model, and then prescribe their slow evolution using quadrupole energy balance. At leading order in the mass ratio, the small body is treated as a structureless test particle minimally coupled to the physical metric $\bar g_{\mu\nu}$. Its trajectory $x^\mu(s)$ is parametrized by proper time $s$, with $\dot x^\mu\equiv\D x^\mu/\D s$ and $\bar g_{\mu\nu}\dot x^\mu\dot x^\nu=-1$. The geodesic Lagrangian is
\begin{equation}
    \mathcal L=\frac12\bar g_{\mu\nu}\dot x^\mu\dot x^\nu.
    \label{eq:geodesic-lagrangian}
\end{equation}
Its Euler--Lagrange equations take the form
\begin{equation}
    \frac{\D}{\D s}\left(\bar g_{\sigma\nu}\dot x^\nu\right)
    =\frac12\partial_\sigma\bar g_{\mu\nu}\dot x^\mu\dot x^\nu.
    \label{eq:geodesic-euler-lagrange}
\end{equation}
These are equivalent to the geodesic equations $\ddot x^\lambda+\bar\Gamma^\lambda_{\mu\nu}\dot x^\mu\dot x^\nu=0$. Stationarity and axisymmetry make $t$ and $\varphi$ cyclic coordinates, giving the specific energy and axial angular momentum
\begin{equation}
    E=-\bar g_{t\nu}\dot x^\nu
      =-\bar g_{tt}\dot t-\bar g_{t\varphi}\dot\varphi,
    \qquad
    L_z=\bar g_{\varphi\nu}\dot x^\nu
      =\bar g_{t\varphi}\dot t+\bar g_{\varphi\varphi}\dot\varphi.
    \label{conserved quantities}
\end{equation}

Following Ref.~\cite{Tahara:2023pyg}, we restrict the conservative dynamics to circular geodesics about the symmetry axis with constant Boyer-Lindquist radius $r(s)=r_0$, constant polar angle $\theta(s)=\theta_a$, and constant coordinate angular velocity $\omega=\D\varphi/\D t=\dot\varphi/\dot t$. Thus $\dot r=\dot\theta=0$ on each such geodesic. Constancy of $\theta$ is a restriction on the orbital family and is not implied by $r={\rm const}$ alone. Independent polar oscillations are excluded. The equilibrium angle $\theta_a$ is determined by polar force balance and can differ from $\pi/2$; the branch considered here reduces to equatorial Kerr circular orbits as $\varepsilon\to0$. To express the equilibrium conditions compactly, define the seed-metric combination
\begin{equation}
    \mathcal G(r,\theta;\omega)
      \equiv g_{tt}+2\omega g_{t\varphi}+\omega^2g_{\varphi\varphi},
    \qquad
    N\equiv-\Omega\mathcal G.
    \label{eq:circular-metric-combination}
\end{equation}
With $\dot r=\dot\theta=0$ and $\dot\varphi=\omega\dot t$, the four-velocity normalization becomes $\Omega\mathcal G\dot t^2=-1$, giving $\dot t=N^{-1/2}$ on the future-directed branch. A timelike orbit therefore requires $N>0$, and its conserved quantities are
\begin{equation}
    E=-\frac{\bar g_{tt}+\omega\bar g_{t\varphi}}{\sqrt N},
    \qquad
    L_z=\frac{\bar g_{t\varphi}+\omega\bar g_{\varphi\varphi}}{\sqrt N}.
    \label{eq:specific-constants}
\end{equation}
On each such circular geodesic, the radial and polar momenta $\bar g_{rr}\dot r$ and $\bar g_{\theta\theta}\dot\theta$ vanish. The left-hand sides of the corresponding components of Eq.~\eqref{eq:geodesic-euler-lagrange} are therefore zero. Substituting $\dot\varphi=\omega\dot t$ and dividing by $\dot t^2/2$ gives
\begin{equation}
    \left.\partial_r(\Omega\mathcal G)\right|_\omega=0,
    \qquad
    \left.\partial_\theta(\Omega\mathcal G)\right|_\omega=0.
    \label{eq:circular-geodesic-conditions}
\end{equation}
The derivatives are taken at fixed $\omega$, before substituting $\theta=\theta_a(r)$ and $\omega=\omega(r)$. At a prescribed radius, these two force-balance equations determine the equilibrium polar angle and angular velocity; together with $N>0$, they characterize a timelike circular geodesic in the restricted family.

Existence must be distinguished from stability. In the effective-potential formulation of Ref.~\cite{Tahara:2023pyg}, the normalization condition becomes
\begin{equation}
    \begin{aligned}
        \bar g_{rr}\dot r^2+\bar g_{\theta\theta}\dot\theta^2
        +V_{\rm eff}&=0,\\
        V_{\rm eff}(r,\theta;E,L_z)
        &\equiv1+\frac{L_z^2\bar g_{tt}+2EL_z\bar g_{t\varphi}
                     +E^2\bar g_{\varphi\varphi}}
                    {\bar g_{tt}\bar g_{\varphi\varphi}-\bar g_{t\varphi}^2}.
    \end{aligned}
    \label{eq:circular-effective-potential}
\end{equation}
The conditions $V_{\rm eff}=\partial_rV_{\rm eff}=\partial_\theta V_{\rm eff}=0$ are equivalent to the circular-orbit conditions above. All potential derivatives are evaluated at fixed $E$ and $L_z$. Linear stability against coupled radial and polar perturbations additionally requires, outside the horizon,
\begin{equation}
    V_{{\rm eff},rr}V_{{\rm eff},\theta\theta}
       -V_{{\rm eff},r\theta}^2>0,
    \qquad
    \frac{V_{{\rm eff},rr}}{\bar g_{rr}}
       +\frac{V_{{\rm eff},\theta\theta}}{\bar g_{\theta\theta}}>0.
\end{equation}
The adiabatic construction uses the stable branch connected continuously to equatorial Kerr circular geodesics. Its inner boundary is a marginally stable circular orbit (MSCO), where one of the two coupled radial--polar oscillation frequencies vanishes while the other remains real and nonzero. Define, along the circular sequence,
\begin{equation}
    \mathcal S(r;\varepsilon)
       \equiv\left[V_{{\rm eff},rr}V_{{\rm eff},\theta\theta}
                      -V_{{\rm eff},r\theta}^{\,2}\right]_{\rm circ}.
\end{equation}
The potential is differentiated at fixed $E$ and $L_z$ before substituting the circular-orbit values of $\theta_a$, $E$, and $L_z$. Marginal stability requires $\mathcal S=0$ with the trace condition in the preceding equation still positive. Because radial and polar perturbations are coupled away from the Kerr equator, imposing $V_{{\rm eff},rr}=0$ alone is generally insufficient. For a sufficiently small deformation there is a single MSCO, which is the innermost stable circular orbit (ISCO). At larger deformations, multiple MSCOs and disconnected stable intervals can occur \cite{Tahara:2023pyg}; an inspiral from large radius then encounters the outer stable branch's inner boundary first, which need not be the globally innermost stable orbit. For example, the exact geometry at $a/M=0.99$ develops three MSCOs above approximately $\varepsilon=0.058$ \cite{Tahara:2023pyg}. Thus, the second-order, single-branch description requires a separate validity check when applied to the high-spin, larger-deformation configurations.

In the Kerr limit, the mixed derivative vanishes at the equator, and the prograde ISCO satisfies $r^2-6Mr+8a\sqrt{Mr}-3a^2=0$. For $0\leq\chi\equiv a/M<1$, its radius is \cite{Bardeen:1972isco}
\begin{equation}
    \begin{aligned}
        \frac{r_{\rm ISCO}^{\rm K}}{M}
          &=3+Z_2-\sqrt{(3-Z_1)(3+Z_1+2Z_2)},\\
        Z_1&=1+(1-\chi^2)^{1/3}
                    \left[(1+\chi)^{1/3}+(1-\chi)^{1/3}\right],
        \qquad Z_2=\sqrt{3\chi^2+Z_1^2}.
    \end{aligned}
\end{equation}
This gives $r_{\rm ISCO}^{\rm K}=6M$ for Schwarzschild and a decreasing prograde ISCO radius as the spin increases. In the deformed spacetime, the marginal-stability condition must instead be evaluated on the displaced circular orbit.

Before solving for the circular branch, its allowed perturbative orders can be fixed by reflection. Let $\mathcal R$ act as $(t,r,\theta,\varphi)\mapsto(t,r,\pi-\theta,\varphi)$. Since the seed metric is reflection symmetric and $\mathcal P$ is odd, the physical metric satisfies
\begin{equation}
    \mathcal R^*\bar g(\varepsilon)=\bar g(-\varepsilon),
    \label{eq:reflection-family}
\end{equation}
where $\mathcal R^*$ denotes the tensor pullback. This relates two members of the family; it does not assert reflection symmetry of either metric at nonzero $\varepsilon$. At fixed $r$, $M$, and $a$, reflection maps a circular geodesic at $\varepsilon$ into one at $-\varepsilon$, preserving its angular velocity and specific energy. Local uniqueness of the branch connected to the Kerr equator therefore implies
\begin{equation}
    \begin{aligned}
        \theta_a(-\varepsilon)&=\pi-\theta_a(\varepsilon),\\
        \omega(-\varepsilon)&=\omega(\varepsilon),
        \qquad E(-\varepsilon)=E(\varepsilon).
    \end{aligned}
    \label{eq:reflection-orbit-parity}
\end{equation}
For an analytic branch, the displacement has only odd powers of $\varepsilon$, whereas the frequency and energy have only even powers. The argument applies to any stationary, axisymmetric metric family obeying Eq.~\eqref{eq:reflection-family}, with the remaining parameters held fixed and a locally unique analytic circular branch through an equatorial orbit at $\varepsilon=0$; a conformal relation to Kerr is not required. It fixes the allowed orders, not their coefficients or observational importance. A related quadratic dependence of the binding energy on reflection-asymmetric multipoles occurs in the near-circular analysis of Ref.~\cite{Fransen:2022jtw}.

We now evaluate these orders explicitly using Eq.~\eqref{eq:circular-geodesic-conditions} at fixed radius. Begin with the general expansions
\begin{equation}
    \theta_a=\frac\pi2+\varepsilon\theta_1
                      +\varepsilon^2\theta_2+\mathcal O(\varepsilon^3),
    \qquad
    \omega=\omega_0+\varepsilon\omega_1
                  +\varepsilon^2\omega_2+\mathcal O(\varepsilon^3).
    \label{eq:circular-orbit-ansatz}
\end{equation}
For the prograde branch with $a\geq0$, used in our numerical calculations, the zeroth-order radial equation gives
\begin{equation}
    \omega_0=\frac{\sqrt M}{r^{3/2}+a\sqrt M}.
    \label{eq:kerr-circular-frequency}
\end{equation}
Let a subscript $0$ denote evaluation at $(r,\theta,\omega)=(r,\pi/2,\omega_0)$, and introduce
\begin{equation}
    p(r)\equiv\left.\partial_\theta\widehat{\mathcal P}\right|_{\theta=\pi/2}
       =\frac{288aM^6}{r^7},
    \qquad p'(r)=-\frac{7p(r)}r.
\end{equation}
Here $\widehat{\mathcal P}=M^4\mathcal P$, as defined in Sec.~\ref{sec2}. At the equator, $\widehat{\mathcal P}=\partial_r\widehat{\mathcal P}=\partial_\theta^2\widehat{\mathcal P}=0$, while $p$ is nonzero for $a\ne0$. Equatorial symmetry of the seed Kerr metric also gives $\mathcal G_{\theta,0}=\mathcal G_{r\theta,0}=\mathcal G_{\theta\theta\theta,0}=0$. Using Eq.~\eqref{parity-break coefficient}, the circular conditions at first order give $\omega_1=0$ and fix $\theta_1$; after imposing these results, their second-order terms determine $\theta_2$ and $\omega_2$:
\begin{equation}
    \begin{aligned}
        \mathcal O(\varepsilon):\quad
        &\mathcal G_{r\omega,0}\omega_1=0,\\
        &\mathcal G_{\theta\theta,0}\theta_1+p\mathcal G_0=0,\\[2pt]
        \mathcal O(\varepsilon^2):\quad
        &\mathcal G_{\theta\theta,0}\theta_2=0,\\
        &\mathcal G_{r\omega,0}\omega_2
          +\tfrac12\mathcal G_{r\theta\theta,0}\theta_1^2
          +p'\mathcal G_0\theta_1=0.
    \end{aligned}
    \label{eq:circular-order-equations}
\end{equation}
All derivatives of $\mathcal G$ in this equation are partial derivatives, treating $r$, $\theta$, and $\omega$ as independent variables. For a timelike seed orbit with $N_0>0$, the nondegeneracy condition $\mathcal G_{r\omega,0}\mathcal G_{\theta\theta,0}\ne0$ ensures a local circular branch for sufficiently small $\varepsilon$. In particular, $\omega_1=\theta_2=0$, and the remaining coefficients are
\begin{equation}
    \begin{aligned}
        \theta_1
        &=-\frac{p\mathcal G_0}{\mathcal G_{\theta\theta,0}}
         =-\frac{144aM^5(r^2-3Mr+2a\sqrt{Mr})}
                 {r^6(r^2+3a^2-4a\sqrt{Mr})},\\
        \omega_2
        &=-\frac{\tfrac12\mathcal G_{r\theta\theta,0}\theta_1^2
                    +p'\mathcal G_0\theta_1}
                  {\mathcal G_{r\omega,0}}
         =\frac{\sqrt M(4r^2-13a\sqrt{Mr}+9a^2)}
                 {\sqrt r(r^{3/2}+a\sqrt M)^2}\,\theta_1^2.
    \end{aligned}
    \label{eq:circular-orbit-coefficients}
\end{equation}
The first expression reproduces the polar displacement derived in Ref.~\cite{Tahara:2023pyg}, with our convention $\theta_a=\pi/2+\varepsilon\theta_1$. \numrev{The second gives the leading frequency correction.} The vanishing equatorial radial gradient of $\Omega$ gives $\omega_1=0$, whereas its nonzero polar gradient sources $\theta_1$. \numrev{These results satisfy the reflection relations in Eq.~\eqref{eq:reflection-orbit-parity}, which also determine the leading omitted orders:}
\begin{equation}
    \theta_a=\frac\pi2+\varepsilon\theta_1+\mathcal O(\varepsilon^3),
    \qquad
    \omega=\omega_0+\varepsilon^2\omega_2+\mathcal O(\varepsilon^4).
\end{equation}
Thus, a calculation through second order retains the first-order displacement and its quadratic effects, even though $\theta_2$ itself vanishes.

The specific energy needed for radiation reaction follows by substituting this orbit into Eq.~\eqref{eq:specific-constants}. To display its second-order content without a lengthy polynomial, define the seed-metric function
\begin{equation}
    E_{\rm K}(r,\theta,\omega)
       \equiv-\frac{g_{tt}+\omega g_{t\varphi}}{\sqrt{-\mathcal G}},
    \qquad E_0(r)\equiv E_{{\rm K},0}.
\end{equation}
Since $E=\sqrt\Omega E_{\rm K}$ and $\Omega[r,\theta_a(r)]=1+\varepsilon^2p\theta_1+\mathcal O(\varepsilon^4)$, one obtains
\begin{equation}
    \begin{aligned}
        E(r)&=E_0(r)+\varepsilon^2E_2(r)+\mathcal O(\varepsilon^4),\\
        E_2&=\left(\partial_\omega E_{\rm K}\right)_0\omega_2
          +\frac12\left(\partial_\theta^2E_{\rm K}\right)_0\theta_1^2
          +\frac12p\theta_1E_0.
    \end{aligned}
    \label{eq:circular-energy-expansion}
\end{equation}
The three terms in $E_2$ account for the frequency correction, \numrev{the displacement in polar angle}, and the conformal factor, respectively. There is no first-order energy correction. When differentiating $E(r)$ along the circular sequence, both $\theta_a(r)$ and $\omega(r)$ must vary with $r$; this total derivative differs from the fixed-$\omega$ partial derivatives used to find each circular orbit.

The absence of a linear ISCO shift is consistent with the general first-order result in Eq.~(38) of Ref.~\cite{Tahara:2023pyg}. That result assumes $\Omega=1+\varepsilon\varpi(r,\theta)$ and depends on the first and second radial derivatives of $\varpi$ at the Kerr ISCO on the equatorial plane. In our model, $\varpi=\widehat{\mathcal P}$ at this order. Since $\widehat{\mathcal P}(r,\pi/2)=0$ for every $r$, both radial derivatives vanish, and so does the linear ISCO correction, as also noted after Eq.~(40) of that reference. The calculation below determines the leading quadratic coefficient, including the effects of the polar displacement and coupled radial--polar perturbations.

To specify this ISCO correction, expand the effective-potential Hessian evaluated on the circular sequence as
\begin{equation}
    \begin{aligned}
        [V_{{\rm eff},rr}]_{\rm circ}
           &=\mathsf A_0+\varepsilon^2\mathsf A_2+\mathcal O(\varepsilon^4),\\
        [V_{{\rm eff},\theta\theta}]_{\rm circ}
           &=\mathsf B_0+\varepsilon^2\mathsf B_2+\mathcal O(\varepsilon^4),\\
        [V_{{\rm eff},r\theta}]_{\rm circ}
           &=\varepsilon\mathsf C_1+\mathcal O(\varepsilon^3).
    \end{aligned}
\end{equation}
Their even and odd orders follow from the combined reflection $\varepsilon\to-\varepsilon$, $\theta\to\pi-\theta$. Thus $\mathcal S=\mathcal S_0+\varepsilon^2\mathcal S_2+\mathcal O(\varepsilon^4)$, with
\begin{equation}
    \mathcal S_0=\mathsf A_0\mathsf B_0,
    \qquad
    \mathcal S_2=\mathsf A_2\mathsf B_0+\mathsf A_0\mathsf B_2
                          -\mathsf C_1^2.
\end{equation}
In particular, defining $D_c\equiv r^2-3Mr+2a\sqrt{Mr}$, the Kerr coefficient is
\begin{equation}
    \mathcal S_0(r)=\frac{4M^2}{r^2\Delta D_c^2}
       \left(r^2-6Mr+8a\sqrt{Mr}-3a^2\right)
       \left(r^2-4a\sqrt{Mr}+3a^2\right).
\end{equation}
The term $-\mathsf C_1^2$ retains the radial--polar coupling at second order. Appendix~\ref{app:isco-coefficients} gives explicit differential expressions for all the Hessian coefficients, including the energy and angular-momentum corrections. For a simple Kerr root, the resulting even ISCO shift is
\begin{equation}
    r_{\rm ISCO}=r_{\rm ISCO}^{\rm K}
       -\varepsilon^2\frac{\mathcal S_2(r_{\rm ISCO}^{\rm K})}
                            {\mathcal S_0'(r_{\rm ISCO}^{\rm K})}
       +\mathcal O(\varepsilon^4).
\end{equation}
Here the prime is a total radial derivative along the Kerr circular sequence. In the numerical construction, we solve the marginal-stability equation truncated through second order, selecting the physical boundary of the stable branch connected to Kerr.

Having specified the conservative orbit, we introduce the radiation prescription. We adopt the leading mass-quadrupole formula \cite{Einstein:1918btx,Thorne:1980ru,Peters:1963ux}. To evaluate the multipoles, we use the Cartesian coordinate identification of the kludge prescription \cite{Babak:2006uv},
\begin{equation}
    \boldsymbol X(t)=\left(
        r\sin\theta_a\cos\Phi,\,
        r\sin\theta_a\sin\Phi,\,
        r\cos\theta_a\right).
    \label{eq:waveform-cartesian-trajectory}
\end{equation}
Here $\Phi$ is the orbital phase. Within this coordinate mapping, each instantaneous orbit is a circle of cylindrical radius $R=r\sin\theta_a$ at height $Z=r\cos\theta_a$. The time-dependent components of its mass quadrupole $I_{ij}=mX_iX_j$ satisfy
\begin{equation}
    \begin{aligned}
        I_{XX}-I_{YY}&=mR^2\cos(2\Phi),
        &2I_{XY}&=mR^2\sin(2\Phi),\\
        I_{XZ}&=mRZ\cos\Phi,
        &I_{YZ}&=mRZ\sin\Phi.
    \end{aligned}
    \label{eq:quadrupole-harmonic-content}
\end{equation}
At fixed radius, $R$ is even and $Z$ is odd in $\varepsilon$. Together with the even angular velocity, this makes the signed first-harmonic amplitudes odd and the second-harmonic amplitudes even within this radiation prescription. This kinematic result requires the circular trajectory and mass-quadrupole approximation, but not the particular conformal factor. With $I_{ij}^{(3)}\equiv\D^3I_{ij}/\D t^3$, the positive outgoing energy flux per unit small-body mass is
\begin{equation}
    \mathcal F
      =\frac{1}{5m}\left\langle
         I_{ij}^{(3)}I^{ij(3)}-\frac13\bigl(I_k{}^{k(3)}\bigr)^2
        \right\rangle.
    \label{eq:quadrupole-energy-flux}
\end{equation}
Spatial indices are contracted with the Euclidean metric, and brackets denote an orbital average. The trace subtraction is equivalent to using the symmetric trace-free quadrupole $Q_{ij}=I_{ij}-\delta_{ij}I_k{}^k/3$. The flux $\mathcal F$ is radiated energy per unit small-body mass and coordinate time $t$. Overdots on the geodesic variables continue to denote proper-time derivatives. At leading adiabatic order, the multipole time derivatives are evaluated on an instantaneous circular orbit, holding $r$, $\theta_a$, and $\omega$ fixed. The slow drift of the orbit is restored through energy balance below. This is a kludge radiation prescription rather than a calculation of black-hole perturbations in the full modified theory.

\numrev{For the off-equatorial circular trajectory, the radiated power separates into contributions from the waveform harmonics at $\omega$ and $2\omega$:}
\begin{equation}
    \mathcal F=\mathcal F_1+\mathcal F_2,
    \qquad
    \mathcal F_1=\frac25mr^4\omega^6\sin^2\theta_a\cos^2\theta_a,
    \qquad
    \mathcal F_2=\frac{32}{5}mr^4\omega^6\sin^4\theta_a.
    \label{eq:harmonic-energy-flux}
\end{equation}
Their ratio is
\begin{equation}
    \frac{\mathcal F_1}{\mathcal F_2}
      =\frac{\cot^2\theta_a}{16}
      =\frac{\varepsilon^2\theta_1^2}{16}
         +\mathcal O(\varepsilon^4).
    \label{eq:first-harmonic-flux-ratio}
\end{equation}
Writing the angular offset as $\delta\theta\equiv\theta_a-\pi/2=\varepsilon\theta_1+\mathcal O(\varepsilon^3)$, the first-harmonic flux is quadratic in $\delta\theta$. \numrev{The first harmonic has half the frequency of the second, and the three time derivatives in the energy-flux formula further suppress its contribution to the radiated power.} Together with the relative amplitudes of the quadrupole components, this gives the factor $1/16$ in Eq.~\eqref{eq:first-harmonic-flux-ratio}. Equivalently, $\mathcal F_1/\mathcal F_2=(Z/R)^2/16$ for any circle evaluated with this mass-quadrupole prescription; the factor is kinematic, while the height and frequency are determined by the background. Substituting the orbital expansions into both harmonics yields the total flux through second order,
\begin{equation}
    \mathcal F=\mathcal F_0
       \left[1+\varepsilon^2\left(
          6\frac{\omega_2}{\omega_0}-\frac{31}{16}\theta_1^2
       \right)\right]+\mathcal O(\varepsilon^4),
    \qquad
    \mathcal F_0\equiv\frac{32}{5}mr^4\omega_0^6.
    \label{eq:quadrupole-flux-expansion}
\end{equation}
The frequency shift contributes $6\omega_2/\omega_0$, while the geometric terms combine as $-2\theta_1^2+\theta_1^2/16=-31\theta_1^2/16$. The specific energy and flux are both even in $\varepsilon$, so their leading deformation effects on the inspiral appear at second order.

The relative size of these flux contributions can also be assessed analytically. For $\theta_1\ne0$, define $C_\omega(r)\equiv\omega_2/(\omega_0\theta_1^2)$ using Eq.~\eqref{eq:circular-orbit-coefficients}. At fixed radius,
\begin{equation}
    \frac{\mathcal F_1}{\mathcal F_0}
       =\frac{\delta\theta^2}{16}+\mathcal O(\varepsilon^4),
    \qquad
    \frac{\mathcal F_2-\mathcal F_0}{\mathcal F_0}
       =(6C_\omega-2)\delta\theta^2+\mathcal O(\varepsilon^4).
\end{equation}
Thus the first-harmonic flux is smaller than the leading correction to the second-harmonic flux when $1/16\ll|6C_\omega-2|$. For example, at $r=6M$ and $a/M=0.5$, $C_\omega\simeq3.50$, giving $\mathcal F_1/|\mathcal F_2-\mathcal F_0|\simeq3.29\times10^{-3}$ at leading order. Both flux contributions scale as $\varepsilon^2$, so their leading-order ratio at fixed radius is independent of $\varepsilon$. Its value depends on the orbital coefficients, and this hierarchy need not hold near a cancellation of $6C_\omega-2$. The first-harmonic flux is included in Eq.~\eqref{eq:quadrupole-flux-expansion}; its effect on the inspiral can also accumulate indirectly in the phase.

Finally, we adopt an adiabatic approximation in which radiation reaction slowly moves the particle along this family of stable circular geodesics. With $M$, $a$, and $\varepsilon$ held fixed, energy balance gives
\begin{equation}
    \frac{\D E}{\D t}=\frac{\D E}{\D r}\frac{\D r}{\D t}=-\mathcal F,
    \qquad
    \frac{\D r}{\D t}=-\frac{\mathcal F(r)}{\D E(r)/\D r}.
    \label{the energy-balance equation}
\end{equation}
The numerical evolution inserts the second-order energy and flux of Eqs.~\eqref{eq:circular-energy-expansion} and \eqref{eq:quadrupole-flux-expansion} into this ratio without further expanding the ratio in $\varepsilon$. This prescription agrees with a strict second-order expansion where that expansion is uniform; it does not supply the missing higher-order dynamics. In particular, a small correction to $E$ need not imply a small correction to $\D E/\D r$ near marginal stability.

Energy balance is compatible with axial angular-momentum balance on this circular sequence. At fixed background parameters, the geodesic conditions imply $\D E/\D r=\omega\,\D L_z/\D r$. Both azimuthal quadrupole harmonics have the same pattern angular velocity $\omega$, so their positive outgoing axial angular-momentum flux per unit small-body mass satisfies $\mathcal F_{L_z}=\mathcal F/\omega$. Consequently, the same radial evolution satisfies $\D L_z/\D t=-\mathcal F_{L_z}$ through the retained order.

The corresponding angular motion is
\begin{equation}
    \frac{\D\Phi}{\D t}=\omega[r(t)],
    \qquad
    \frac{\D\theta_a}{\D t}
      =\frac{\D\theta_a}{\D r}\frac{\D r}{\D t},
    \label{eq:adiabatic-angular-evolution}
\end{equation}
where $\Phi$ is the accumulated orbital phase. The second relation describes the slow drift of the equilibrium polar angle as the radius changes. The condition $\dot\theta=0$ applies to each conservative circular geodesic, while $\D\theta_a/\D t$ can be nonzero on the radiation-reaction timescale. The construction assumes that this timescale is much longer than the orbital period and that the evolution remains on the stable circular branch, with independent polar oscillations excluded. Near the ISCO the restoring force of the marginal mode vanishes, and a transition treatment is needed to continue into plunge \cite{Ori:2000transition}. We therefore terminate the modeled inspiral at $r_{\rm end}=r_{\rm ISCO}+\delta r$, with a small positive radial offset; its numerical value is a cutoff choice rather than a prediction of the transition dynamics. The energy-balance prescription alone does not resolve the transition or plunge. The resulting functions $r(t)$, $\theta_a[r(t)]$, and $\Phi(t)$ supply the trajectory used for the waveform and dephasing calculation in the next subsection.

\subsection{\texorpdfstring{\numrev{Waveform, dephasing, and detector response}}{Waveform, dephasing, and detector response}}

The adiabatic orbit obtained above specifies the instantaneous radius $r(t)$, polar angle $\theta_a[r(t)]$, and angular velocity $\omega[r(t)]$. Its accumulated orbital phase is
\begin{equation}
    \Phi(t)=\Phi_0+\int_0^t\omega[r(t')]\,\D t',
    \label{eq:waveform-orbital-phase}
\end{equation}
where $\Phi_0$ is the initial phase. For a common initial phase, the dephasing of the second GW harmonic relative to the Kerr reference is
\begin{equation}
    \Delta\Phi(t)=2\int_0^t
       \left\{\omega_\varepsilon[r_\varepsilon(t')]
              -\omega_0[r_{\rm Kerr}(t')]\right\}\,\D t'.
    \label{dephase}
\end{equation}
The two frequencies are evaluated along their respective inspiral trajectories. This quantity measures the source phase shift produced by the modified orbital evolution.

To construct the polarizations from the trajectory in Eq.~\eqref{eq:waveform-cartesian-trajectory}, choose the black-hole symmetry axis as the $Z$ axis and place the line of sight in the $X$--$Z$ plane. The viewing inclination $\iota$, distinct from the orbital polar angle $\theta_a$, is the angle between the black-hole angular-momentum direction $\boldsymbol\ell$ and the GW propagation direction $\boldsymbol n$: $\cos\iota=\boldsymbol\ell\cdot\boldsymbol n$. The direction $\boldsymbol s$ from the observer to the source is opposite to $\boldsymbol n$, so $\cos\iota=-\boldsymbol\ell\cdot\boldsymbol s$. We choose
\begin{equation}
    \boldsymbol n=(\sin\iota,0,\cos\iota),\qquad
    \boldsymbol p=(0,1,0),\qquad
    \boldsymbol q=(-\cos\iota,0,\sin\iota).
    \label{eq:transverse-basis}
\end{equation}

This transverse basis satisfies $\boldsymbol p\times\boldsymbol q=\boldsymbol n$. With $e^+_{ij}=p_ip_j-q_iq_j$ and $e^\times_{ij}=p_iq_j+q_ip_j$, the polarization tensors satisfy $n_i e_A^{ij}=0$, $\delta_{ij}e_A^{ij}=0$, and $e_A^{ij}e_{B\,ij}=2\delta_{AB}$, where $A,B\in\{+,\times\}$. Extracting $h_A=\tfrac12 e_A^{ij}h_{ij}^{\rm TT}$ from the standard far-zone transverse-traceless (TT) quadrupole field \cite{Flanagan:2005yc} therefore gives
\begin{equation}
    h_A(t)=\frac{1}{D}e_A^{ij}I_{ij}^{(2)}(t-D).
    \label{eq:waveform-tt-projection}
\end{equation}
The contraction with $e_A^{ij}$ removes the trace and longitudinal components of $I_{ij}^{(2)}$, so the TT projection is already implicit in this expression. Here $D$ is the source-observer distance. As in the flux calculation, the derivatives are evaluated on an instantaneous circular orbit with $r$, $\theta_a$, and $\omega$ held fixed; these quantities subsequently follow their adiabatic evolution. Suppressing the common constant propagation delay, the resulting waveforms are
\begin{align}
    h_+(t)
    &=\frac{2mr^2\omega^2}{D}
      \Big[\sin^2\theta_a(1+\cos^2\iota)\cos(2\Phi)
          -\sin\theta_a\cos\theta_a\sin\iota\cos\iota\cos\Phi\Big],
    \label{eq:off-equatorial-hplus}\\
    h_\times(t)
    &=\frac{2mr^2\omega^2}{D}
      \Big[2\sin^2\theta_a\cos\iota\sin(2\Phi)
          -\sin\theta_a\cos\theta_a\sin\iota\sin\Phi\Big].
    \label{eq:off-equatorial-hcross}
\end{align}
The basis in Eq.~\eqref{eq:transverse-basis} fixes the polarization signs. The first harmonic comes from the mixed quadrupole components $I_{XZ}$ and $I_{YZ}$ and is proportional to $\sin\theta_a\cos\theta_a$; it vanishes for an equatorial orbit. The second harmonic reduces to the usual circular-orbit waveform in the Kerr limit.

Using the orbital coefficients in Eq.~\eqref{eq:circular-orbit-coefficients}, define
\begin{equation}
    \mathcal A_0\equiv\frac{2mr^2\omega_0^2}{D},
    \qquad B_2\equiv2\frac{\omega_2}{\omega_0}-\theta_1^2.
    \label{eq:waveform-amplitude-coefficients}
\end{equation}
\numrev{Expanding the instantaneous amplitude factors through second order yields}
\begin{align}
    h_+
    &=\mathcal A_0
      \Big[(1+\varepsilon^2B_2)(1+\cos^2\iota)\cos(2\Phi)
          +\varepsilon\theta_1\sin\iota\cos\iota\cos\Phi\Big]
      +\mathcal O(\varepsilon^3),
    \label{eq:hplus-second-order}\\
    h_\times
    &=\mathcal A_0
      \Big[2(1+\varepsilon^2B_2)\cos\iota\sin(2\Phi)
          +\varepsilon\theta_1\sin\iota\sin\Phi\Big]
      +\mathcal O(\varepsilon^3).
    \label{eq:hcross-second-order}
\end{align}
The coefficient $B_2$ combines the frequency shift with the change in the cylindrical orbital radius $r\sin\theta_a$ about the black-hole spin axis in the Cartesian mapping of Eq.~\eqref{eq:waveform-cartesian-trajectory}. The accumulated phase $\Phi(t)$ is computed from the corrected inspiral and retained inside the trigonometric functions without a Taylor expansion, preserving the secular phase accumulation over the observation.

The odd/even amplitude ordering above concerns mass-quadrupole radiation from this circular branch. Evenness of the orbital frequency does not by itself exclude linear corrections to other radiative multipoles at $2\omega$. Moreover, unequal-mass circular binaries in GR already have first-harmonic emission from the current quadrupole, including the $(\ell,m)=(2,1)$ mode \cite{Blanchet:2008modes}. The additional first harmonic considered here is therefore not, by itself, a unique signature of equatorial reflection symmetry breaking in a complete waveform.

The suppression of the first harmonic has a geometric origin. Within the Cartesian mapping, the orbit lies at height $Z=r\cos\theta_a\simeq-r\delta\theta$, so the mixed quadrupole components $I_{XZ}$ and $I_{YZ}$ are linear in the small angular offset. For generic viewing directions, the first-to-second harmonic amplitude ratio is therefore of order $|\delta\theta|$. Moreover, Eq.~\eqref{eq:circular-orbit-coefficients} gives $\theta_1\simeq-144(a/M)(M/r)^6$ at large radius. The small displacement suppresses the direct first-harmonic amplitude and its radiated power. Comparing instantaneous amplitudes alone, however, does not justify neglecting this linear term relative to a quadratic amplitude correction.

The second-order orbital effect can become more prominent through secular phase accumulation. Over an interval $T_{\rm obs}$ in which the radius and coefficients change little, the direct frequency correction gives the scale estimate
\begin{equation}
    \Delta\Phi_\omega\simeq
       2\pi N_{\rm GW}C_\omega\delta\theta^2,
    \qquad
    N_{\rm GW}\simeq\frac{\omega_0T_{\rm obs}}{\pi},
\end{equation}
where $N_{\rm GW}$ counts cycles of the second harmonic. This estimate illustrates the cycle enhancement of the direct frequency correction; the full dephasing in Eq.~\eqref{dephase} also includes changes in the inspiral trajectory through energy balance. The instantaneous relative strain amplitude of the additional first harmonic has no analogous enhancement.

The first-harmonic energy flux $\mathcal F_1$, however, modifies the inspiral rate and thereby contributes to the accumulated phase of the dominant second harmonic. This feedback begins at $\mathcal O(\varepsilon^2)$ and is retained through the total flux in Eq.~\eqref{eq:quadrupole-flux-expansion}. Cycle enhancement therefore distinguishes accumulated phase effects from direct amplitude effects; it does not by itself determine the relative importance of the phase feedback from $\mathcal F_1$ and the other second-order corrections. That comparison requires the flux coefficients and the full inspiral evolution. The observable hierarchy also depends on viewing geometry, detector weighting, and observation duration. Both harmonics are retained in the detector strain and mismatch calculations.

This waveform prescription uses the GR radiation formula with the same $1/D$ scaling and retarded time for both polarizations. It describes the effect of the deformed background on the source trajectory, without solving the sourced perturbation equations of the complete theory or including a helicity-dependent propagation transfer function. The additional harmonic and accumulated source dephasing therefore do not by themselves demonstrate propagation birefringence \cite{Ng:2023jjt}. A separate, limited result is that the model in Eq.~\eqref{eq:exact-conformal-factor} has the same quadratic vacuum action as GR about Minkowski spacetime, as shown in Appendix~\ref{app:flat-vacuum-waves}. This result concerns linear waves in flat vacuum; radiation on the curved black-hole background and propagation on a cosmological background require their own perturbation analyses.

For one LISA-like channel with opening angle $\beta=\pi/3$, the long-wavelength detector response is
\begin{equation}
    h(t)=\frac{\sqrt3}{2}
       \bigl[F_+(t)h_+(t)+F_\times(t)h_\times(t)\bigr].
    \label{eq:detector-response}
\end{equation}
For detector-frame source angles $(\vartheta_D,\varphi_D)$ and polarization angle $\psi_D$, the antenna functions are \cite{Barack:2003fp}
\begin{equation}
    \begin{aligned}
        F_+&=\tfrac12(1+\cos^2\vartheta_D)\cos2\varphi_D\cos2\psi_D
               -\cos\vartheta_D\sin2\varphi_D\sin2\psi_D,\\
        F_\times&=\tfrac12(1+\cos^2\vartheta_D)\cos2\varphi_D\sin2\psi_D
               +\cos\vartheta_D\sin2\varphi_D\cos2\psi_D.
    \end{aligned}
    \label{eq:pattern-functions}
\end{equation}
Their time dependence follows from the detector motion relative to the fixed ecliptic source position $(\theta_s,\phi_s)$ and black-hole axis $(\theta_L,\phi_L)$ \cite{Barack:2003fp,Cutler:1997ta}. The corresponding standard angle transformations are collected in Appendix~\ref{app:detector-angles}. We retain this time-dependent antenna modulation while neglecting the Doppler time delay and finite-arm transfer effects. The resulting strain combines the modified orbital phase, harmonic amplitudes, and detector response for the waveform comparisons below.

\subsection{Mismatch and Fisher information matrix}
\label{subsec:mismatch-fisher}

We compare the detector strains using the noise-weighted mismatch, which includes the accumulated phase, relative harmonic amplitudes, and antenna modulation. For two strains $h_a(t)$ and $h_b(t)$, the inner product is
\begin{equation}
    \langle h_a|h_b\rangle
    =4\,\mathrm{Re}
      \int_{f_{\min}}^{f_{\max}}
      \frac{\tilde h_a(f)\tilde h_b^*(f)}
        {S_n(f)}\,\D f,
    \label{eq:noise-weighted-inner-product}
\end{equation}
where a tilde denotes the Fourier transform and $S_n(f)$ is the one-sided noise power spectral density of the adopted detector channel \cite{LISA:2017pwj}. The strains are evaluated over a common observation interval with the same sampling, windowing, and frequency-band prescriptions. The normalized overlap and mismatch are
\begin{equation}
    \mathcal O(h_a,h_b)
    =\frac{\langle h_a|h_b\rangle}
    {\sqrt{\langle h_a|h_a\rangle\langle h_b|h_b\rangle}},
    \label{eq:normalized-overlap}
\end{equation}
and
\begin{equation}
    \mathcal M(h_a,h_b)
    =1-\mathcal O(h_a,h_b).
    \label{eq:waveform-mismatch}
\end{equation}
Here $h_a=h_\varepsilon$ is the deformed signal of Eq.~\eqref{eq:detector-response}, and $h_b=h_{\mathrm{Kerr}}$ is its $\varepsilon=0$ reference. We align the initial orbital phases and apply the prescribed orbital-data convention consistently to both inspirals. No relative time or phase shift is optimized, and the Kerr parameters are held fixed. Consequently, $-1\leq\mathcal O\leq1$ and $0\leq\mathcal M\leq2$; values above unity correspond to a negative overlap. This fixed-parameter comparison measures waveform differences before allowing a Kerr template to adjust its parameters. The normalization removes any overall positive amplitude rescaling of either strain.

Two auxiliary comparisons distinguish the mechanisms. Let $h_1$ and $h_2$ denote the first- and second-harmonic components of the deformed strain. Define $\mathcal M_{\rm first}\equiv\mathcal M(h_2+h_1,h_2)$, using the same trajectory and phase in both signals. The total energy flux, including $\mathcal F_1$, is kept in this trajectory. For weak $h_1$ approximately orthogonal to $h_2$ in the noise-weighted inner product, $\mathcal M_{\rm first}\simeq\langle h_1|h_1\rangle/(2\langle h_2|h_2\rangle)=\mathcal O(\varepsilon^2\theta_1^2)$ for generic viewing geometry.

Define $\mathcal M_{\rm phase}$ by comparing the Kerr second-harmonic strain with a signal having the same amplitude factors and detector response but the phase shift $\Delta\Phi(t)$ of Eq.~\eqref{dephase}. For small phase differences, $\mathcal M_{\rm phase}=\mathcal O[(\Delta\Phi)^2]$. The full $\Delta\Phi$ includes all retained frequency and energy-balance corrections, including the phase feedback from $\mathcal F_1$. Thus ``first'' and ``phase'' label direct strain and phase effects, not the effects of $\mathcal F_1$ and $\mathcal F_2$, respectively.

Using only the direct-frequency-shift estimate $\Delta\Phi_\omega$ above gives an auxiliary phase mismatch $\mathcal M_{\rm phase,\omega}$, with
\begin{equation}
    \frac{\mathcal M_{\rm phase,\omega}}{\mathcal M_{\rm first}}
    \propto\left(2\pi N_{\rm GW}C_\omega\delta\theta\right)^2.
\end{equation}
The proportionality factor depends on viewing geometry, detector noise, and time weighting. The factor $N_{\rm GW}^2$ can make the $\mathcal O(\varepsilon^4)$ phase contribution exceed the $\mathcal O(\varepsilon^2)$ direct harmonic contribution at finite small deformation. Neither the powers of $\varepsilon$ alone nor the energy-flux factor $1/16$ fixes the noise-weighted mismatch hierarchy.
These diagnostics do not sum to the full mismatch, which also includes second-harmonic amplitude changes and possible cross terms. The numerical results use the full strains and total energy flux.

The optimal signal-to-noise ratio (SNR) is
\begin{equation}
    \rho=\sqrt{\langle h|h\rangle}.
    \label{eq:optimal-snr}
\end{equation}
For two signals rescaled to the same SNR, their residual $\Delta h=h_a-h_b$ satisfies
\begin{equation}
    \langle\Delta h|\Delta h\rangle
    =2\rho^2\mathcal M.
    \label{eq:mismatch-residual}
\end{equation}

The unit-residual condition $\langle\Delta h|\Delta h\rangle\sim1$ therefore corresponds to the characteristic mismatch scale
\begin{equation}
    \mathcal M_{\mathrm{unit}}
    \sim\frac{1}{2\rho^2}.
    \label{eq:mismatch-threshold}
\end{equation}
This unit-residual scale for equal-SNR signals \cite{Flanagan:1997kp,Lindblom:2008cm} differs from joint parameter-estimation criteria. In the local Gaussian approximation, a joint confidence region involving $N$ identifiable parameter directions is characterized by a chi-square quantile $q_N(p)$, defined by $\Pr(\chi_N^2\leq q_N(p))=p$. With a consistent treatment of normalization and nuisance-parameter fitting, the corresponding small-difference mismatch scale is $q_N(p)/(2\rho^2)$ \cite{Baird:2012confidence}; the frequently used $N/(2\rho^2)$ instead describes a typical multivariate statistical scale \cite{Chatziioannou:2017faithfulness}. The relevant dimension depends on the parameter subspace and fitting procedure, and is not determined solely by the number of parameters varied in a waveform scan.

For the numerical comparisons, we choose $\mathcal M=0.01$ as a one-percent overlap-loss benchmark. \numrev{For the prescribed Kerr template with fixed phase, the signed expected matched-filter SNR is} \cite{Lindblom:2008cm}
\begin{equation}
    \rho_{\rm rec}
    =\frac{\langle h_\varepsilon|h_{\mathrm{Kerr}}\rangle}
          {\sqrt{\langle h_{\mathrm{Kerr}}|h_{\mathrm{Kerr}}\rangle}}
    =\rho_\varepsilon(1-\mathcal M),
\end{equation}
where $\rho_\varepsilon$ is the optimal SNR of the deformed signal. Thus, this comparison level corresponds to an overlap of $0.99$ and a 1\% reduction in the expected matched-filter SNR. \numrev{This benchmark is chosen independently of Eq.~\eqref{eq:mismatch-threshold}.} A statistical interpretation requires specifying the signal SNR, the relevant parameter space, the confidence level, and the template-fitting procedure, including adjustment of the Kerr parameters and relative time and phase.

We also consider the local sensitivity to $\varepsilon$ using the Fisher matrix
\begin{equation}
    \Gamma_{ab}
    =
    \left\langle
        \frac{\partial h}{\partial\lambda^a}
        \,\middle|\,
        \frac{\partial h}{\partial\lambda^b}
    \right\rangle,
    \label{eq:fisher-information-matrix}
\end{equation}
evaluated at a fiducial set of independent parameters $\boldsymbol\lambda_0$. This forecast assumes stationary Gaussian noise and a locally linear waveform dependence, with negligible prior information. When only $\varepsilon$ is varied, the relevant element is
\begin{equation}
    \Gamma_{\varepsilon\varepsilon}
    =
    4\int_{f_{\min}}^{f_{\max}}
    \frac{
        \left|
        \partial_\varepsilon\tilde h(f)
        \right|^2
       }{S_n(f)}\,\D f,
    \label{eq:fisher-epsilon}
\end{equation}
and the conditional uncertainty is
\begin{equation}
    \sigma_{\varepsilon,\mathrm{cond}}
    =\Gamma_{\varepsilon\varepsilon}^{-1/2}.
    \label{eq:conditional-epsilon-error}
\end{equation}
\numrev{For $\varepsilon\neq0$, the corresponding fractional uncertainty is $\sigma_{\varepsilon,\mathrm{cond}}/|\varepsilon|$.} If other parameters are inferred simultaneously, their correlations must be included, and the marginalized uncertainty becomes
\begin{equation}
    \sigma_{\varepsilon,\mathrm{marg}}
    =\sqrt{(\Gamma^{-1})^{\varepsilon\varepsilon}},
    \label{eq:fisher-marginalized-error}
\end{equation}
provided the full Fisher matrix is nonsingular and the local Gaussian approximation is valid. The single-parameter estimate therefore describes sensitivity with the other parameters known, rather than the precision of a full EMRI parameter inference.

Unlike the normalized mismatch, the Fisher uncertainty depends on the physical strain amplitude. At fixed intrinsic parameters, source orientation, and observation duration, $h\propto D^{-1}$ implies $\Gamma_{\varepsilon\varepsilon}\propto D^{-2}$ and $\sigma_{\varepsilon,\mathrm{cond}}\propto D\propto\rho^{-1}$. Absolute uncertainties must therefore be specified together with the adopted source distance or SNR and detector noise model.

The derivative $\partial_\varepsilon h$ must include the dependence of the orbital trajectory, accumulated phase, and polarization amplitudes on $\varepsilon$. In a finite-difference evaluation, the perturbed inspirals must be regenerated with a consistent orbital initial- or boundary-data prescription, and their strains compared on a common time grid and observation interval. The estimate requires convergence checks with respect to the parameter step and waveform sampling. The noise curve and Fourier frequencies must also be expressed in consistent physical units.

The fiducial deformation matters for this local forecast. The second-order orbital corrections give an even phase dependence on $\varepsilon$, so that $\partial_\varepsilon\Phi$ vanishes at the Kerr limit. The first harmonic, however, has an amplitude linear in $\varepsilon$ and can contribute to $\partial_\varepsilon h$ even there. Thus, its small contribution to the radiated power does not by itself justify omitting it from a Fisher derivative near $\varepsilon=0$. \numrev{A phase-dominated measurement primarily constrains $|\varepsilon|$; a local Fisher width around one nonzero value does not resolve a possible $\varepsilon\leftrightarrow-\varepsilon$ ambiguity.} A forecast at nonzero $\varepsilon$ should not be identified directly with an upper limit obtained for a Kerr signal.


\section{Numerical results}
\label{sec4}

We now examine how the parity-violating conformal deformation affects the inspiral dynamics, accumulated phase, polarization waveforms, and detector strain. Our calculation retains corrections through second order in $\varepsilon$, including the displacement of the circular orbit away from the equatorial plane, the orbital-frequency shift, and the conformal-factor contribution to the orbital energy, together with their effects on the quadrupole flux and adiabatic evolution. The additional first harmonic generated by the off-equatorial motion is also retained in the waveform. We compare the conformal Kerr inspirals with the $\varepsilon=0$ reference using the adiabatic quadrupole model of Sec.~\ref{sec3}. The mismatch characterizes the difference between prescribed signals, while the single-parameter Fisher analysis estimates the local sensitivity to $\varepsilon$.

We take $m=10M_\odot$ and $M=10^6M_\odot$, corresponding to a mass ratio $\mu=10^{-5}$, and vary $a/M$ and $\varepsilon$ as indicated in the figures. The dephasing, polarization, mismatch, and Fisher calculations use a one-year inspiral ending just outside the adopted stability boundary. For each value of $\varepsilon$, including the Kerr reference, we impose
\begin{equation}
    r_\varepsilon(T_{\rm obs})
       =r_{\rm ISCO}(a,\varepsilon)+0.07M,
    \qquad \Phi_\varepsilon(0)=0,
    \qquad T_{\rm obs}=1\,{\rm yr}.
    \label{eq:numerical-terminal-data}
\end{equation}
Numerically, the adopted duration is $T_{\rm obs}/M\simeq6.41\times10^6$. The radius denoted by $r_{\rm ISCO}$ in this numerical prescription is the inner boundary of the stable branch reached from large radius, obtained from the marginal-stability equation truncated through second order. We integrate the radial energy-balance equation backward from the prescribed endpoint to determine the initial radius $r_{\rm ini}(a,\varepsilon)$. Equivalently, on the stable inward-evolving branch it satisfies
\begin{equation}
    T_{\rm obs}
       =\int_{r_{\rm ISCO}(a,\varepsilon)+0.07M}^{r_{\rm ini}(a,\varepsilon)}
          \frac{\D E(r;a,\varepsilon)/\D r}
               {\mathcal F(r;a,\varepsilon)}\,\D r.
    \label{eq:numerical-initial-radius}
\end{equation}
The initial polar angle and angular velocity are $\theta_a[r_{\rm ini}(a,\varepsilon)]$ and $\omega[r_{\rm ini}(a,\varepsilon)]$, respectively, and the initial radial drift is fixed by Eq.~\eqref{the energy-balance equation}. Thus, the initial orbit is specified by the circular branch and the terminal condition. Independent polar oscillations and eccentricity are excluded.

The Kerr reference is constructed separately with $\varepsilon=0$ and the same mass and spin. The two signals terminate at the same time and at the same offset from their respective ISCOs; their coordinate endpoint radii, as well as their initial radii, generally differ. Their initial orbital phases are aligned by $\Phi(0)=0$; the terminal phases are not aligned. The comparison therefore describes the year before each signal's own near-ISCO endpoint and includes this boundary-data convention as well as the modified dynamics. The positive offset is a numerical cutoff choice; the transition and plunge are not modeled. For the central finite-difference Fisher derivative, each perturbed waveform is regenerated with the same one-year terminal prescription and interpolated onto a common time grid over the shared waveform interval. The initial radius is consequently a derived, $\varepsilon$-dependent quantity; the forecast is conditional on this terminal-data prescription.

For the polarization and detector-strain calculations, we adopt $D=1\,\mathrm{Gpc}$ and $\iota=2\pi/3$. The fixed ecliptic sky and black-hole-axis angles and the initial detector phases are
\begin{equation}
    \begin{aligned}
        (\theta_s,\phi_s)&=(\pi/2,\pi/2),&
        (\theta_L,\phi_L)&=(\pi/4,\pi/4),\\
        \bar\phi_0&=0,&\bar\alpha_0&=0.
    \end{aligned}
    \label{eq:numerical-viewing-data}
\end{equation}
These angles imply $\cos\iota=-\boldsymbol\ell\cdot\boldsymbol s=-1/2$, consistent with $\iota=2\pi/3$. The waveform uses the circular-branch polar angle $\theta_a[r(t)]=\pi/2+\varepsilon\theta_1[r(t)]+\mathcal O(\varepsilon^3)$. The detector arm-opening angle is $\beta=\pi/3$.
Cosmological redshift is neglected in this illustrative asymptotically flat waveform model; $D$ sets the strain amplitude.

The physical-unit conversion, noise fit, windowing, and sampling requirements used for the Fourier-domain comparisons are specified in Appendix~\ref{app:numerical-conventions}.


\subsection{Gravitational-wave dephasing}
\label{subsec:results-dephasing}

\begin{figure}[!tp]
    \centering
    \begin{subfigure}{0.45\textwidth}
        \centering
        \includegraphics[width=\textwidth]{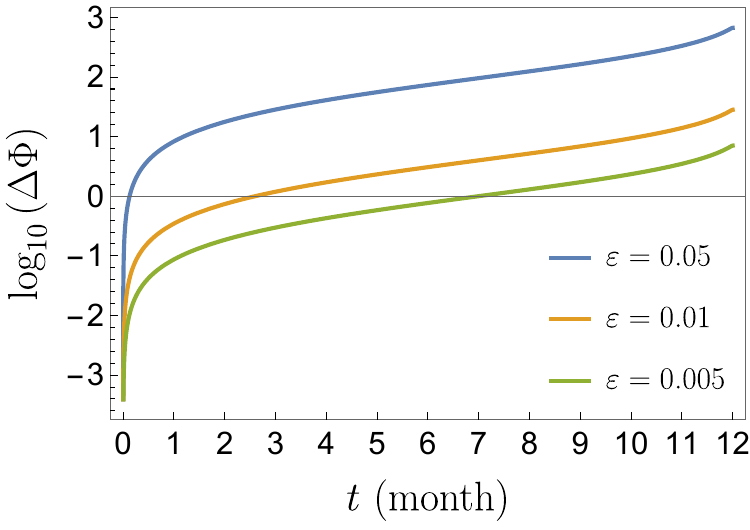}
        \caption{$a/M=0.9$}
        \label{fig:1a}
    \end{subfigure}
    \hfill
    \begin{subfigure}{0.45\textwidth}
        \centering
        \includegraphics[width=\textwidth]{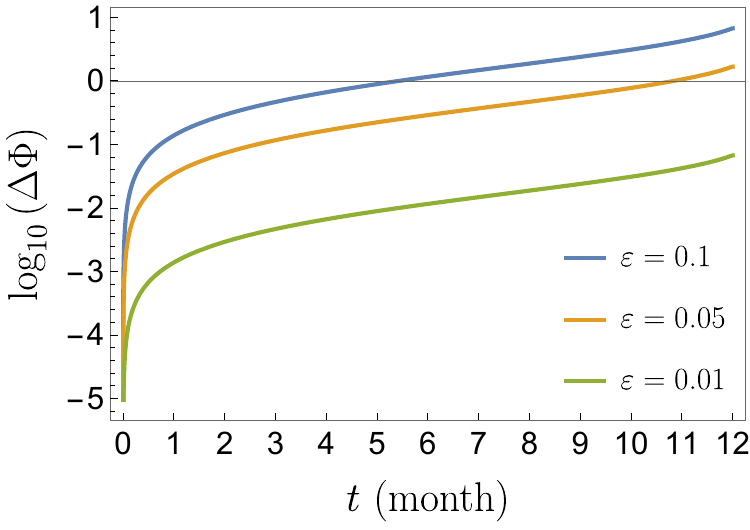}
        \caption{$a/M=0.6$}
        \label{fig:1b}
    \end{subfigure}
    \hfill
    \begin{subfigure}{0.45\textwidth}
        \centering
        \includegraphics[width=\textwidth]{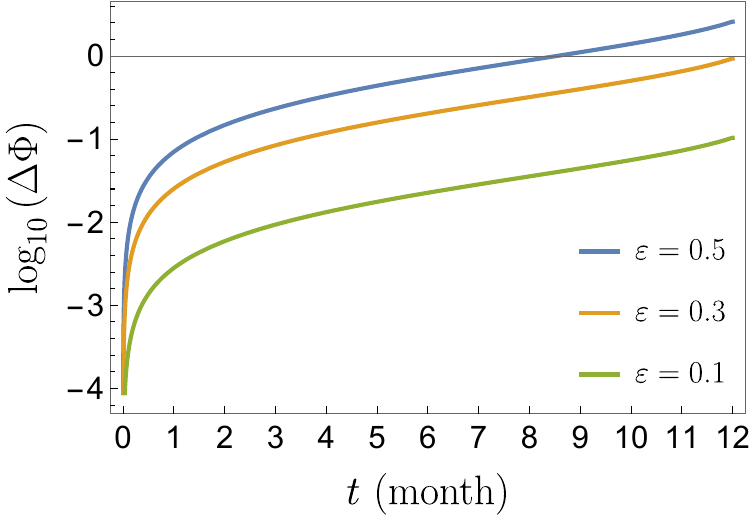}
        \caption{$a/M=0.3$}
        \label{fig:1c}
    \end{subfigure}
    \caption{Magnitude of the accumulated second-harmonic phase difference between the deformed and Kerr inspirals, with aligned initial phases. Panels (a), (b), and (c) correspond to $a/M=0.9$, $0.6$, and $0.3$, respectively; the deformations are given in the legends. The vertical axis represents $\log_{10}|\Delta\Phi|$, with the phase difference in radians. The horizontal line marks the one-radian comparison benchmark.}
    \label{fig:dephasing}
\end{figure}

Figure~\ref{fig:dephasing} shows the magnitude of the second-harmonic phase difference, $|\Delta\Phi|$, defined by Eq.~\eqref{dephase}. Each frequency is evaluated along its corresponding inspiral, so the phase difference includes both the local frequency correction and the change in the radial evolution. Taking the magnitude removes the sign convention associated with the ordering of the two signals. The one-radian line provides a reference scale for the accumulated phase difference; assessing its observational significance requires a noise-weighted waveform comparison, as discussed in Sec.~\ref{subsec:results-mismatch}.

The two phase histories use the terminal-data prescription specified at the beginning of this section. In particular, the plotted dephasing compares inspirals starting at their separately determined initial radii, with $\Delta\Phi(0)=0$.


For $a/M=0.9$, the one-radian level is reached at approximately $0.1$, $2.6$, and $7.0$ months for $\varepsilon=0.05$, $0.01$, and $0.005$, respectively. At $a/M=0.6$, the crossings occur much later: approximately $5.4$ months for $\varepsilon=0.1$ and $10.8$ months for $\varepsilon=0.05$, while $\varepsilon=0.01$ remains below the reference throughout the year. In particular, the same deformation $\varepsilon=0.05$ reaches the benchmark much earlier at $a/M=0.9$ than at $a/M=0.6$.

At $a/M=0.3$, the $\varepsilon=0.5$ curve reaches the benchmark at approximately $8.5$ months, while $\varepsilon=0.3$ approaches it but remains slightly below it at the end of the year; $\varepsilon=0.1$ also remains below it. These larger fiducial deformations require particular care with the perturbative approximation: the conformal-factor and orbital corrections must remain small over the entire inspiral, as discussed in Sec.~\ref{sec2}. Their plotted phase shifts alone do not establish the accuracy of the truncation.

The phase effect begins at $\mathcal O(\varepsilon^2)$ because both the circular frequency and the leading inspiral correction are even in $\varepsilon$. Small changes in the instantaneous frequency and inspiral rate nevertheless accumulate over many orbital cycles. The stronger effect at high spin is consistent with the prograde stable branch extending to smaller radii, where the curvature-dependent deformation and orbital response become more pronounced. Within the configurations shown, rapid rotation therefore enhances the accumulated source phase shift. This is a property of the prescribed inspirals; assigning a detection significance also requires detector noise and parameter fitting. The plotted phases describe the leading deformation effects within the second-order orbital model; establishing their quantitative accuracy additionally requires control of accumulated errors from omitted higher-order orbital terms, particularly at larger spins and deformations.


\subsection{Polarization waveforms}
\label{subsec:results-polarizations}

\begin{figure}[!tp]
    \centering
    \begin{subfigure}{0.45\textwidth}
        \centering
        \includegraphics[width=\textwidth]{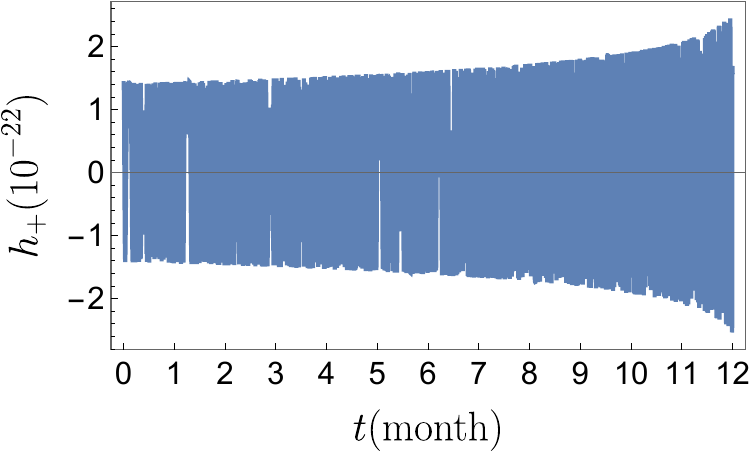}
        \caption{$a/M=0.6$, $\varepsilon=0.05$}
        \label{fig:2a}
    \end{subfigure}
    \hfill
    \begin{subfigure}{0.45\textwidth}
        \centering
        \includegraphics[width=\textwidth]{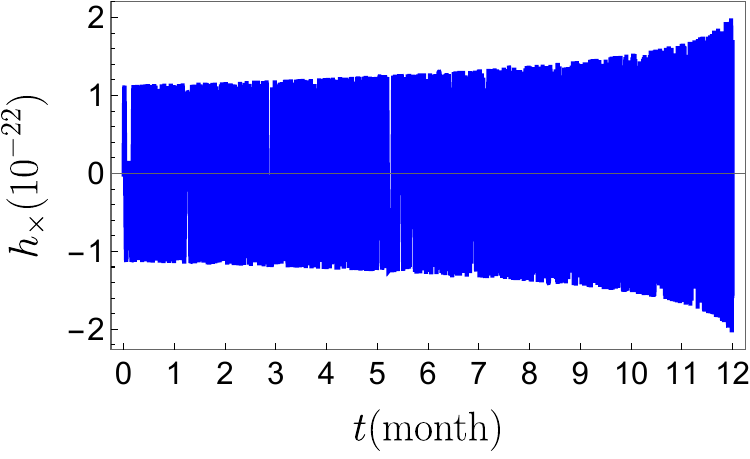}
        \caption{$a/M=0.6$, $\varepsilon=0.05$}
        \label{fig:2b}
    \end{subfigure}
    \hfill
    \begin{subfigure}{0.45\textwidth}
        \centering
        \includegraphics[width=\textwidth]{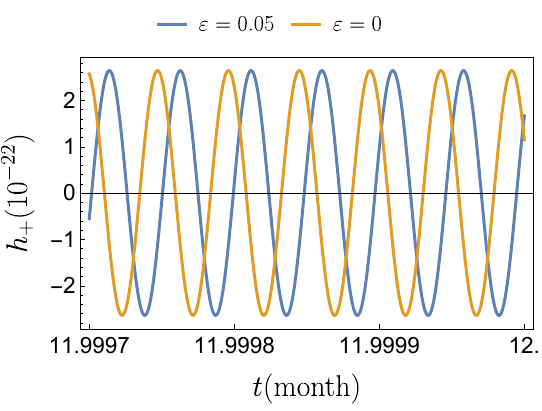}
        \caption{$a/M=0.6$}
        \label{fig:2c}
    \end{subfigure}
    \hfill
    \begin{subfigure}{0.45\textwidth}
        \centering
        \includegraphics[width=\textwidth]{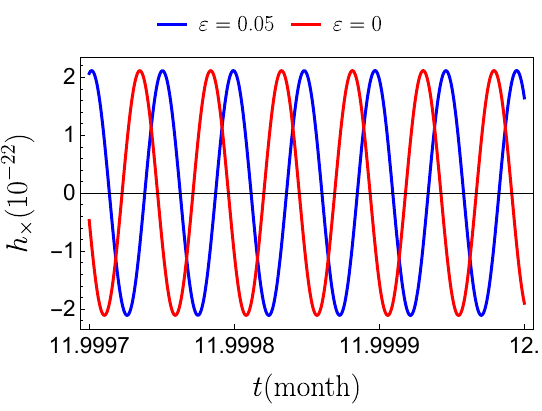}
        \caption{$a/M=0.6$}
        \label{fig:2d}
    \end{subfigure}
    \caption{\rev{Polarization waveforms for $a/M=0.6$. Panels (a) and (b) show the plus and cross polarizations over the modeled year for $\varepsilon=0.05$. Panels (c) and (d) compare the corresponding late-time polarizations with the Kerr signal, $\varepsilon=0$, over $t\in[11.9997,12]$ months. The enlarged views show a phase offset with similar peak amplitudes at the displayed resolution.}}
    \label{fig:polarizations}
\end{figure}

Figure~\ref{fig:polarizations} illustrates how the accumulated phase shift appears in the polarizations for $a/M=0.6$ and $\varepsilon=0.05$, using $\iota=2\pi/3$, $D=1\,\mathrm{Gpc}$, and the zero initial orbital phase specified above. The displayed strains are in units of $10^{-22}$. Panels (a) and (b) show the full-year plus and cross polarizations, respectively; panels (c) and (d) compare their late-time oscillations with the Kerr reference over $t\in[11.9997,12]$ months.

Both full-year signals have a growing amplitude envelope as the small body moves inward, consistent with the quadrupole amplitude scale $mr^2\omega^2/D$. The dense oscillations obscure individual cycles on this timescale, whereas the enlarged views reveal a clear phase offset in both polarizations. The similar peak amplitudes suggest that the visible difference is dominated by the accumulated phase offset, although the model also includes amplitude corrections.

The off-equatorial orbit contributes a first harmonic, but these time-domain plots do not isolate it. Its small direct contribution in the configurations studied, discussed in Sec.~\ref{sec3}, supports examining the accumulated second-order orbital effects. The polarization comparison illustrates emission within the adopted quadrupole prescription; it does not test a helicity-dependent propagation law. The absolute strain scale depends on the source distance and viewing geometry, while the phase offset and normalized waveform comparison are insensitive to a common distance rescaling.

\FloatBarrier

\subsection{Mismatch}
\label{subsec:results-mismatch}

We next compare the detector strains through the noise-weighted mismatch of Eq.~\eqref{eq:waveform-mismatch}. As specified in Sec.~\ref{subsec:mismatch-fisher}, the initial phases are aligned, with no optimization over relative time, phase, or Kerr parameters. Each pair of one-year inspirals is generated once using the \numrev{terminal-data prescription} above and sampled over their common integration interval. The duration-dependent mismatches use successive initial segments of these same signals; the initial radii are kept at their one-year values when the segment is shortened. The parameter map uses the full one-year signals. We use the one-percent overlap-loss benchmark $\mathcal M=0.01$ defined in Sec.~\ref{subsec:mismatch-fisher}.

\begin{figure}[!tp]
    \centering
    \includegraphics[width=0.6\textwidth]{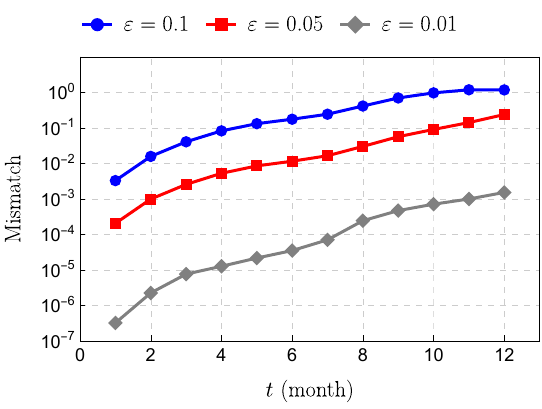}
    \caption{\rev{Mismatch as a function of observation duration for $a/M=0.6$ and the deformations indicated in the legend. Each strain is compared with its Kerr reference without optimizing time, phase, or source parameters. The $\mathcal M=0.01$ level marks the one-percent overlap-loss benchmark defined in Sec.~\ref{subsec:mismatch-fisher}.}}
    \label{fig:mismatch-time}
\end{figure}

Figure~\ref{fig:mismatch-time} shows an overall increase in mismatch as the observation is extended. The $\varepsilon=0.1$ case exceeds the benchmark between the first and second months, and $\varepsilon=0.05$ crosses it between the fifth and sixth months; $\varepsilon=0.01$ remains below it throughout the year. The final mismatches are approximately $1.2$, $0.24$, and $1.5\times10^{-3}$, respectively. The largest deformation reaches a mismatch of order unity near ten months and varies slightly thereafter. Such behavior is allowed by the fixed-phase definition: the overlap can become negative, and the mismatch need not grow monotonically with duration.

A fixed-radius example illustrates why the powers of $\varepsilon$ alone do not determine the hierarchy between $\mathcal M_{\rm phase,\omega}\propto\varepsilon^4N_{\rm GW}^2$ and $\mathcal M_{\rm first}$, defined in Sec.~\ref{subsec:mismatch-fisher}. 
For illustration, hold the radius fixed at $r=6M$, with $M=10^6M_\odot$, $a/M=0.6$, $\varepsilon=0.01$, and $T_{\rm obs}=1\,\mathrm{yr}$. Compare the deformed and Kerr plus polarizations at $\iota=2\pi/3$ using constant noise weighting, aligned initial phases, and no detector modulation. Equation~\eqref{eq:circular-orbit-coefficients} gives $C_\omega\simeq3.42$, $N_{\rm GW}\simeq1.33\times10^5$, $\delta\theta\simeq-1.24\times10^{-5}$, and $\Delta\Phi_\omega(T_{\rm obs})\simeq4.42\times10^{-4}$. The frequency shift gives $\Delta\Phi_\omega(t)=\Delta\Phi_\omega(T_{\rm obs})t/T_{\rm obs}$, while the first-to-second harmonic amplitude ratio is $\sqrt3|\delta\theta|/5$ from Eq.~\eqref{eq:hplus-second-order}. 
Compare equal-amplitude signals $A\cos\phi(t)$ and $A\cos[\phi(t)+\Delta\Phi_\omega(t)]$, where $\phi$ is the Kerr second-harmonic phase. With constant amplitude and noise weighting, averaging over many cycles gives $\mathcal O\simeq\langle\cos\Delta\Phi_\omega\rangle_t$, where $\langle f\rangle_t\equiv T_{\rm obs}^{-1}\int_0^{T_{\rm obs}}f(t)\,\D t$.

\begin{samepage}
For $|\Delta\Phi_\omega|\ll1$, the expansion $1-\cos x\simeq x^2/2$ gives half the mean squared phase difference, without phase optimization. Using its linear growth yields
\begin{equation}
    \begin{aligned}
        \mathcal M_{\rm phase,\omega}
        &\simeq\frac{1}{2T_{\rm obs}}
          \int_0^{T_{\rm obs}}[\Delta\Phi_\omega(t)]^2\,\D t
         =\frac{[\Delta\Phi_\omega(T_{\rm obs})]^2}{6}
         \simeq3.27\times10^{-8},\\
        \mathcal M_{\rm first}
        &\simeq\frac12\left(\frac{\sqrt3\,\delta\theta}{5}\right)^2
         \simeq9.27\times10^{-12}.
    \end{aligned}
\end{equation}
\end{samepage}
Here cycle enhancement makes $\mathcal M_{\rm phase,\omega}$ about $3.5\times10^3$ times larger than $\mathcal M_{\rm first}$ despite its higher power of $\varepsilon$. Figure~\ref{fig:mismatch-time} includes the full inspiral, detector response, and noise weighting.

\begin{figure}[!tp]
    \centering
    \begin{subfigure}[t]{0.49\textwidth}
        \centering
        \includegraphics[width=\textwidth]{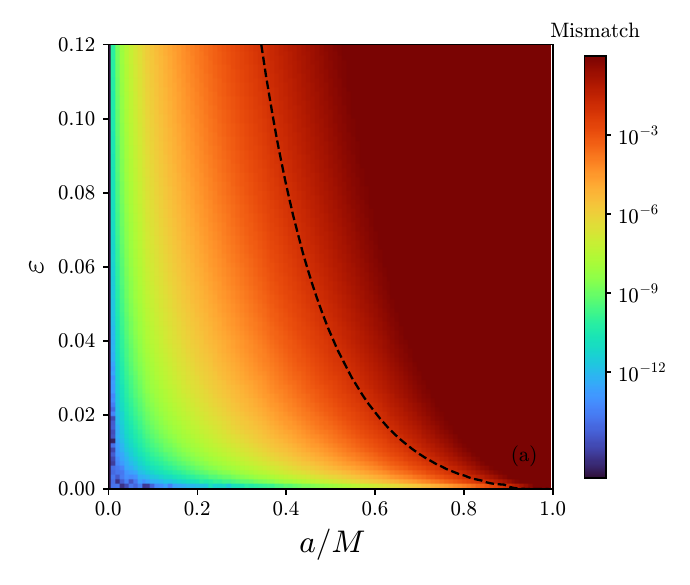}
        \phantomsubcaption
        \label{fig:4a}
    \end{subfigure}
    \hfill
    \begin{subfigure}[t]{0.49\textwidth}
        \centering
        \includegraphics[width=\textwidth]{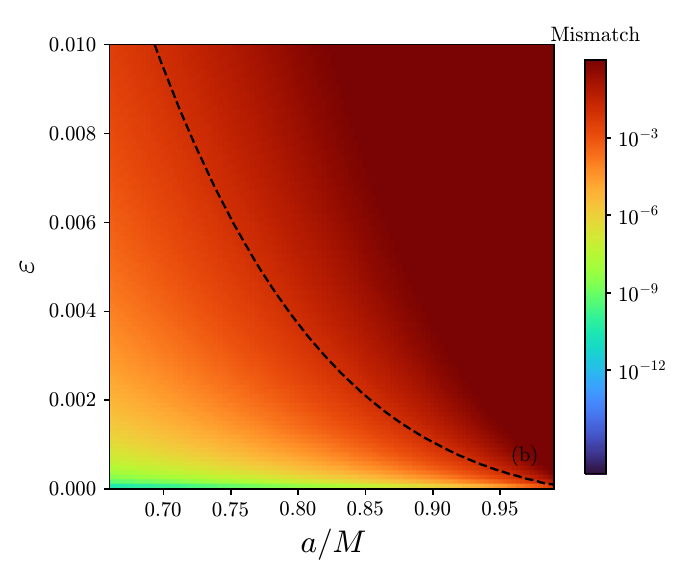}
        \phantomsubcaption
        \label{fig:4b}
    \end{subfigure}
    \caption{\numrev{Fixed-parameter mismatch after a one-year observation in the $(a/M,\varepsilon)$ plane, with $\iota=2\pi/3$ and the initial and terminal data of Eqs.~\eqref{eq:numerical-terminal-data}--\eqref{eq:numerical-viewing-data}. Panel (a) shows the full region considered, and panel (b) enlarges the high-spin, small-deformation region with a finer sampling grid. The color scale is logarithmic in mismatch. The dashed contour marks $\mathcal M=0.01$; the region toward larger spins and deformations exceeds this comparison level. No optimization over time, phase, or Kerr parameters is performed.}}
    \label{fig:mismatch-map}
\end{figure}

The one-year contour in Fig.~\ref{fig:mismatch-map} shows that larger spins allow smaller deformations to reach the same waveform-difference benchmark. It passes through approximately $\varepsilon=0.048$ at $a/M=0.48$ and $\varepsilon=0.021$ at $a/M=0.6$. In the enlarged high-spin region, the corresponding values are approximately $3.7\times10^{-3}$, $1.1\times10^{-3}$, and $4.1\times10^{-4}$ at $a/M=0.8$, $0.9$, and $0.95$, respectively. The small-deformation crossings are read from panel (b): the coarser grid in panel (a) limits the resolution of its interpolated contour close to $\varepsilon=0$.

These contour readings characterize the prescribed one-year comparison. Their statistical interpretation requires parameter fitting and a specified SNR, while their numerical precision requires the resolution checks of Eq.~\eqref{eq:numerical-sampling-condition}.

\FloatBarrier

\subsection{Uncertainty in the parity-violating parameter}
\label{subsec:results-fisher}

\begin{figure}[!tp]
	\centering
	\includegraphics[width=0.72\textwidth]{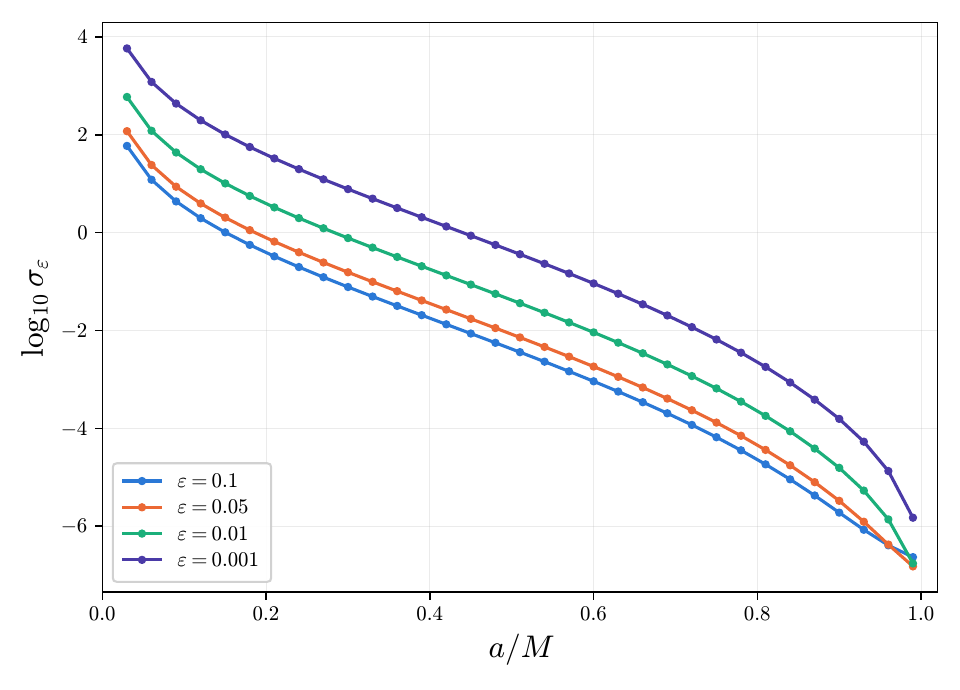}
	\caption{\numrev{Conditional Fisher uncertainty versus $a/M$ for four fiducial deformations, $\varepsilon=0.1$, $0.05$, $0.01$, and $0.001$, shown together in a single plot. The signals use $T_{\rm obs}=1\,\mathrm{yr}$, $D=1\,\mathrm{Gpc}$, $\iota=2\pi/3$, and the boundary and viewing data specified above; the other parameters are held fixed. The vertical axis represents $\log_{10}\sigma_{\varepsilon,\mathrm{cond}}$. A fractional uncertainty of unity corresponds to $\sigma_{\varepsilon,\mathrm{cond}}=|\varepsilon|$.}}
	\label{fig:fisher}
\end{figure}

Finally, Fig.~\ref{fig:fisher} presents the single-parameter Fisher uncertainty $\sigma_{\varepsilon,\mathrm{cond}}=\Gamma_{\varepsilon\varepsilon}^{-1/2}$ of Sec.~\ref{subsec:mismatch-fisher}. This is a local estimate around each nonzero fiducial deformation, with the remaining parameters held fixed. It measures sensitivity to a variation of $\varepsilon$ but does not by itself quantify the statistical significance of a departure from Kerr.

All four curves in Fig.~\ref{fig:fisher} show an overall reduction in the formal uncertainty as spin increases. At $a/M=0.6$, the plotted values are approximately $9.2\times10^{-4}$, $1.8\times10^{-3}$, $9.2\times10^{-3}$, and $9.2\times10^{-2}$ for $\varepsilon=0.1$, $0.05$, $0.01$, and $0.001$, respectively. At $a/M=0.9$, they decrease to approximately $1.9\times10^{-6}$, $3.3\times10^{-6}$, $1.6\times10^{-5}$, and $1.6\times10^{-4}$. These are conditional uncertainties at the adopted source distance and strain normalization.

The moderate-spin ordering with fiducial deformation is consistent with the second-order phase correction. In a phase-dominated regime, $\partial_\varepsilon\Delta\Phi\propto\varepsilon$, giving $\sigma_{\varepsilon,\mathrm{cond}}\propto|\varepsilon|^{-1}$ at fixed amplitude. The values at $a/M=0.6$ approximately follow this scaling. At the largest spins, the curves for the three larger deformations approach and cross, so this simple ordering does not hold throughout the plot. As explained in Sec.~\ref{subsec:mismatch-fisher}, the linear first harmonic can affect the derivative near $\varepsilon=0$; the phase-based scaling should not be extrapolated to the Kerr limit.

At $a/M=0.6$, the fractional uncertainties are approximately $0.9\%$, $3.7\%$, and $92\%$ for $\varepsilon=0.1$, $0.05$, and $0.01$. For $\varepsilon=0.001$, the absolute uncertainty is about $90$ times the fiducial parameter. The last two cases therefore require particular care with the local linear and Gaussian approximations. Large formal errors, including those at low spin, should be treated as indications of weak local sensitivity until checked against the nonlinear likelihood \cite{Vallisneri:2007ev}. A Fisher width alone does not establish a reliable confidence interval in that regime.

\numrev{The absolute forecasts depend on the distance or SNR and detector noise model: at fixed waveform shape, $\sigma_{\varepsilon,\mathrm{cond}}\propto D\propto\rho^{-1}$. The highest-spin, larger-deformation portions of Figs.~\ref{fig:mismatch-map} and \ref{fig:fisher} are interpreted as illustrative trends of the second-order model; their quantitative accuracy requires a separate assessment of orbital truncation errors as well as temporal resolution. A dedicated treatment of near-extremal inspirals, including higher-order orbital effects and the coupled stability of the circular branch, is left for future work. Parameter correlations and systematic waveform errors are excluded. These local forecasts therefore remain conditional on the waveform, terminal-data prescription, and numerical convergence; observational constraints require a full parameter inference.}

\FloatBarrier

\section{\texorpdfstring{\numrev{Conclusions}}{Conclusions}}
\label{sec5}

We quantified how equatorial reflection symmetry breaking in the parity-violating conformal Kerr spacetime enters EMRI signals within an adiabatic quadrupole model. The calculation follows the prograde circular-geodesic branch connected to equatorial Kerr orbits and retains corrections through second order in $\varepsilon$. It includes the polar displacement, orbital-frequency shift, and conformal-factor contribution to the orbital energy, together with their effects on the energy flux and inspiral evolution. Both waveform harmonics are retained.

Reflection relates solutions with opposite signs of $\varepsilon$, forcing the polar displacement to be odd and the frequency and energy to be even on a locally unique analytic circular branch. This argument extends beyond conformal Kerr to metric families with the same reflection property. For the off-equatorial circular branch studied here, the mass-quadrupole prescription gives a first harmonic linear in $\varepsilon$ and quadratic corrections to the second harmonic and inspiral evolution. The latter accumulate in the phase through changes in both $\omega$ and $r(t)$ and can become more prominent than the direct first-harmonic contribution, as found for the configurations studied. The allowed perturbative orders follow from symmetry; the numerical hierarchy depends on the orbital coefficients, viewing geometry, detector weighting, and observation duration.

In the configurations considered, higher spin produces larger accumulated phase differences and generally larger noise-weighted mismatches. For one-year inspirals with $m=10M_\odot$ and $M=10^6M_\odot$, the fixed-parameter mismatch reaches the one-percent overlap-loss benchmark $\mathcal M=0.01$ at approximately $\varepsilon=10^{-3}$ for $a/M=0.9$. The comparison holds the Kerr parameters fixed and does not optimize relative time or phase. Within this model, the single-parameter Fisher trends suggest improved local sensitivity at high spin, conditional on the other parameters and the adopted strain normalization. They describe precision around nonzero fiducial deformations; an upper limit for a Kerr signal requires a separate analysis near $\varepsilon=0$.

We apply a source-phase approach, previously explored in dynamical Chern-Simons EMRI studies \cite{Sopuerta:2009cs,Canizares:2012is}, to a concrete black-hole background that breaks equatorial reflection symmetry. The characteristic ingredients are the off-equatorial circular branch and the accumulated second-order frequency and energy-balance corrections; the exact background allows their spin dependence to be studied without a slow-rotation expansion. These source effects complement propagation-birefringence tests. The distinction between source-waveform corrections and propagation effects has also been made explicit in dCS binary studies \cite{Li:2022dcswaveform}.

The present estimates rely on the weak-deformation expansion, quadrupole radiation prescription, and restricted orbital family. Quantitative forecasts also require checks of waveform sampling, Fisher steps, the near-ISCO cutoff, and the validity of the circular branch at high spin. The near-extremal, larger-deformation regime requires a dedicated analysis of orbital truncation errors and coupled stability beyond the scope of this work. More accurate radiation modeling from the sourced perturbation equations and a multiparameter inference would address waveform errors and parameter correlations. Extending the calculation to eccentricity and independent polar motion would broaden its applicability. Helicity-dependent generation and cosmological propagation require their own perturbation analyses. Observational constraints on $\varepsilon$ or $\alpha/\Lambda^4$ would require these validated waveform and inference models.

\paragraph{AI use} During the preparation of this manuscript, we used ChatGPT (OpenAI, GPT-6.1 Sol) to assist with language editing, organizing the presentation of analytical derivations, checking mathematical expressions, and improving physical explanations. All AI-assisted content was reviewed and verified by the authors against their independent calculations. All theoretical and numerical results were obtained independently by the authors, who take full responsibility for the manuscript.

\begin{acknowledgments}
	\numrev{We thank Tieguang Zi for helpful discussions. This work was supported in part by the National Natural Science Foundation of China under Grant No.~12205104.}
\end{acknowledgments}
\appendix
\section{Coefficients of the marginal-stability condition}
\label{app:isco-coefficients}

To evaluate the ISCO correction in Sec.~\ref{sec3}, define the seed-metric potential
\begin{equation}
    U(r,\theta;E,L_z)\equiv
       \frac{L_z^2g_{tt}+2EL_zg_{t\varphi}+E^2g_{\varphi\varphi}}
            {g_{tt}g_{\varphi\varphi}-g_{t\varphi}^2}.
    \label{eq:isco-seed-potential}
\end{equation}
The physical effective potential is $V_{\rm eff}=1+U/\Omega$. With $P\equiv\widehat{\mathcal P}$ for brevity, its expansion at fixed $r$, $\theta$, $E$, and $L_z$ is
\begin{equation}
    V_{\rm eff}=1+U-\varepsilon PU+\varepsilon^2P^2U
                   +\mathcal O(\varepsilon^3).
    \label{eq:isco-potential-expansion}
\end{equation}
We first differentiate this potential at fixed $E$ and $L_z$, and only then substitute the circular sequence
\begin{equation}
    \theta_a=\frac\pi2+\varepsilon\theta_1+\mathcal O(\varepsilon^3),
    \quad E=E_0+\varepsilon^2E_2+\mathcal O(\varepsilon^4),
    \quad L_z=L_0+\varepsilon^2L_2+\mathcal O(\varepsilon^4).
    \label{eq:isco-circular-constants-expansion}
\end{equation}
Here $E_2$ is given by Eq.~\eqref{eq:circular-energy-expansion}. The angular-momentum coefficient follows analogously from $L_z=\sqrt\Omega L_{\rm K}$:
\begin{equation}
    \begin{aligned}
        L_{\rm K}(r,\theta,\omega)
           &\equiv\frac{g_{t\varphi}+\omega g_{\varphi\varphi}}
                            {\sqrt{-\mathcal G}},
        \qquad L_0\equiv L_{{\rm K},0},\\
        L_2&=(\partial_\omega L_{\rm K})_0\omega_2
             +\frac12(\partial_\theta^2L_{\rm K})_0\theta_1^2
             +\frac12p\theta_1L_0.
    \end{aligned}
    \label{eq:isco-angular-momentum-coefficient}
\end{equation}
\numrev{For derivatives of $U$ below, a subscript $0$ denotes evaluation at $(\theta,E,L_z)=(\pi/2,E_0,L_0)$ at the specified radius. For $L_{\rm K}$, it denotes evaluation at $(\theta,\omega)=(\pi/2,\omega_0)$. The $r$ and $\theta$ derivatives of $U$ are taken at fixed $E$ and $L_z$.} At the seed circular orbit, $U_0=-1$ and $U_{r,0}=U_{\theta,0}=0$. The needed derivatives of $P$ are
\begin{equation}
    \begin{aligned}
        p&\equiv P_{\theta,0}=\frac{288aM^6}{r^7},
        &p'&=-\frac{7p}{r},
        &p''&=\frac{56p}{r^2},\\
        p_3&\equiv P_{\theta\theta\theta,0}
                    =-p\left(1+\frac{56a^2}{r^2}\right).
    \end{aligned}
    \label{eq:isco-pontryagin-derivatives}
\end{equation}
The zeroth-order Hessian entries are
\begin{equation}
    \begin{aligned}
        \mathsf A_0=U_{rr,0}
           &=\frac{2M(r^2-6Mr+8a\sqrt{Mr}-3a^2)}{r\Delta D_c},\\
        \mathsf B_0=U_{\theta\theta,0}
           &=\frac{2M(r^2-4a\sqrt{Mr}+3a^2)}{rD_c}.
    \end{aligned}
    \label{eq:isco-kerr-hessian}
\end{equation}
Here $D_c=r^2-3Mr+2a\sqrt{Mr}$ as in the main text. Expanding the differentiated potential on the displaced orbit gives
\begin{equation}
    \begin{aligned}
        \mathsf A_2={}&E_2(\partial_EU_{rr})_0
                 +L_2(\partial_{L_z}U_{rr})_0
                 +\frac12\theta_1^2U_{rr\theta\theta,0}
                 +\theta_1(p''-p\mathsf A_0),\\
        \mathsf B_2={}&E_2(\partial_EU_{\theta\theta})_0
                 +L_2(\partial_{L_z}U_{\theta\theta})_0
                 +\frac12\theta_1^2U_{\theta\theta\theta\theta,0}
                 +\theta_1(p_3-3p\mathsf B_0)-2p^2,\\
        \mathsf C_1={}&\theta_1U_{r\theta\theta,0}+p'.
    \end{aligned}
    \label{eq:isco-deformed-hessian}
\end{equation}
These expressions include the changes in $E$ and $L_z$, the displacement in $\theta$, and the conformal factor. In particular, the $-2p^2$ term comes from $\partial_\theta^2(P^2U)_0$. \numrev{The orbital coefficients multiply the potential derivatives after differentiation; allowing $E(r)$ or $L_z(r)$ to vary while forming the Hessian would mix variations along the circular sequence with the fixed-$E$, fixed-$L_z$ variations required for stability.}

At a nonextremal Kerr ISCO, $\mathsf A_0=0$ and $\mathsf B_0>0$. The second-order radial shift can therefore also be written as
\begin{equation}
    r_{\rm ISCO}=r_{\rm ISCO}^{\rm K}
       -\varepsilon^2\left.
            \frac{\mathsf A_2-\mathsf C_1^2/\mathsf B_0}
                 {\D\mathsf A_0/\D r}
          \right|_{r=r_{\rm ISCO}^{\rm K}}
       +\mathcal O(\varepsilon^4).
    \label{eq:isco-coupled-radius-shift}
\end{equation}
The denominator is the total derivative of $\mathsf A_0(r)$ along the Kerr circular sequence. The term $\mathsf C_1^2/\mathsf B_0$ shows explicitly how polar coupling modifies the radial marginal-stability condition. This expansion assumes a simple root and does not describe the appearance of multiple stable branches at larger deformations.

\section{\texorpdfstring{\rev{Linear tensor waves in flat vacuum}}{Linear tensor waves in flat vacuum}}
\label{app:flat-vacuum-waves}

\rev{
\numrev{The equality of the linear tensor-wave dynamics with GR in flat vacuum, discussed in Sec.~\ref{sec3}, follows from the complete pure-gravity action of Ref.~\cite{Tahara:2023pyg} for the conformal factor in Eq.~\eqref{eq:exact-conformal-factor}. Up to boundary terms, the action is}
\begin{equation}
    S[\bar g]=\frac{M_{\rm Pl}^2}{2}
    \int \D^4x\sqrt{-\bar g}\left[
        \bar\Omega\bar R
        +\frac{3}{2\bar\Omega}
         \bar\nabla_\mu\bar\Omega\,\bar\nabla^\mu\bar\Omega
    \right].
    \label{eq:flat-vacuum-action}
\end{equation}
Here $M_{\rm Pl}$ is the reduced Planck mass, $\bar{\mathcal P}$ is constructed from the physical metric $\bar g_{\mu\nu}$, and $\bar\Omega(\bar{\mathcal P})\equiv\Omega[\mathcal P(\bar{\mathcal P})]^{-1}$ is the inverse conformal factor.

Write $\bar g_{\mu\nu}=\eta_{\mu\nu}+\lambda h_{\mu\nu}$, where $\lambda$ counts the wave amplitude independently of the deformation parameter $\varepsilon$. The vanishing background Riemann tensor implies $\bar R_{\mu\nu\rho\sigma}=\mathcal O(\lambda)$ and $\bar R=\mathcal O(\lambda)$. Since the Pontryagin density contains two curvature tensors, $\bar{\mathcal P}=\mathcal O(\lambda^2)$. At fixed couplings, the inverse conformal factor has the expansion
\begin{equation}
    \bar\Omega
    =1-\frac{\alpha\bar{\mathcal P}}{\Lambda^4}
       +\mathcal O\!\left[
           \left(\frac{\alpha\bar{\mathcal P}}{\Lambda^4}\right)^2
        \right],
    \label{eq:inverse-factor-flat-vacuum}
\end{equation}
so $\bar\Omega-1=\mathcal O(\lambda^2)$. The two corrections to the Einstein-Hilbert integrand therefore begin at
\begin{equation}
    \begin{aligned}
        (\bar\Omega-1)\bar R
          &=\mathcal O(\lambda^3),\\
        \frac{1}{\bar\Omega}
        \bar\nabla_\mu\bar\Omega\,\bar\nabla^\mu\bar\Omega
          &=\mathcal O(\lambda^4).
    \end{aligned}
    \label{eq:flat-vacuum-correction-orders}
\end{equation}
Derivatives do not change the amplitude order, and $\sqrt{-\bar g}=1+\mathcal O(\lambda)$ cannot lower these orders. Thus $S^{(2)}=S_{\rm EH}^{(2)}$: the quadratic vacuum action, whose variation determines the linear wave equation, is the same as in GR. Linear tensor waves consequently have no amplitude or phase birefringence in this limit.

This argument uses an amplitude expansion of the complete action and does not replace the theory by a finite higher-derivative truncation. On a curved background, nonzero background curvature can produce a first-order perturbation of $\bar{\mathcal P}$, so the above counting does not apply. The flat-vacuum result therefore does not settle the full radiation or propagation problem in the backgrounds relevant to an EMRI observation.
}

\section{\texorpdfstring{\rev{Detector angles in the ecliptic frame}}{Detector angles in the ecliptic frame}}
\label{app:detector-angles}

\rev{
\numrev{The standard angle transformations used in the long-wavelength response \cite{Barack:2003fp,Cutler:1997ta} can be written compactly using unit vectors for the ecliptic sky direction $\boldsymbol s$, black-hole symmetry axis $\boldsymbol\ell$, and detector-plane normal $\boldsymbol z_D$:}
\begin{equation}
    \begin{aligned}
        \boldsymbol s&=(\sin\theta_s\cos\phi_s,\,
                        \sin\theta_s\sin\phi_s,\,\cos\theta_s),\\
        \boldsymbol\ell&=(\sin\theta_L\cos\phi_L,\,
                          \sin\theta_L\sin\phi_L,\,\cos\theta_L),\\
        \boldsymbol z_D(t)&=\left(-\frac{\sqrt3}{2}\cos\gamma(t),\,
                                  -\frac{\sqrt3}{2}\sin\gamma(t),\,\frac12\right).
    \end{aligned}
    \label{eq:detector-ecliptic-vectors}
\end{equation}
Here $\boldsymbol s$ points from the detector to the source. The vectors $\boldsymbol s$ and $\boldsymbol n$ in Eq.~\eqref{eq:transverse-basis} describe the same line of sight with opposite orientations, but their components are given in the ecliptic and black-hole frames, respectively; in a common frame, $\boldsymbol n=-\boldsymbol s$. \numrev{The inclination therefore obeys $\cos\iota=\boldsymbol\ell\cdot\boldsymbol n=-\boldsymbol\ell\cdot\boldsymbol s$.} We define $\gamma(t)\equiv\bar\phi_0+2\pi t/T$, with $T=1\,\mathrm{yr}$. The constants $\bar\phi_0$ and $\bar\alpha_0$ specify the detector's initial orbital and rotational phases. The detector-frame source angles are
\begin{equation}
    \begin{aligned}
        \cos\vartheta_D(t)
        &=\boldsymbol s\cdot\boldsymbol z_D(t),\\
        \varphi_D(t)
        &=\bar\alpha_0+\frac{2\pi t}{T}
          +\tan^{-1}\!\left[
             \frac{\sqrt3\cos\theta_s+\sin\theta_s\cos[\gamma(t)-\phi_s]}
                  {2\sin\theta_s\sin[\gamma(t)-\phi_s]}
          \right].
    \end{aligned}
    \label{eq:detector-ecliptic-angles}
\end{equation}
With the polarization convention of Eq.~\eqref{eq:transverse-basis}, the polarization angle is
\begin{equation}
    \tan\psi_D(t)
      =\frac{\boldsymbol\ell\cdot\boldsymbol z_D(t)
              -(\boldsymbol\ell\cdot\boldsymbol s)\cos\vartheta_D(t)}
             {\boldsymbol s\cdot[\boldsymbol\ell\times\boldsymbol z_D(t)]}.
    \label{eq:detector-polarization-angle}
\end{equation}
Substitution into Eq.~\eqref{eq:pattern-functions} gives the time-dependent antenna functions used in the detector strain.
}

\section{\texorpdfstring{\numrev{Numerical conventions}}{Numerical conventions}}
\label{app:numerical-conventions}

\numrev{
For the Fourier-domain comparisons, geometric time is converted to seconds before applying the noise weighting. With $\tau\equiv t/M$ and $t_M\equiv GM_{\rm BH}/c^3$, the conversions are
\begin{equation}
    \begin{aligned}
        t_{\rm phys}&=t_M\tau,
        & f_{\rm phys}&=\hat f/t_M,\\
        \tilde h_{\rm phys}(f_{\rm phys})
          &=t_M\tilde h_\tau(\hat f),
        & t_M&\simeq4.9255\,\mathrm{s}
            \quad(M_{\rm BH}=10^6M_\odot).
    \end{aligned}
    \label{eq:fourier-physical-units}
\end{equation}
Here $\hat f$ is the dimensionless frequency conjugate to $\tau$, and $\tilde h_\tau$ is the Fourier transform with respect to $\tau$. The frequency measure transforms as $\D f_{\rm phys}=\D\hat f/t_M$. Thus, consistent noise weighting requires converting the Fourier amplitude and integration measure as well as the frequency entering the PSD. The analytic noise fit used for the numerical weighting is
\begin{equation}
    \begin{aligned}
        S_n(f)=\Bigg[&1.097\times10^{-49}
                        \left(\frac{f}{\mathrm{Hz}}\right)^{-4}
                       +3.228\times10^{-42}\\
                     &+7.682\times10^{-40}
                        \left(\frac{f}{\mathrm{Hz}}\right)^2
                       +1.791\times10^{-44}
                        \left(\frac{f}{\mathrm{Hz}}\right)^{-7/3}
                \Bigg]\,\mathrm{Hz}^{-1},
    \end{aligned}
    \label{eq:numerical-noise-fit}
\end{equation}
with $f$ in physical units and a lower integration cutoff of $10^{-4}\,\mathrm{Hz}$. The strain segments and Fisher derivatives are multiplied by a Tukey window with parameter $0.1$ before transformation; each duration-dependent comparison uses its own segment window.

The time sampling must resolve the highest retained harmonic in every waveform entering a comparison or finite difference. In particular, with $\hat\omega_j=M\omega_j$ and a physical sampling interval $\Delta t_{\rm phys}$, it must satisfy
\begin{equation}
    f_{\rm Nyq}\equiv\frac{1}{2\Delta t_{\rm phys}}
       >\max_{j,t}\frac{\hat\omega_j(t)}{\pi t_M},
    \label{eq:numerical-sampling-condition}
\end{equation}
where $j$ labels all waveforms entering the comparison or finite difference, and the right-hand side is the largest instantaneous second-harmonic frequency. Meeting this condition is necessary; convergence of the overlaps and Fisher derivatives under sampling refinement, taper changes, and finite-difference step variation is also required for quantitative error estimates.
}

\bibliographystyle{apsrev4-2}

\begin{thebibliography}{96}%
	\makeatletter
	\providecommand \@ifxundefined [1]{%
		\@ifx{#1\undefined}
	}%
	\providecommand \@ifnum [1]{%
		\ifnum #1\expandafter \@firstoftwo
		\else \expandafter \@secondoftwo
		\fi
	}%
	\providecommand \@ifx [1]{%
		\ifx #1\expandafter \@firstoftwo
		\else \expandafter \@secondoftwo
		\fi
	}%
	\providecommand \natexlab [1]{#1}%
	\providecommand \enquote  [1]{``#1''}%
	\providecommand \bibnamefont  [1]{#1}%
	\providecommand \bibfnamefont [1]{#1}%
	\providecommand \citenamefont [1]{#1}%
	\providecommand \href@noop [0]{\@secondoftwo}%
	\providecommand \href [0]{\begingroup \@sanitize@url \@href}%
	\providecommand \@href[1]{\@@startlink{#1}\@@href}%
	\providecommand \@@href[1]{\endgroup#1\@@endlink}%
	\providecommand \@sanitize@url [0]{\catcode `\\12\catcode `\$12\catcode
		`\&12\catcode `\#12\catcode `\^12\catcode `\_12\catcode `\%12\relax}%
	\providecommand \@@startlink[1]{}%
	\providecommand \@@endlink[0]{}%
	\providecommand \url  [0]{\begingroup\@sanitize@url \@url }%
	\providecommand \@url [1]{\endgroup\@href {#1}{\urlprefix }}%
	\providecommand \urlprefix  [0]{URL }%
	\providecommand \Eprint [0]{\href }%
	\providecommand \doibase [0]{https://doi.org/}%
	\providecommand \selectlanguage [0]{\@gobble}%
	\providecommand \bibinfo  [0]{\@secondoftwo}%
	\providecommand \bibfield  [0]{\@secondoftwo}%
	\providecommand \translation [1]{[#1]}%
	\providecommand \BibitemOpen [0]{}%
	\providecommand \bibitemStop [0]{}%
	\providecommand \bibitemNoStop [0]{.\EOS\space}%
	\providecommand \EOS [0]{\spacefactor3000\relax}%
	\providecommand \BibitemShut  [1]{\csname bibitem#1\endcsname}%
	\let\auto@bib@innerbib\@empty
	\bibitem [{\citenamefont {Abbott}\ \emph {et~al.}(2016)\citenamefont {Abbott}
		\emph {et~al.}}]{LIGOScientific:2016aoc}%
	\BibitemOpen
	\bibfield  {author} {\bibinfo {author} {\bibfnamefont {B.~P.}\ \bibnamefont
			{Abbott}} \emph {et~al.} (\bibinfo {collaboration} {LIGO Scientific,
			Virgo}),\ }\href {https://doi.org/10.1103/PhysRevLett.116.061102} {\bibfield
		{journal} {\bibinfo  {journal} {Phys. Rev. Lett.}\ }\textbf {\bibinfo
			{volume} {116}},\ \bibinfo {pages} {061102} (\bibinfo {year} {2016})},\
	\Eprint {https://arxiv.org/abs/1602.03837} {arXiv:1602.03837 [gr-qc]}
	\BibitemShut {NoStop}%
	\bibitem [{\citenamefont {Abbott}\ \emph {et~al.}(2025)\citenamefont {Abbott}
		\emph {et~al.}}]{LIGOScientific:2021grtests}%
	\BibitemOpen
	\bibfield  {author} {\bibinfo {author} {\bibfnamefont {R.}~\bibnamefont
			{Abbott}} \emph {et~al.} (\bibinfo {collaboration} {LIGO Scientific, Virgo,
			KAGRA}),\ }\href {https://doi.org/10.1103/PhysRevD.112.084080} {\bibfield
		{journal} {\bibinfo  {journal} {Phys. Rev. D}\ }\textbf {\bibinfo {volume}
			{112}},\ \bibinfo {pages} {084080} (\bibinfo {year} {2025})},\ \Eprint
	{https://arxiv.org/abs/2112.06861} {arXiv:2112.06861 [gr-qc]} \BibitemShut
	{NoStop}%
	\bibitem [{\citenamefont {Carter}(1971)}]{Carter:1971zc}%
	\BibitemOpen
	\bibfield  {author} {\bibinfo {author} {\bibfnamefont {B.}~\bibnamefont
			{Carter}},\ }\href {https://doi.org/10.1103/PhysRevLett.26.331} {\bibfield
		{journal} {\bibinfo  {journal} {Phys. Rev. Lett.}\ }\textbf {\bibinfo
			{volume} {26}},\ \bibinfo {pages} {331} (\bibinfo {year} {1971})}\BibitemShut
	{NoStop}%
	\bibitem [{\citenamefont {Chru{\'s}ciel}\ \emph {et~al.}(2012)\citenamefont
		{Chru{\'s}ciel}, \citenamefont {Costa},\ and\ \citenamefont
		{Heusler}}]{Chrusciel:2012uniqueness}%
	\BibitemOpen
	\bibfield  {author} {\bibinfo {author} {\bibfnamefont {P.~T.}\ \bibnamefont
			{Chru{\'s}ciel}}, \bibinfo {author} {\bibfnamefont {J.~L.}\ \bibnamefont
			{Costa}},\ and\ \bibinfo {author} {\bibfnamefont {M.}~\bibnamefont
			{Heusler}},\ }\href {https://doi.org/10.12942/lrr-2012-7} {\bibfield
		{journal} {\bibinfo  {journal} {Living Rev. Rel.}\ }\textbf {\bibinfo
			{volume} {15}},\ \bibinfo {pages} {7} (\bibinfo {year} {2012})},\ \Eprint
	{https://arxiv.org/abs/1205.6112} {arXiv:1205.6112 [gr-qc]} \BibitemShut
	{NoStop}%
	\bibitem [{\citenamefont {Berti}\ \emph {et~al.}(2015)\citenamefont {Berti}
		\emph {et~al.}}]{Berti:2015grtests}%
	\BibitemOpen
	\bibfield  {author} {\bibinfo {author} {\bibfnamefont {E.}~\bibnamefont
			{Berti}} \emph {et~al.},\ }\href
	{https://doi.org/10.1088/0264-9381/32/24/243001} {\bibfield  {journal}
		{\bibinfo  {journal} {Class. Quant. Grav.}\ }\textbf {\bibinfo {volume}
			{32}},\ \bibinfo {pages} {243001} (\bibinfo {year} {2015})},\ \Eprint
	{https://arxiv.org/abs/1501.07274} {arXiv:1501.07274 [gr-qc]} \BibitemShut
	{NoStop}%
	\bibitem [{\citenamefont {Babak}\ \emph {et~al.}(2017)\citenamefont {Babak},
		\citenamefont {Gair}, \citenamefont {Sesana}, \citenamefont {Barausse},
		\citenamefont {Sopuerta}, \citenamefont {Berry}, \citenamefont {Berti},
		\citenamefont {Amaro-Seoane}, \citenamefont {Petiteau},\ and\ \citenamefont
		{Klein}}]{Babak:2017tow}%
	\BibitemOpen
	\bibfield  {author} {\bibinfo {author} {\bibfnamefont {S.}~\bibnamefont
			{Babak}}, \bibinfo {author} {\bibfnamefont {J.}~\bibnamefont {Gair}},
		\bibinfo {author} {\bibfnamefont {A.}~\bibnamefont {Sesana}}, \bibinfo
		{author} {\bibfnamefont {E.}~\bibnamefont {Barausse}}, \bibinfo {author}
		{\bibfnamefont {C.~F.}\ \bibnamefont {Sopuerta}}, \bibinfo {author}
		{\bibfnamefont {C.~P.~L.}\ \bibnamefont {Berry}}, \bibinfo {author}
		{\bibfnamefont {E.}~\bibnamefont {Berti}}, \bibinfo {author} {\bibfnamefont
			{P.}~\bibnamefont {Amaro-Seoane}}, \bibinfo {author} {\bibfnamefont
			{A.}~\bibnamefont {Petiteau}},\ and\ \bibinfo {author} {\bibfnamefont
			{A.}~\bibnamefont {Klein}},\ }\href
	{https://doi.org/10.1103/PhysRevD.95.103012} {\bibfield  {journal} {\bibinfo
			{journal} {Phys. Rev. D}\ }\textbf {\bibinfo {volume} {95}},\ \bibinfo
		{pages} {103012} (\bibinfo {year} {2017})},\ \Eprint
	{https://arxiv.org/abs/1703.09722} {arXiv:1703.09722 [gr-qc]} \BibitemShut
	{NoStop}%
	\bibitem [{\citenamefont {Amaro-Seoane}\ \emph {et~al.}(2017)\citenamefont
		{Amaro-Seoane} \emph {et~al.}}]{LISA:2017pwj}%
	\BibitemOpen
	\bibfield  {author} {\bibinfo {author} {\bibfnamefont {P.}~\bibnamefont
			{Amaro-Seoane}} \emph {et~al.} (\bibinfo {collaboration} {LISA}),\
	}\href@noop {} {\  (\bibinfo {year} {2017})},\ \Eprint
	{https://arxiv.org/abs/1702.00786} {arXiv:1702.00786 [astro-ph.IM]}
	\BibitemShut {NoStop}%
	\bibitem [{\citenamefont {Luo}\ \emph {et~al.}(2016)\citenamefont {Luo} \emph
		{et~al.}}]{TianQin:2015yph}%
	\BibitemOpen
	\bibfield  {author} {\bibinfo {author} {\bibfnamefont {J.}~\bibnamefont
			{Luo}} \emph {et~al.} (\bibinfo {collaboration} {TianQin}),\ }\href
	{https://doi.org/10.1088/0264-9381/33/3/035010} {\bibfield  {journal}
		{\bibinfo  {journal} {Class. Quant. Grav.}\ }\textbf {\bibinfo {volume}
			{33}},\ \bibinfo {pages} {035010} (\bibinfo {year} {2016})},\ \Eprint
	{https://arxiv.org/abs/1512.02076} {arXiv:1512.02076 [astro-ph.IM]}
	\BibitemShut {NoStop}%
	\bibitem [{\citenamefont {Ruan}\ \emph {et~al.}(2020)\citenamefont {Ruan},
		\citenamefont {Guo}, \citenamefont {Cai},\ and\ \citenamefont
		{Zhang}}]{Ruan:2018tsw}%
	\BibitemOpen
	\bibfield  {author} {\bibinfo {author} {\bibfnamefont {W.-H.}\ \bibnamefont
			{Ruan}}, \bibinfo {author} {\bibfnamefont {Z.-K.}\ \bibnamefont {Guo}},
		\bibinfo {author} {\bibfnamefont {R.-G.}\ \bibnamefont {Cai}},\ and\ \bibinfo
		{author} {\bibfnamefont {Y.-Z.}\ \bibnamefont {Zhang}},\ }\href
	{https://doi.org/10.1142/S0217751X2050075X} {\bibfield  {journal} {\bibinfo
			{journal} {Int. J. Mod. Phys. A}\ }\textbf {\bibinfo {volume} {35}},\
		\bibinfo {pages} {2050075} (\bibinfo {year} {2020})},\ \Eprint
	{https://arxiv.org/abs/1807.09495} {arXiv:1807.09495 [gr-qc]} \BibitemShut
	{NoStop}%
	\bibitem [{\citenamefont {Ryan}(1995)}]{Ryan:1995wh}%
	\BibitemOpen
	\bibfield  {author} {\bibinfo {author} {\bibfnamefont {F.~D.}\ \bibnamefont
			{Ryan}},\ }\href {https://doi.org/10.1103/PhysRevD.52.5707} {\bibfield
		{journal} {\bibinfo  {journal} {Phys. Rev. D}\ }\textbf {\bibinfo {volume}
			{52}},\ \bibinfo {pages} {5707} (\bibinfo {year} {1995})}\BibitemShut
	{NoStop}%
	\bibitem [{\citenamefont {Glampedakis}\ and\ \citenamefont
		{Babak}(2006)}]{Glampedakis:2005cf}%
	\BibitemOpen
	\bibfield  {author} {\bibinfo {author} {\bibfnamefont {K.}~\bibnamefont
			{Glampedakis}}\ and\ \bibinfo {author} {\bibfnamefont {S.}~\bibnamefont
			{Babak}},\ }\href {https://doi.org/10.1088/0264-9381/23/12/013} {\bibfield
		{journal} {\bibinfo  {journal} {Class. Quant. Grav.}\ }\textbf {\bibinfo
			{volume} {23}},\ \bibinfo {pages} {4167} (\bibinfo {year} {2006})},\ \Eprint
	{https://arxiv.org/abs/gr-qc/0510057} {arXiv:gr-qc/0510057} \BibitemShut
	{NoStop}%
	\bibitem [{\citenamefont {Barack}\ and\ \citenamefont
		{Cutler}(2007)}]{Barack:2006pq}%
	\BibitemOpen
	\bibfield  {author} {\bibinfo {author} {\bibfnamefont {L.}~\bibnamefont
			{Barack}}\ and\ \bibinfo {author} {\bibfnamefont {C.}~\bibnamefont
			{Cutler}},\ }\href {https://doi.org/10.1103/PhysRevD.75.042003} {\bibfield
		{journal} {\bibinfo  {journal} {Phys. Rev. D}\ }\textbf {\bibinfo {volume}
			{75}},\ \bibinfo {pages} {042003} (\bibinfo {year} {2007})},\ \Eprint
	{https://arxiv.org/abs/gr-qc/0612029} {arXiv:gr-qc/0612029} \BibitemShut
	{NoStop}%
	\bibitem [{\citenamefont {Moore}\ \emph {et~al.}(2017)\citenamefont {Moore},
		\citenamefont {Chua},\ and\ \citenamefont {Gair}}]{Moore:2017lxy}%
	\BibitemOpen
	\bibfield  {author} {\bibinfo {author} {\bibfnamefont {C.~J.}\ \bibnamefont
			{Moore}}, \bibinfo {author} {\bibfnamefont {A.~J.~K.}\ \bibnamefont {Chua}},\
		and\ \bibinfo {author} {\bibfnamefont {J.~R.}\ \bibnamefont {Gair}},\ }\href
	{https://doi.org/10.1088/1361-6382/aa85fa} {\bibfield  {journal} {\bibinfo
			{journal} {Class. Quant. Grav.}\ }\textbf {\bibinfo {volume} {34}},\ \bibinfo
		{pages} {195009} (\bibinfo {year} {2017})},\ \Eprint
	{https://arxiv.org/abs/1707.00712} {arXiv:1707.00712 [gr-qc]} \BibitemShut
	{NoStop}%
	\bibitem [{\citenamefont {Zi}\ \emph {et~al.}(2021)\citenamefont {Zi},
		\citenamefont {Zhang}, \citenamefont {Fan}, \citenamefont {Zhang},
		\citenamefont {Hu}, \citenamefont {Shi},\ and\ \citenamefont
		{Mei}}]{Zi:2021pdp}%
	\BibitemOpen
	\bibfield  {author} {\bibinfo {author} {\bibfnamefont {T.-G.}\ \bibnamefont
			{Zi}}, \bibinfo {author} {\bibfnamefont {J.-D.}\ \bibnamefont {Zhang}},
		\bibinfo {author} {\bibfnamefont {H.-M.}\ \bibnamefont {Fan}}, \bibinfo
		{author} {\bibfnamefont {X.-T.}\ \bibnamefont {Zhang}}, \bibinfo {author}
		{\bibfnamefont {Y.-M.}\ \bibnamefont {Hu}}, \bibinfo {author} {\bibfnamefont
			{C.}~\bibnamefont {Shi}},\ and\ \bibinfo {author} {\bibfnamefont
			{J.}~\bibnamefont {Mei}},\ }\href
	{https://doi.org/10.1103/PhysRevD.104.064008} {\bibfield  {journal} {\bibinfo
			{journal} {Phys. Rev. D}\ }\textbf {\bibinfo {volume} {104}},\ \bibinfo
		{pages} {064008} (\bibinfo {year} {2021})},\ \Eprint
	{https://arxiv.org/abs/2104.06047} {arXiv:2104.06047 [gr-qc]} \BibitemShut
	{NoStop}%
	\bibitem [{\citenamefont {C{\'a}rdenas-Avenda{\~{n}}o}\ and\ \citenamefont
		{Sopuerta}(2024)}]{Cardenas-Avendano:2024mqp}%
	\BibitemOpen
	\bibfield  {author} {\bibinfo {author} {\bibfnamefont {A.}~\bibnamefont
			{C{\'a}rdenas-Avenda{\~{n}}o}}\ and\ \bibinfo {author} {\bibfnamefont
			{C.~F.}\ \bibnamefont {Sopuerta}},\ }\bibinfo {title} {Testing gravity with
		extreme-mass-ratio inspirals},\ in\ \href
	{https://doi.org/10.1007/978-981-97-2871-8_8} {\emph {\bibinfo {booktitle}
			{Recent Progress on Gravity Tests: Challenges and Future Perspectives}}},\
	\bibinfo {editor} {edited by\ \bibinfo {editor} {\bibfnamefont
			{C.}~\bibnamefont {Bambi}}\ and\ \bibinfo {editor} {\bibfnamefont
			{A.}~\bibnamefont {C{\'a}rdenas-Avenda{\~{n}}o}}}\ (\bibinfo  {publisher}
	{Springer Nature Singapore},\ \bibinfo {address} {Singapore},\ \bibinfo
	{year} {2024})\ pp.\ \bibinfo {pages} {275--359}\BibitemShut {NoStop}%
	\bibitem [{\citenamefont {Cardoso}\ and\ \citenamefont
		{Pani}(2019)}]{Cardoso:2019rvt}%
	\BibitemOpen
	\bibfield  {author} {\bibinfo {author} {\bibfnamefont {V.}~\bibnamefont
			{Cardoso}}\ and\ \bibinfo {author} {\bibfnamefont {P.}~\bibnamefont {Pani}},\
	}\href {https://doi.org/10.1007/s41114-019-0020-4} {\bibfield  {journal}
		{\bibinfo  {journal} {Living Rev. Rel.}\ }\textbf {\bibinfo {volume} {22}},\
		\bibinfo {pages} {4} (\bibinfo {year} {2019})},\ \Eprint
	{https://arxiv.org/abs/1904.05363} {arXiv:1904.05363 [gr-qc]} \BibitemShut
	{NoStop}%
	\bibitem [{\citenamefont {Zi}\ \emph {et~al.}(2024)\citenamefont {Zi},
		\citenamefont {Ye},\ and\ \citenamefont {Li}}]{Zi:2023geb}%
	\BibitemOpen
	\bibfield  {author} {\bibinfo {author} {\bibfnamefont {T.}~\bibnamefont
			{Zi}}, \bibinfo {author} {\bibfnamefont {C.-Q.}\ \bibnamefont {Ye}},\ and\
		\bibinfo {author} {\bibfnamefont {P.-C.}\ \bibnamefont {Li}},\ }\href
	{https://doi.org/10.1088/1475-7516/2024/10/066} {\bibfield  {journal}
		{\bibinfo  {journal} {JCAP}\ }\textbf {\bibinfo {volume} {2024}}\bibfield
		{number} {\bibinfo  {number} { (10)},\ \bibinfo {pages} {066}},\ }\Eprint
	{https://arxiv.org/abs/2311.15532} {arXiv:2311.15532 [gr-qc]} \BibitemShut
	{NoStop}%
	\bibitem [{\citenamefont {Datta}\ \emph {et~al.}(2024)\citenamefont {Datta},
		\citenamefont {Brito}, \citenamefont {Hughes}, \citenamefont {Klinger},\ and\
		\citenamefont {Pani}}]{Datta:2024vll}%
	\BibitemOpen
	\bibfield  {author} {\bibinfo {author} {\bibfnamefont {S.}~\bibnamefont
			{Datta}}, \bibinfo {author} {\bibfnamefont {R.}~\bibnamefont {Brito}},
		\bibinfo {author} {\bibfnamefont {S.~A.}\ \bibnamefont {Hughes}}, \bibinfo
		{author} {\bibfnamefont {T.}~\bibnamefont {Klinger}},\ and\ \bibinfo {author}
		{\bibfnamefont {P.}~\bibnamefont {Pani}},\ }\href
	{https://doi.org/10.1103/PhysRevD.110.024048} {\bibfield  {journal} {\bibinfo
			{journal} {Phys. Rev. D}\ }\textbf {\bibinfo {volume} {110}},\ \bibinfo
		{pages} {024048} (\bibinfo {year} {2024})},\ \Eprint
	{https://arxiv.org/abs/2404.04013} {arXiv:2404.04013 [gr-qc]} \BibitemShut
	{NoStop}%
	\bibitem [{\citenamefont {Xia}\ \emph {et~al.}(2026)\citenamefont {Xia},
		\citenamefont {Long}, \citenamefont {Pan}, \citenamefont {Jing},\ and\
		\citenamefont {Qian}}]{Xia:2026aty}%
	\BibitemOpen
	\bibfield  {author} {\bibinfo {author} {\bibfnamefont {Z.-W.}\ \bibnamefont
			{Xia}}, \bibinfo {author} {\bibfnamefont {S.}~\bibnamefont {Long}}, \bibinfo
		{author} {\bibfnamefont {Q.}~\bibnamefont {Pan}}, \bibinfo {author}
		{\bibfnamefont {J.}~\bibnamefont {Jing}},\ and\ \bibinfo {author}
		{\bibfnamefont {W.-L.}\ \bibnamefont {Qian}},\ }\href@noop {} {\  (\bibinfo
		{year} {2026})},\ \Eprint {https://arxiv.org/abs/2602.11039}
	{arXiv:2602.11039 [gr-qc]} \BibitemShut {NoStop}%
	\bibitem [{\citenamefont {Zi}\ and\ \citenamefont {Li}(2024)}]{Zi:2023qfk}%
	\BibitemOpen
	\bibfield  {author} {\bibinfo {author} {\bibfnamefont {T.}~\bibnamefont
			{Zi}}\ and\ \bibinfo {author} {\bibfnamefont {P.-C.}\ \bibnamefont {Li}},\
	}\href {https://doi.org/10.1103/PhysRevD.109.064089} {\bibfield  {journal}
		{\bibinfo  {journal} {Phys. Rev. D}\ }\textbf {\bibinfo {volume} {109}},\
		\bibinfo {pages} {064089} (\bibinfo {year} {2024})},\ \Eprint
	{https://arxiv.org/abs/2311.07279} {arXiv:2311.07279 [gr-qc]} \BibitemShut
	{NoStop}%
	\bibitem [{\citenamefont {Fu}\ \emph {et~al.}(2025)\citenamefont {Fu},
		\citenamefont {Liu}, \citenamefont {Wang}, \citenamefont {Wu},\ and\
		\citenamefont {Zhang}}]{Fu:2024cfk}%
	\BibitemOpen
	\bibfield  {author} {\bibinfo {author} {\bibfnamefont {G.}~\bibnamefont
			{Fu}}, \bibinfo {author} {\bibfnamefont {Y.}~\bibnamefont {Liu}}, \bibinfo
		{author} {\bibfnamefont {B.}~\bibnamefont {Wang}}, \bibinfo {author}
		{\bibfnamefont {J.-P.}\ \bibnamefont {Wu}},\ and\ \bibinfo {author}
		{\bibfnamefont {C.}~\bibnamefont {Zhang}},\ }\href
	{https://doi.org/10.1103/PhysRevD.111.084066} {\bibfield  {journal} {\bibinfo
			{journal} {Phys. Rev. D}\ }\textbf {\bibinfo {volume} {111}},\ \bibinfo
		{pages} {084066} (\bibinfo {year} {2025})},\ \Eprint
	{https://arxiv.org/abs/2409.08138} {arXiv:2409.08138 [gr-qc]} \BibitemShut
	{NoStop}%
	\bibitem [{\citenamefont {Zi}\ and\ \citenamefont {Kumar}(2025)}]{Zi:2024jla}%
	\BibitemOpen
	\bibfield  {author} {\bibinfo {author} {\bibfnamefont {T.}~\bibnamefont
			{Zi}}\ and\ \bibinfo {author} {\bibfnamefont {S.}~\bibnamefont {Kumar}},\
	}\href {https://doi.org/10.1140/epjc/s10052-025-14330-7} {\bibfield
		{journal} {\bibinfo  {journal} {Eur. Phys. J. C}\ }\textbf {\bibinfo {volume}
			{85}},\ \bibinfo {pages} {592} (\bibinfo {year} {2025})},\ \Eprint
	{https://arxiv.org/abs/2409.17765} {arXiv:2409.17765 [gr-qc]} \BibitemShut
	{NoStop}%
	\bibitem [{\citenamefont {Zare}\ \emph {et~al.}(2026)\citenamefont {Zare},
		\citenamefont {Zhu}, \citenamefont {Nieto}, \citenamefont {Lu},\ and\
		\citenamefont {Hassanabadi}}]{Zare:2025aek}%
	\BibitemOpen
	\bibfield  {author} {\bibinfo {author} {\bibfnamefont {S.}~\bibnamefont
			{Zare}}, \bibinfo {author} {\bibfnamefont {T.}~\bibnamefont {Zhu}}, \bibinfo
		{author} {\bibfnamefont {L.~M.}\ \bibnamefont {Nieto}}, \bibinfo {author}
		{\bibfnamefont {S.}~\bibnamefont {Lu}},\ and\ \bibinfo {author}
		{\bibfnamefont {H.}~\bibnamefont {Hassanabadi}},\ }\href
	{https://doi.org/10.1088/1475-7516/2026/01/059} {\bibfield  {journal}
		{\bibinfo  {journal} {JCAP}\ }\textbf {\bibinfo {volume} {2026}}\bibfield
		{number} {\bibinfo  {number} { (01)},\ \bibinfo {pages} {059}},\ }\Eprint
	{https://arxiv.org/abs/2510.05166} {arXiv:2510.05166 [gr-qc]} \BibitemShut
	{NoStop}%
	\bibitem [{\citenamefont {Zhao}\ \emph
		{et~al.}(2025{\natexlab{a}})\citenamefont {Zhao}, \citenamefont {Tang},\ and\
		\citenamefont {Xu}}]{Zhao:2025sck}%
	\BibitemOpen
	\bibfield  {author} {\bibinfo {author} {\bibfnamefont {L.}~\bibnamefont
			{Zhao}}, \bibinfo {author} {\bibfnamefont {M.}~\bibnamefont {Tang}},\ and\
		\bibinfo {author} {\bibfnamefont {Z.}~\bibnamefont {Xu}},\ }\href
	{https://doi.org/10.1088/1475-7516/2025/10/002} {\bibfield  {journal}
		{\bibinfo  {journal} {JCAP}\ }\textbf {\bibinfo {volume} {2025}}\bibfield
		{number} {\bibinfo  {number} { (10)},\ \bibinfo {pages} {002}},\ }\Eprint
	{https://arxiv.org/abs/2503.06503} {arXiv:2503.06503 [gr-qc]} \BibitemShut
	{NoStop}%
	\bibitem [{\citenamefont {Tan}\ \emph {et~al.}(2024)\citenamefont {Tan},
		\citenamefont {Zhang}, \citenamefont {Fan},\ and\ \citenamefont
		{Mei}}]{Tan:2024utr}%
	\BibitemOpen
	\bibfield  {author} {\bibinfo {author} {\bibfnamefont {J.}~\bibnamefont
			{Tan}}, \bibinfo {author} {\bibfnamefont {J.-d.}\ \bibnamefont {Zhang}},
		\bibinfo {author} {\bibfnamefont {H.-M.}\ \bibnamefont {Fan}},\ and\ \bibinfo
		{author} {\bibfnamefont {J.}~\bibnamefont {Mei}},\ }\href
	{https://doi.org/10.1140/epjc/s10052-024-13178-7} {\bibfield  {journal}
		{\bibinfo  {journal} {Eur. Phys. J. C}\ }\textbf {\bibinfo {volume} {84}},\
		\bibinfo {pages} {824} (\bibinfo {year} {2024})},\ \Eprint
	{https://arxiv.org/abs/2402.05752} {arXiv:2402.05752 [gr-qc]} \BibitemShut
	{NoStop}%
	\bibitem [{\citenamefont {Babichev}\ \emph {et~al.}(2024)\citenamefont
		{Babichev}, \citenamefont {Charmousis}, \citenamefont {Doneva}, \citenamefont
		{Gyulchev},\ and\ \citenamefont {Yazadjiev}}]{Babichev:2024hjf}%
	\BibitemOpen
	\bibfield  {author} {\bibinfo {author} {\bibfnamefont {E.}~\bibnamefont
			{Babichev}}, \bibinfo {author} {\bibfnamefont {C.}~\bibnamefont
			{Charmousis}}, \bibinfo {author} {\bibfnamefont {D.~D.}\ \bibnamefont
			{Doneva}}, \bibinfo {author} {\bibfnamefont {G.~N.}\ \bibnamefont
			{Gyulchev}},\ and\ \bibinfo {author} {\bibfnamefont {S.~S.}\ \bibnamefont
			{Yazadjiev}},\ }\href {https://doi.org/10.1088/1475-7516/2024/06/065}
	{\bibfield  {journal} {\bibinfo  {journal} {JCAP}\ }\textbf {\bibinfo
			{volume} {2024}}\bibfield  {number} {\bibinfo  {number} { (06)},\ \bibinfo
			{pages} {065}},\ }\Eprint {https://arxiv.org/abs/2403.16192}
	{arXiv:2403.16192 [gr-qc]} \BibitemShut {NoStop}%
	\bibitem [{\citenamefont {Kumar}\ \emph {et~al.}(2025)\citenamefont {Kumar},
		\citenamefont {Zi},\ and\ \citenamefont {Bhattacharyya}}]{Kumar:2025jsi}%
	\BibitemOpen
	\bibfield  {author} {\bibinfo {author} {\bibfnamefont {S.}~\bibnamefont
			{Kumar}}, \bibinfo {author} {\bibfnamefont {T.}~\bibnamefont {Zi}},\ and\
		\bibinfo {author} {\bibfnamefont {A.}~\bibnamefont {Bhattacharyya}},\
	}\href@noop {} {\  (\bibinfo {year} {2025})},\ \Eprint
	{https://arxiv.org/abs/2507.03380} {arXiv:2507.03380 [gr-qc]} \BibitemShut
	{NoStop}%
	\bibitem [{\citenamefont {LaHaye}\ \emph {et~al.}(2026)\citenamefont {LaHaye},
		\citenamefont {Weller}, \citenamefont {Li}, \citenamefont {Bourg},
		\citenamefont {Chen},\ and\ \citenamefont {Yang}}]{LaHaye:2025ley}%
	\BibitemOpen
	\bibfield  {author} {\bibinfo {author} {\bibfnamefont {M.}~\bibnamefont
			{LaHaye}}, \bibinfo {author} {\bibfnamefont {C.}~\bibnamefont {Weller}},
		\bibinfo {author} {\bibfnamefont {D.}~\bibnamefont {Li}}, \bibinfo {author}
		{\bibfnamefont {P.}~\bibnamefont {Bourg}}, \bibinfo {author} {\bibfnamefont
			{Y.}~\bibnamefont {Chen}},\ and\ \bibinfo {author} {\bibfnamefont
			{H.}~\bibnamefont {Yang}},\ }\href {https://doi.org/10.1103/dy1z-36w5}
	{\bibfield  {journal} {\bibinfo  {journal} {Phys. Rev. D}\ }\textbf {\bibinfo
			{volume} {113}},\ \bibinfo {pages} {024069} (\bibinfo {year} {2026})},\
	\Eprint {https://arxiv.org/abs/2510.16102} {arXiv:2510.16102 [gr-qc]}
	\BibitemShut {NoStop}%
	\bibitem [{\citenamefont {Lu}\ \emph {et~al.}(2026)\citenamefont {Lu},
		\citenamefont {Lin}, \citenamefont {Zhu}, \citenamefont {Liu},\ and\
		\citenamefont {Zhang}}]{Lu:2025xlp}%
	\BibitemOpen
	\bibfield  {author} {\bibinfo {author} {\bibfnamefont {S.}~\bibnamefont
			{Lu}}, \bibinfo {author} {\bibfnamefont {H.-J.}\ \bibnamefont {Lin}},
		\bibinfo {author} {\bibfnamefont {T.}~\bibnamefont {Zhu}}, \bibinfo {author}
		{\bibfnamefont {Y.-X.}\ \bibnamefont {Liu}},\ and\ \bibinfo {author}
		{\bibfnamefont {X.}~\bibnamefont {Zhang}},\ }\href
	{https://doi.org/10.1140/epjc/s10052-026-15466-w} {\bibfield  {journal}
		{\bibinfo  {journal} {Eur. Phys. J. C}\ }\textbf {\bibinfo {volume} {86}},\
		\bibinfo {pages} {283} (\bibinfo {year} {2026})},\ \Eprint
	{https://arxiv.org/abs/2512.11911} {arXiv:2512.11911 [gr-qc]} \BibitemShut
	{NoStop}%
	\bibitem [{\citenamefont {Long}\ \emph {et~al.}(2026)\citenamefont {Long},
		\citenamefont {Xia}, \citenamefont {Gong}, \citenamefont {Cao}, \citenamefont
		{Pan},\ and\ \citenamefont {Jing}}]{Long:2026dcb}%
	\BibitemOpen
	\bibfield  {author} {\bibinfo {author} {\bibfnamefont {S.}~\bibnamefont
			{Long}}, \bibinfo {author} {\bibfnamefont {Z.-w.}\ \bibnamefont {Xia}},
		\bibinfo {author} {\bibfnamefont {H.}~\bibnamefont {Gong}}, \bibinfo {author}
		{\bibfnamefont {Z.}~\bibnamefont {Cao}}, \bibinfo {author} {\bibfnamefont
			{Q.}~\bibnamefont {Pan}},\ and\ \bibinfo {author} {\bibfnamefont
			{J.}~\bibnamefont {Jing}},\ }\href@noop {} {\  (\bibinfo {year} {2026})},\
	\Eprint {https://arxiv.org/abs/2605.05362} {arXiv:2605.05362 [gr-qc]}
	\BibitemShut {NoStop}%
	\bibitem [{\citenamefont {Du}\ \emph {et~al.}(2026)\citenamefont {Du},
		\citenamefont {Zhou}, \citenamefont {Xiong}, \citenamefont {Zi},\ and\
		\citenamefont {Li}}]{Du:2026xre}%
	\BibitemOpen
	\bibfield  {author} {\bibinfo {author} {\bibfnamefont {X.-D.}\ \bibnamefont
			{Du}}, \bibinfo {author} {\bibfnamefont {T.}~\bibnamefont {Zhou}}, \bibinfo
		{author} {\bibfnamefont {W.}~\bibnamefont {Xiong}}, \bibinfo {author}
		{\bibfnamefont {T.}~\bibnamefont {Zi}},\ and\ \bibinfo {author}
		{\bibfnamefont {P.-C.}\ \bibnamefont {Li}},\ }\href@noop {} {\  (\bibinfo
		{year} {2026})},\ \Eprint {https://arxiv.org/abs/2608.22249}
	{arXiv:2608.22249 [gr-qc]} \BibitemShut {NoStop}%
	\bibitem [{\citenamefont {Maselli}\ \emph {et~al.}(2020)\citenamefont
		{Maselli}, \citenamefont {Franchini}, \citenamefont {Gualtieri},\ and\
		\citenamefont {Sotiriou}}]{Maselli:2020zgv}%
	\BibitemOpen
	\bibfield  {author} {\bibinfo {author} {\bibfnamefont {A.}~\bibnamefont
			{Maselli}}, \bibinfo {author} {\bibfnamefont {N.}~\bibnamefont {Franchini}},
		\bibinfo {author} {\bibfnamefont {L.}~\bibnamefont {Gualtieri}},\ and\
		\bibinfo {author} {\bibfnamefont {T.~P.}\ \bibnamefont {Sotiriou}},\ }\href
	{https://doi.org/10.1103/PhysRevLett.125.141101} {\bibfield  {journal}
		{\bibinfo  {journal} {Phys. Rev. Lett.}\ }\textbf {\bibinfo {volume} {125}},\
		\bibinfo {pages} {141101} (\bibinfo {year} {2020})},\ \Eprint
	{https://arxiv.org/abs/2004.11895} {arXiv:2004.11895 [gr-qc]} \BibitemShut
	{NoStop}%
	\bibitem [{\citenamefont {Barsanti}\ \emph {et~al.}(2022)\citenamefont
		{Barsanti}, \citenamefont {Franchini}, \citenamefont {Gualtieri},
		\citenamefont {Maselli},\ and\ \citenamefont {Sotiriou}}]{Barsanti:2022ana}%
	\BibitemOpen
	\bibfield  {author} {\bibinfo {author} {\bibfnamefont {S.}~\bibnamefont
			{Barsanti}}, \bibinfo {author} {\bibfnamefont {N.}~\bibnamefont {Franchini}},
		\bibinfo {author} {\bibfnamefont {L.}~\bibnamefont {Gualtieri}}, \bibinfo
		{author} {\bibfnamefont {A.}~\bibnamefont {Maselli}},\ and\ \bibinfo {author}
		{\bibfnamefont {T.~P.}\ \bibnamefont {Sotiriou}},\ }\href
	{https://doi.org/10.1103/PhysRevD.106.044029} {\bibfield  {journal} {\bibinfo
			{journal} {Phys. Rev. D}\ }\textbf {\bibinfo {volume} {106}},\ \bibinfo
		{pages} {044029} (\bibinfo {year} {2022})},\ \Eprint
	{https://arxiv.org/abs/2203.05003} {arXiv:2203.05003 [gr-qc]} \BibitemShut
	{NoStop}%
	\bibitem [{\citenamefont {Zi}\ \emph {et~al.}(2023)\citenamefont {Zi},
		\citenamefont {Zhou}, \citenamefont {Wang}, \citenamefont {Li}, \citenamefont
		{Zhang},\ and\ \citenamefont {Chen}}]{Zi:2022hcc}%
	\BibitemOpen
	\bibfield  {author} {\bibinfo {author} {\bibfnamefont {T.}~\bibnamefont
			{Zi}}, \bibinfo {author} {\bibfnamefont {Z.}~\bibnamefont {Zhou}}, \bibinfo
		{author} {\bibfnamefont {H.-T.}\ \bibnamefont {Wang}}, \bibinfo {author}
		{\bibfnamefont {P.-C.}\ \bibnamefont {Li}}, \bibinfo {author} {\bibfnamefont
			{J.-d.}\ \bibnamefont {Zhang}},\ and\ \bibinfo {author} {\bibfnamefont
			{B.}~\bibnamefont {Chen}},\ }\href
	{https://doi.org/10.1103/PhysRevD.107.023005} {\bibfield  {journal} {\bibinfo
			{journal} {Phys. Rev. D}\ }\textbf {\bibinfo {volume} {107}},\ \bibinfo
		{pages} {023005} (\bibinfo {year} {2023})},\ \Eprint
	{https://arxiv.org/abs/2205.00425} {arXiv:2205.00425 [gr-qc]} \BibitemShut
	{NoStop}%
	\bibitem [{\citenamefont {Zi}\ and\ \citenamefont {Li}(2023)}]{Zi:2023omh}%
	\BibitemOpen
	\bibfield  {author} {\bibinfo {author} {\bibfnamefont {T.}~\bibnamefont
			{Zi}}\ and\ \bibinfo {author} {\bibfnamefont {P.-C.}\ \bibnamefont {Li}},\
	}\href {https://doi.org/10.1103/PhysRevD.108.084001} {\bibfield  {journal}
		{\bibinfo  {journal} {Phys. Rev. D}\ }\textbf {\bibinfo {volume} {108}},\
		\bibinfo {pages} {084001} (\bibinfo {year} {2023})},\ \Eprint
	{https://arxiv.org/abs/2306.02683} {arXiv:2306.02683 [gr-qc]} \BibitemShut
	{NoStop}%
	\bibitem [{\citenamefont {Hu}\ \emph {et~al.}(2026)\citenamefont {Hu},
		\citenamefont {Du}, \citenamefont {Li},\ and\ \citenamefont
		{Zi}}]{Hu:2026way}%
	\BibitemOpen
	\bibfield  {author} {\bibinfo {author} {\bibfnamefont {J.-L.}\ \bibnamefont
			{Hu}}, \bibinfo {author} {\bibfnamefont {X.-D.}\ \bibnamefont {Du}}, \bibinfo
		{author} {\bibfnamefont {P.-C.}\ \bibnamefont {Li}},\ and\ \bibinfo {author}
		{\bibfnamefont {T.}~\bibnamefont {Zi}},\ }\href@noop {} {\  (\bibinfo {year}
		{2026})},\ \Eprint {https://arxiv.org/abs/2607.05909} {arXiv:2607.05909
		[gr-qc]} \BibitemShut {NoStop}%
	\bibitem [{\citenamefont {Kocsis}\ \emph {et~al.}(2011)\citenamefont {Kocsis},
		\citenamefont {Yunes},\ and\ \citenamefont {Loeb}}]{Kocsis:2011dr}%
	\BibitemOpen
	\bibfield  {author} {\bibinfo {author} {\bibfnamefont {B.}~\bibnamefont
			{Kocsis}}, \bibinfo {author} {\bibfnamefont {N.}~\bibnamefont {Yunes}},\ and\
		\bibinfo {author} {\bibfnamefont {A.}~\bibnamefont {Loeb}},\ }\href
	{https://doi.org/10.1103/PhysRevD.86.049907} {\bibfield  {journal} {\bibinfo
			{journal} {Phys. Rev. D}\ }\textbf {\bibinfo {volume} {84}},\ \bibinfo
		{pages} {024032} (\bibinfo {year} {2011})},\ \Eprint
	{https://arxiv.org/abs/1104.2322} {arXiv:1104.2322 [astro-ph.GA]}
	\BibitemShut {NoStop}%
	\bibitem [{\citenamefont {Speri}\ \emph {et~al.}(2023)\citenamefont {Speri},
		\citenamefont {Antonelli}, \citenamefont {Sberna}, \citenamefont {Babak},
		\citenamefont {Barausse}, \citenamefont {Gair},\ and\ \citenamefont
		{Katz}}]{Speri:2022upm}%
	\BibitemOpen
	\bibfield  {author} {\bibinfo {author} {\bibfnamefont {L.}~\bibnamefont
			{Speri}}, \bibinfo {author} {\bibfnamefont {A.}~\bibnamefont {Antonelli}},
		\bibinfo {author} {\bibfnamefont {L.}~\bibnamefont {Sberna}}, \bibinfo
		{author} {\bibfnamefont {S.}~\bibnamefont {Babak}}, \bibinfo {author}
		{\bibfnamefont {E.}~\bibnamefont {Barausse}}, \bibinfo {author}
		{\bibfnamefont {J.~R.}\ \bibnamefont {Gair}},\ and\ \bibinfo {author}
		{\bibfnamefont {M.~L.}\ \bibnamefont {Katz}},\ }\href
	{https://doi.org/10.1103/PhysRevX.13.021035} {\bibfield  {journal} {\bibinfo
			{journal} {Phys. Rev. X}\ }\textbf {\bibinfo {volume} {13}},\ \bibinfo
		{pages} {021035} (\bibinfo {year} {2023})},\ \Eprint
	{https://arxiv.org/abs/2207.10086} {arXiv:2207.10086 [gr-qc]} \BibitemShut
	{NoStop}%
	\bibitem [{\citenamefont {Hegade K.~R.}\ \emph {et~al.}(2025)\citenamefont
		{Hegade K.~R.}, \citenamefont {Gammie},\ and\ \citenamefont
		{Yunes}}]{HegadeKR:2025dur}%
	\BibitemOpen
	\bibfield  {author} {\bibinfo {author} {\bibfnamefont {A.}~\bibnamefont
			{Hegade K.~R.}}, \bibinfo {author} {\bibfnamefont {C.~F.}\ \bibnamefont
			{Gammie}},\ and\ \bibinfo {author} {\bibfnamefont {N.}~\bibnamefont
			{Yunes}},\ }\href {https://doi.org/10.1103/9src-p7sp} {\bibfield  {journal}
		{\bibinfo  {journal} {Phys. Rev. D}\ }\textbf {\bibinfo {volume} {112}},\
		\bibinfo {pages} {124012} (\bibinfo {year} {2025})},\ \Eprint
	{https://arxiv.org/abs/2509.20457} {arXiv:2509.20457 [gr-qc]} \BibitemShut
	{NoStop}%
	\bibitem [{\citenamefont {Lui}\ \emph {et~al.}(2026{\natexlab{a}})\citenamefont
		{Lui}, \citenamefont {Chung},\ and\ \citenamefont
		{Torres-Orjuela}}]{Lui:2026azr}%
	\BibitemOpen
	\bibfield  {author} {\bibinfo {author} {\bibfnamefont {L.}~\bibnamefont
			{Lui}}, \bibinfo {author} {\bibfnamefont {A.~K.-W.}\ \bibnamefont {Chung}},\
		and\ \bibinfo {author} {\bibfnamefont {A.}~\bibnamefont {Torres-Orjuela}},\
	}\href@noop {} {\  (\bibinfo {year} {2026}{\natexlab{a}})},\ \Eprint
	{https://arxiv.org/abs/2608.28037} {arXiv:2608.28037 [gr-qc]} \BibitemShut
	{NoStop}%
	\bibitem [{\citenamefont {Dai}\ \emph {et~al.}(2024)\citenamefont {Dai},
		\citenamefont {Gong}, \citenamefont {Zhao},\ and\ \citenamefont
		{Jiang}}]{Dai:2023cft}%
	\BibitemOpen
	\bibfield  {author} {\bibinfo {author} {\bibfnamefont {N.}~\bibnamefont
			{Dai}}, \bibinfo {author} {\bibfnamefont {Y.}~\bibnamefont {Gong}}, \bibinfo
		{author} {\bibfnamefont {Y.}~\bibnamefont {Zhao}},\ and\ \bibinfo {author}
		{\bibfnamefont {T.}~\bibnamefont {Jiang}},\ }\href
	{https://doi.org/10.1103/PhysRevD.110.084080} {\bibfield  {journal} {\bibinfo
			{journal} {Phys. Rev. D}\ }\textbf {\bibinfo {volume} {110}},\ \bibinfo
		{pages} {084080} (\bibinfo {year} {2024})},\ \Eprint
	{https://arxiv.org/abs/2301.05088} {arXiv:2301.05088 [gr-qc]} \BibitemShut
	{NoStop}%
	\bibitem [{\citenamefont {Zhao}\ \emph
		{et~al.}(2025{\natexlab{b}})\citenamefont {Zhao}, \citenamefont {Dai},\ and\
		\citenamefont {Gong}}]{Zhao:2024bpp}%
	\BibitemOpen
	\bibfield  {author} {\bibinfo {author} {\bibfnamefont {Y.}~\bibnamefont
			{Zhao}}, \bibinfo {author} {\bibfnamefont {N.}~\bibnamefont {Dai}},\ and\
		\bibinfo {author} {\bibfnamefont {Y.}~\bibnamefont {Gong}},\ }\href
	{https://doi.org/10.1093/mnras/staf1471} {\bibfield  {journal} {\bibinfo
			{journal} {Mon. Not. Roy. Astron. Soc.}\ }\textbf {\bibinfo {volume} {543}},\
		\bibinfo {pages} {2326} (\bibinfo {year} {2025}{\natexlab{b}})},\ \Eprint
	{https://arxiv.org/abs/2410.06882} {arXiv:2410.06882 [gr-qc]} \BibitemShut
	{NoStop}%
	\bibitem [{\citenamefont {Mitra}\ \emph {et~al.}(2025)\citenamefont {Mitra},
		\citenamefont {Speeney}, \citenamefont {Chakraborty},\ and\ \citenamefont
		{Berti}}]{Mitra:2025tag}%
	\BibitemOpen
	\bibfield  {author} {\bibinfo {author} {\bibfnamefont {S.}~\bibnamefont
			{Mitra}}, \bibinfo {author} {\bibfnamefont {N.}~\bibnamefont {Speeney}},
		\bibinfo {author} {\bibfnamefont {S.}~\bibnamefont {Chakraborty}},\ and\
		\bibinfo {author} {\bibfnamefont {E.}~\bibnamefont {Berti}},\ }\href
	{https://doi.org/10.1103/ycl1-kx7d} {\bibfield  {journal} {\bibinfo
			{journal} {Phys. Rev. D}\ }\textbf {\bibinfo {volume} {112}},\ \bibinfo
		{pages} {044030} (\bibinfo {year} {2025})},\ \Eprint
	{https://arxiv.org/abs/2505.04697} {arXiv:2505.04697 [gr-qc]} \BibitemShut
	{NoStop}%
	\bibitem [{\citenamefont {Vicente}\ \emph {et~al.}(2025)\citenamefont
		{Vicente}, \citenamefont {Karydas},\ and\ \citenamefont
		{Bertone}}]{Vicente:2025gsg}%
	\BibitemOpen
	\bibfield  {author} {\bibinfo {author} {\bibfnamefont {R.}~\bibnamefont
			{Vicente}}, \bibinfo {author} {\bibfnamefont {T.~K.}\ \bibnamefont
			{Karydas}},\ and\ \bibinfo {author} {\bibfnamefont {G.}~\bibnamefont
			{Bertone}},\ }\href {https://doi.org/10.1103/s4wh-x6c4} {\bibfield  {journal}
		{\bibinfo  {journal} {Phys. Rev. Lett.}\ }\textbf {\bibinfo {volume} {135}},\
		\bibinfo {pages} {211401} (\bibinfo {year} {2025})},\ \Eprint
	{https://arxiv.org/abs/2505.09715} {arXiv:2505.09715 [gr-qc]} \BibitemShut
	{NoStop}%
	\bibitem [{\citenamefont {Wang}\ \emph {et~al.}(2025)\citenamefont {Wang},
		\citenamefont {Tang}, \citenamefont {Han},\ and\ \citenamefont
		{Liang}}]{Wang:2025rrj}%
	\BibitemOpen
	\bibfield  {author} {\bibinfo {author} {\bibfnamefont {Y.}~\bibnamefont
			{Wang}}, \bibinfo {author} {\bibfnamefont {R.}~\bibnamefont {Tang}}, \bibinfo
		{author} {\bibfnamefont {W.}~\bibnamefont {Han}},\ and\ \bibinfo {author}
		{\bibfnamefont {E.}~\bibnamefont {Liang}},\ }\href
	{https://doi.org/10.3390/sym17111878} {\bibfield  {journal} {\bibinfo
			{journal} {Symmetry}\ }\textbf {\bibinfo {volume} {17}},\ \bibinfo {pages}
		{1878} (\bibinfo {year} {2025})}\BibitemShut {NoStop}%
	\bibitem [{\citenamefont {Haroon}\ and\ \citenamefont
		{Zhu}(2025)}]{Haroon:2025rzx}%
	\BibitemOpen
	\bibfield  {author} {\bibinfo {author} {\bibfnamefont {S.}~\bibnamefont
			{Haroon}}\ and\ \bibinfo {author} {\bibfnamefont {T.}~\bibnamefont {Zhu}},\
	}\href {https://doi.org/10.1103/ckdt-wtsl} {\bibfield  {journal} {\bibinfo
			{journal} {Phys. Rev. D}\ }\textbf {\bibinfo {volume} {112}},\ \bibinfo
		{pages} {044046} (\bibinfo {year} {2025})},\ \Eprint
	{https://arxiv.org/abs/2502.09171} {arXiv:2502.09171 [gr-qc]} \BibitemShut
	{NoStop}%
	\bibitem [{\citenamefont {Rahman}\ and\ \citenamefont
		{Takahashi}(2026)}]{Rahman:2025mip}%
	\BibitemOpen
	\bibfield  {author} {\bibinfo {author} {\bibfnamefont {M.}~\bibnamefont
			{Rahman}}\ and\ \bibinfo {author} {\bibfnamefont {T.}~\bibnamefont
			{Takahashi}},\ }\href {https://doi.org/10.1103/76m9-rpzj} {\bibfield
		{journal} {\bibinfo  {journal} {Phys. Rev. D}\ }\textbf {\bibinfo {volume}
			{113}},\ \bibinfo {pages} {044033} (\bibinfo {year} {2026})},\ \Eprint
	{https://arxiv.org/abs/2507.06923} {arXiv:2507.06923 [gr-qc]} \BibitemShut
	{NoStop}%
	\bibitem [{\citenamefont {Das}\ \emph {et~al.}(2026)\citenamefont {Das},
		\citenamefont {Dalui}, \citenamefont {Lee},\ and\ \citenamefont
		{Cai}}]{Das:2025vja}%
	\BibitemOpen
	\bibfield  {author} {\bibinfo {author} {\bibfnamefont {S.}~\bibnamefont
			{Das}}, \bibinfo {author} {\bibfnamefont {S.}~\bibnamefont {Dalui}}, \bibinfo
		{author} {\bibfnamefont {B.-H.}\ \bibnamefont {Lee}},\ and\ \bibinfo {author}
		{\bibfnamefont {Y.-F.}\ \bibnamefont {Cai}},\ }\href
	{https://doi.org/10.1007/s11433-026-3011-6} {\bibfield  {journal} {\bibinfo
			{journal} {Sci. China Phys. Mech. Astron.}\ }\textbf {\bibinfo {volume}
			{69}},\ \bibinfo {pages} {290413} (\bibinfo {year} {2026})},\ \Eprint
	{https://arxiv.org/abs/2511.03657} {arXiv:2511.03657 [gr-qc]} \BibitemShut
	{NoStop}%
	\bibitem [{\citenamefont {Das}\ \emph {et~al.}(2025)\citenamefont {Das},
		\citenamefont {Dalui}, \citenamefont {Lee},\ and\ \citenamefont
		{Cai}}]{Das:2025eiv}%
	\BibitemOpen
	\bibfield  {author} {\bibinfo {author} {\bibfnamefont {S.}~\bibnamefont
			{Das}}, \bibinfo {author} {\bibfnamefont {S.}~\bibnamefont {Dalui}}, \bibinfo
		{author} {\bibfnamefont {B.-H.}\ \bibnamefont {Lee}},\ and\ \bibinfo {author}
		{\bibfnamefont {Y.-F.}\ \bibnamefont {Cai}},\ }\href@noop {} {\  (\bibinfo
		{year} {2025})},\ \Eprint {https://arxiv.org/abs/2512.04848}
	{arXiv:2512.04848 [gr-qc]} \BibitemShut {NoStop}%
	\bibitem [{\citenamefont {Azreg-A{\"\i}nou}\ \emph {et~al.}(2026)\citenamefont
		{Azreg-A{\"\i}nou}, \citenamefont {Jamil},\ and\ \citenamefont
		{Saridakis}}]{Azreg-Ainou:2026xcc}%
	\BibitemOpen
	\bibfield  {author} {\bibinfo {author} {\bibfnamefont {M.}~\bibnamefont
			{Azreg-A{\"\i}nou}}, \bibinfo {author} {\bibfnamefont {M.}~\bibnamefont
			{Jamil}},\ and\ \bibinfo {author} {\bibfnamefont {E.~N.}\ \bibnamefont
			{Saridakis}},\ }\href@noop {} {\  (\bibinfo {year} {2026})},\ \Eprint
	{https://arxiv.org/abs/2602.19541} {arXiv:2602.19541 [gr-qc]} \BibitemShut
	{NoStop}%
	\bibitem [{\citenamefont {Zhao}\ and\ \citenamefont
		{Gong}(2026)}]{Zhao:2026yis}%
	\BibitemOpen
	\bibfield  {author} {\bibinfo {author} {\bibfnamefont {Y.}~\bibnamefont
			{Zhao}}\ and\ \bibinfo {author} {\bibfnamefont {Y.}~\bibnamefont {Gong}},\
	}\href@noop {} {\  (\bibinfo {year} {2026})},\ \Eprint
	{https://arxiv.org/abs/2602.12022} {arXiv:2602.12022 [gr-qc]} \BibitemShut
	{NoStop}%
	\bibitem [{\citenamefont {Fu}\ \emph {et~al.}(2026)\citenamefont {Fu},
		\citenamefont {Liu}, \citenamefont {Wu}, \citenamefont {Wang},\ and\
		\citenamefont {Yue}}]{Fu:2026gcu}%
	\BibitemOpen
	\bibfield  {author} {\bibinfo {author} {\bibfnamefont {G.}~\bibnamefont
			{Fu}}, \bibinfo {author} {\bibfnamefont {Y.}~\bibnamefont {Liu}}, \bibinfo
		{author} {\bibfnamefont {J.-P.}\ \bibnamefont {Wu}}, \bibinfo {author}
		{\bibfnamefont {B.}~\bibnamefont {Wang}},\ and\ \bibinfo {author}
		{\bibfnamefont {R.-H.}\ \bibnamefont {Yue}},\ }\href@noop {} {\  (\bibinfo
		{year} {2026})},\ \Eprint {https://arxiv.org/abs/2608.15099}
	{arXiv:2608.15099 [gr-qc]} \BibitemShut {NoStop}%
	\bibitem [{\citenamefont {Barausse}\ \emph {et~al.}(2014)\citenamefont
		{Barausse}, \citenamefont {Cardoso},\ and\ \citenamefont
		{Pani}}]{Barausse:2014tra}%
	\BibitemOpen
	\bibfield  {author} {\bibinfo {author} {\bibfnamefont {E.}~\bibnamefont
			{Barausse}}, \bibinfo {author} {\bibfnamefont {V.}~\bibnamefont {Cardoso}},\
		and\ \bibinfo {author} {\bibfnamefont {P.}~\bibnamefont {Pani}},\ }\href
	{https://doi.org/10.1103/PhysRevD.89.104059} {\bibfield  {journal} {\bibinfo
			{journal} {Phys. Rev. D}\ }\textbf {\bibinfo {volume} {89}},\ \bibinfo
		{pages} {104059} (\bibinfo {year} {2014})},\ \Eprint
	{https://arxiv.org/abs/1404.7149} {arXiv:1404.7149 [gr-qc]} \BibitemShut
	{NoStop}%
	\bibitem [{\citenamefont {Cardoso}\ \emph {et~al.}(2022)\citenamefont
		{Cardoso}, \citenamefont {Destounis}, \citenamefont {Duque}, \citenamefont
		{Panosso~Macedo},\ and\ \citenamefont {Maselli}}]{Cardoso:2022whc}%
	\BibitemOpen
	\bibfield  {author} {\bibinfo {author} {\bibfnamefont {V.}~\bibnamefont
			{Cardoso}}, \bibinfo {author} {\bibfnamefont {K.}~\bibnamefont {Destounis}},
		\bibinfo {author} {\bibfnamefont {F.}~\bibnamefont {Duque}}, \bibinfo
		{author} {\bibfnamefont {R.}~\bibnamefont {Panosso~Macedo}},\ and\ \bibinfo
		{author} {\bibfnamefont {A.}~\bibnamefont {Maselli}},\ }\href
	{https://doi.org/10.1103/PhysRevLett.129.241103} {\bibfield  {journal}
		{\bibinfo  {journal} {Phys. Rev. Lett.}\ }\textbf {\bibinfo {volume} {129}},\
		\bibinfo {pages} {241103} (\bibinfo {year} {2022})},\ \Eprint
	{https://arxiv.org/abs/2210.01133} {arXiv:2210.01133 [gr-qc]} \BibitemShut
	{NoStop}%
	\bibitem [{\citenamefont {Jiang}\ and\ \citenamefont
		{Han}(2024)}]{Jiang:2024lwg}%
	\BibitemOpen
	\bibfield  {author} {\bibinfo {author} {\bibfnamefont {Y.}~\bibnamefont
			{Jiang}}\ and\ \bibinfo {author} {\bibfnamefont {W.-B.}\ \bibnamefont
			{Han}},\ }\href {https://doi.org/10.1007/s11433-024-2366-5} {\bibfield
		{journal} {\bibinfo  {journal} {Sci. China Phys. Mech. Astron.}\ }\textbf
		{\bibinfo {volume} {67}},\ \bibinfo {pages} {270411} (\bibinfo {year}
		{2024})}\BibitemShut {NoStop}%
	\bibitem [{\citenamefont {Dyson}\ \emph {et~al.}(2025)\citenamefont {Dyson},
		\citenamefont {Spieksma}, \citenamefont {Brito}, \citenamefont {van~de
			Meent},\ and\ \citenamefont {Dolan}}]{Dyson:2025dlj}%
	\BibitemOpen
	\bibfield  {author} {\bibinfo {author} {\bibfnamefont {C.}~\bibnamefont
			{Dyson}}, \bibinfo {author} {\bibfnamefont {T.~F.~M.}\ \bibnamefont
			{Spieksma}}, \bibinfo {author} {\bibfnamefont {R.}~\bibnamefont {Brito}},
		\bibinfo {author} {\bibfnamefont {M.}~\bibnamefont {van~de Meent}},\ and\
		\bibinfo {author} {\bibfnamefont {S.}~\bibnamefont {Dolan}},\ }\href
	{https://doi.org/10.1103/PhysRevLett.134.211403} {\bibfield  {journal}
		{\bibinfo  {journal} {Phys. Rev. Lett.}\ }\textbf {\bibinfo {volume} {134}},\
		\bibinfo {pages} {211403} (\bibinfo {year} {2025})},\ \Eprint
	{https://arxiv.org/abs/2501.09806} {arXiv:2501.09806 [gr-qc]} \BibitemShut
	{NoStop}%
	\bibitem [{\citenamefont {Polcar}\ and\ \citenamefont
		{Witzany}(2025)}]{Polcar:2025yto}%
	\BibitemOpen
	\bibfield  {author} {\bibinfo {author} {\bibfnamefont {L.}~\bibnamefont
			{Polcar}}\ and\ \bibinfo {author} {\bibfnamefont {V.}~\bibnamefont
			{Witzany}},\ }\href {https://doi.org/10.1103/79w6-s8sk} {\bibfield  {journal}
		{\bibinfo  {journal} {Phys. Rev. D}\ }\textbf {\bibinfo {volume} {112}},\
		\bibinfo {pages} {104003} (\bibinfo {year} {2025})},\ \Eprint
	{https://arxiv.org/abs/2507.15720} {arXiv:2507.15720 [gr-qc]} \BibitemShut
	{NoStop}%
	\bibitem [{\citenamefont {Lui}\ \emph {et~al.}(2026{\natexlab{b}})\citenamefont
		{Lui}, \citenamefont {Drummond},\ and\ \citenamefont
		{Torres-Orjuela}}]{Lui:2026uai}%
	\BibitemOpen
	\bibfield  {author} {\bibinfo {author} {\bibfnamefont {L.}~\bibnamefont
			{Lui}}, \bibinfo {author} {\bibfnamefont {L.~V.}\ \bibnamefont {Drummond}},\
		and\ \bibinfo {author} {\bibfnamefont {A.}~\bibnamefont {Torres-Orjuela}},\
	}\href {https://doi.org/10.1103/dy3h-65ty} {\bibfield  {journal} {\bibinfo
			{journal} {Phys. Rev. Lett.}\ }\textbf {\bibinfo {volume} {137}},\ \bibinfo
		{pages} {121403} (\bibinfo {year} {2026}{\natexlab{b}})},\ \Eprint
	{https://arxiv.org/abs/2606.01569} {arXiv:2606.01569 [gr-qc]} \BibitemShut
	{NoStop}%
	\bibitem [{\citenamefont {Kejriwal}\ \emph {et~al.}(2026)\citenamefont
		{Kejriwal}, \citenamefont {Barausse},\ and\ \citenamefont
		{Chua}}]{Kejriwal:2025jao}%
	\BibitemOpen
	\bibfield  {author} {\bibinfo {author} {\bibfnamefont {S.}~\bibnamefont
			{Kejriwal}}, \bibinfo {author} {\bibfnamefont {E.}~\bibnamefont {Barausse}},\
		and\ \bibinfo {author} {\bibfnamefont {A.~J.~K.}\ \bibnamefont {Chua}},\
	}\href {https://doi.org/10.1103/rnvp-jbj4} {\bibfield  {journal} {\bibinfo
			{journal} {Phys. Rev. D}\ }\textbf {\bibinfo {volume} {113}},\ \bibinfo
		{pages} {064001} (\bibinfo {year} {2026})},\ \Eprint
	{https://arxiv.org/abs/2510.17398} {arXiv:2510.17398 [gr-qc]} \BibitemShut
	{NoStop}%
	\bibitem [{\citenamefont {Copparoni}\ \emph {et~al.}(2026)\citenamefont
		{Copparoni}, \citenamefont {Chandramouli},\ and\ \citenamefont
		{Barausse}}]{Copparoni:2025vty}%
	\BibitemOpen
	\bibfield  {author} {\bibinfo {author} {\bibfnamefont {L.}~\bibnamefont
			{Copparoni}}, \bibinfo {author} {\bibfnamefont {R.~S.}\ \bibnamefont
			{Chandramouli}},\ and\ \bibinfo {author} {\bibfnamefont {E.}~\bibnamefont
			{Barausse}},\ }\href {https://doi.org/10.1103/pqcz-cvsv} {\bibfield
		{journal} {\bibinfo  {journal} {Phys. Rev. Lett.}\ }\textbf {\bibinfo
			{volume} {137}},\ \bibinfo {pages} {021405} (\bibinfo {year} {2026})},\
	\Eprint {https://arxiv.org/abs/2510.06948} {arXiv:2510.06948 [gr-qc]}
	\BibitemShut {NoStop}%
	\bibitem [{\citenamefont {Zi}\ \emph {et~al.}(2026)\citenamefont {Zi},
		\citenamefont {Rahman},\ and\ \citenamefont {Kumar}}]{Zi:2026zpw}%
	\BibitemOpen
	\bibfield  {author} {\bibinfo {author} {\bibfnamefont {T.}~\bibnamefont
			{Zi}}, \bibinfo {author} {\bibfnamefont {M.}~\bibnamefont {Rahman}},\ and\
		\bibinfo {author} {\bibfnamefont {S.}~\bibnamefont {Kumar}},\ }\href@noop {}
	{\  (\bibinfo {year} {2026})},\ \Eprint {https://arxiv.org/abs/2601.03374}
	{arXiv:2601.03374 [gr-qc]} \BibitemShut {NoStop}%
	\bibitem [{\citenamefont {Arun}\ \emph {et~al.}(2022)\citenamefont {Arun} \emph
		{et~al.}}]{LISA:2022kgy}%
	\BibitemOpen
	\bibfield  {author} {\bibinfo {author} {\bibfnamefont {K.~G.}\ \bibnamefont
			{Arun}} \emph {et~al.} (\bibinfo {collaboration} {LISA}),\ }\href
	{https://doi.org/10.1007/s41114-022-00036-9} {\bibfield  {journal} {\bibinfo
			{journal} {Living Rev. Rel.}\ }\textbf {\bibinfo {volume} {25}},\ \bibinfo
		{pages} {4} (\bibinfo {year} {2022})},\ \Eprint
	{https://arxiv.org/abs/2205.01597} {arXiv:2205.01597 [gr-qc]} \BibitemShut
	{NoStop}%
	\bibitem [{\citenamefont {Luo}\ \emph {et~al.}(2026)\citenamefont {Luo} \emph
		{et~al.}}]{Luo:2025ewp}%
	\BibitemOpen
	\bibfield  {author} {\bibinfo {author} {\bibfnamefont {J.}~\bibnamefont
			{Luo}} \emph {et~al.},\ }\href {https://doi.org/10.1007/s41114-025-00063-2}
	{\bibfield  {journal} {\bibinfo  {journal} {Living Rev. Rel.}\ }\textbf
		{\bibinfo {volume} {29}},\ \bibinfo {pages} {1} (\bibinfo {year} {2026})},\
	\Eprint {https://arxiv.org/abs/2502.20138} {arXiv:2502.20138 [gr-qc]}
	\BibitemShut {NoStop}%
	\bibitem [{\citenamefont {Alexander}\ and\ \citenamefont
		{Yunes}(2009)}]{Alexander:2009cs}%
	\BibitemOpen
	\bibfield  {author} {\bibinfo {author} {\bibfnamefont {S.}~\bibnamefont
			{Alexander}}\ and\ \bibinfo {author} {\bibfnamefont {N.}~\bibnamefont
			{Yunes}},\ }\href {https://doi.org/10.1016/j.physrep.2009.07.002} {\bibfield
		{journal} {\bibinfo  {journal} {Phys. Rept.}\ }\textbf {\bibinfo {volume}
			{480}},\ \bibinfo {pages} {1} (\bibinfo {year} {2009})},\ \Eprint
	{https://arxiv.org/abs/0907.2562} {arXiv:0907.2562 [hep-th]} \BibitemShut
	{NoStop}%
	\bibitem [{\citenamefont {Qiao}\ \emph {et~al.}(2023)\citenamefont {Qiao},
		\citenamefont {Li}, \citenamefont {Zhu}, \citenamefont {Ji}, \citenamefont
		{Li},\ and\ \citenamefont {Zhao}}]{Qiao:2022mln}%
	\BibitemOpen
	\bibfield  {author} {\bibinfo {author} {\bibfnamefont {J.}~\bibnamefont
			{Qiao}}, \bibinfo {author} {\bibfnamefont {Z.}~\bibnamefont {Li}}, \bibinfo
		{author} {\bibfnamefont {T.}~\bibnamefont {Zhu}}, \bibinfo {author}
		{\bibfnamefont {R.}~\bibnamefont {Ji}}, \bibinfo {author} {\bibfnamefont
			{G.}~\bibnamefont {Li}},\ and\ \bibinfo {author} {\bibfnamefont
			{W.}~\bibnamefont {Zhao}},\ }\href
	{https://doi.org/10.3389/fspas.2022.1109086} {\bibfield  {journal} {\bibinfo
			{journal} {Front. Astron. Space Sci.}\ }\textbf {\bibinfo {volume} {9}},\
		\bibinfo {pages} {1109086} (\bibinfo {year} {2023})},\ \Eprint
	{https://arxiv.org/abs/2211.16825} {arXiv:2211.16825 [gr-qc]} \BibitemShut
	{NoStop}%
	\bibitem [{\citenamefont {Alexander}\ and\ \citenamefont
		{Yunes}(2018)}]{Alexander:2017jmt}%
	\BibitemOpen
	\bibfield  {author} {\bibinfo {author} {\bibfnamefont {S.~H.}\ \bibnamefont
			{Alexander}}\ and\ \bibinfo {author} {\bibfnamefont {N.}~\bibnamefont
			{Yunes}},\ }\href {https://doi.org/10.1103/PhysRevD.97.064033} {\bibfield
		{journal} {\bibinfo  {journal} {Phys. Rev. D}\ }\textbf {\bibinfo {volume}
			{97}},\ \bibinfo {pages} {064033} (\bibinfo {year} {2018})},\ \Eprint
	{https://arxiv.org/abs/1712.01853} {arXiv:1712.01853 [gr-qc]} \BibitemShut
	{NoStop}%
	\bibitem [{\citenamefont {Zhao}\ \emph {et~al.}(2020)\citenamefont {Zhao},
		\citenamefont {Zhu}, \citenamefont {Qiao},\ and\ \citenamefont
		{Wang}}]{Zhao:2019xmm}%
	\BibitemOpen
	\bibfield  {author} {\bibinfo {author} {\bibfnamefont {W.}~\bibnamefont
			{Zhao}}, \bibinfo {author} {\bibfnamefont {T.}~\bibnamefont {Zhu}}, \bibinfo
		{author} {\bibfnamefont {J.}~\bibnamefont {Qiao}},\ and\ \bibinfo {author}
		{\bibfnamefont {A.}~\bibnamefont {Wang}},\ }\href
	{https://doi.org/10.1103/PhysRevD.101.024002} {\bibfield  {journal} {\bibinfo
			{journal} {Phys. Rev. D}\ }\textbf {\bibinfo {volume} {101}},\ \bibinfo
		{pages} {024002} (\bibinfo {year} {2020})},\ \Eprint
	{https://arxiv.org/abs/1909.10887} {arXiv:1909.10887 [gr-qc]} \BibitemShut
	{NoStop}%
	\bibitem [{\citenamefont {Yunes}\ and\ \citenamefont
		{Finn}(2009)}]{Yunes:2008bu}%
	\BibitemOpen
	\bibfield  {author} {\bibinfo {author} {\bibfnamefont {N.}~\bibnamefont
			{Yunes}}\ and\ \bibinfo {author} {\bibfnamefont {L.~S.}\ \bibnamefont
			{Finn}},\ }\href {https://doi.org/10.1088/1742-6596/154/1/012041} {\bibfield
		{journal} {\bibinfo  {journal} {J. Phys. Conf. Ser.}\ }\textbf {\bibinfo
			{volume} {154}},\ \bibinfo {pages} {012041} (\bibinfo {year} {2009})},\
	\Eprint {https://arxiv.org/abs/0811.0181} {arXiv:0811.0181 [gr-qc]}
	\BibitemShut {NoStop}%
	\bibitem [{\citenamefont {Wang}\ \emph {et~al.}(2022)\citenamefont {Wang},
		\citenamefont {Brown}, \citenamefont {Shao},\ and\ \citenamefont
		{Zhao}}]{Wang:2021gqm}%
	\BibitemOpen
	\bibfield  {author} {\bibinfo {author} {\bibfnamefont {Y.-F.}\ \bibnamefont
			{Wang}}, \bibinfo {author} {\bibfnamefont {S.~M.}\ \bibnamefont {Brown}},
		\bibinfo {author} {\bibfnamefont {L.}~\bibnamefont {Shao}},\ and\ \bibinfo
		{author} {\bibfnamefont {W.}~\bibnamefont {Zhao}},\ }\href
	{https://doi.org/10.1103/PhysRevD.106.084005} {\bibfield  {journal} {\bibinfo
			{journal} {Phys. Rev. D}\ }\textbf {\bibinfo {volume} {106}},\ \bibinfo
		{pages} {084005} (\bibinfo {year} {2022})},\ \Eprint
	{https://arxiv.org/abs/2109.09718} {arXiv:2109.09718 [astro-ph.HE]}
	\BibitemShut {NoStop}%
	\bibitem [{\citenamefont {Ng}\ \emph {et~al.}(2023)\citenamefont {Ng},
		\citenamefont {Isi}, \citenamefont {Wong},\ and\ \citenamefont
		{Farr}}]{Ng:2023jjt}%
	\BibitemOpen
	\bibfield  {author} {\bibinfo {author} {\bibfnamefont {T.~C.~K.}\
			\bibnamefont {Ng}}, \bibinfo {author} {\bibfnamefont {M.}~\bibnamefont
			{Isi}}, \bibinfo {author} {\bibfnamefont {K.~W.~K.}\ \bibnamefont {Wong}},\
		and\ \bibinfo {author} {\bibfnamefont {W.~M.}\ \bibnamefont {Farr}},\ }\href
	{https://doi.org/10.1103/PhysRevD.108.084068} {\bibfield  {journal} {\bibinfo
			{journal} {Phys. Rev. D}\ }\textbf {\bibinfo {volume} {108}},\ \bibinfo
		{pages} {084068} (\bibinfo {year} {2023})},\ \Eprint
	{https://arxiv.org/abs/2305.05844} {arXiv:2305.05844 [gr-qc]} \BibitemShut
	{NoStop}%
	\bibitem [{\citenamefont {Gong}\ \emph {et~al.}(2022)\citenamefont {Gong},
		\citenamefont {Zhu}, \citenamefont {Niu}, \citenamefont {Wu}, \citenamefont
		{Cui}, \citenamefont {Zhang}, \citenamefont {Zhao},\ and\ \citenamefont
		{Wang}}]{Gong:2021jgg}%
	\BibitemOpen
	\bibfield  {author} {\bibinfo {author} {\bibfnamefont {C.}~\bibnamefont
			{Gong}}, \bibinfo {author} {\bibfnamefont {T.}~\bibnamefont {Zhu}}, \bibinfo
		{author} {\bibfnamefont {R.}~\bibnamefont {Niu}}, \bibinfo {author}
		{\bibfnamefont {Q.}~\bibnamefont {Wu}}, \bibinfo {author} {\bibfnamefont
			{J.-L.}\ \bibnamefont {Cui}}, \bibinfo {author} {\bibfnamefont
			{X.}~\bibnamefont {Zhang}}, \bibinfo {author} {\bibfnamefont
			{W.}~\bibnamefont {Zhao}},\ and\ \bibinfo {author} {\bibfnamefont
			{A.}~\bibnamefont {Wang}},\ }\href
	{https://doi.org/10.1103/PhysRevD.105.044034} {\bibfield  {journal} {\bibinfo
			{journal} {Phys. Rev. D}\ }\textbf {\bibinfo {volume} {105}},\ \bibinfo
		{pages} {044034} (\bibinfo {year} {2022})},\ \Eprint
	{https://arxiv.org/abs/2112.06446} {arXiv:2112.06446 [gr-qc]} \BibitemShut
	{NoStop}%
	\bibitem [{\citenamefont {Zhu}\ \emph {et~al.}(2024)\citenamefont {Zhu},
		\citenamefont {Zhao}, \citenamefont {Yan}, \citenamefont {Wang},
		\citenamefont {Gong},\ and\ \citenamefont {Wang}}]{Zhu:2023rrx}%
	\BibitemOpen
	\bibfield  {author} {\bibinfo {author} {\bibfnamefont {T.}~\bibnamefont
			{Zhu}}, \bibinfo {author} {\bibfnamefont {W.}~\bibnamefont {Zhao}}, \bibinfo
		{author} {\bibfnamefont {J.-M.}\ \bibnamefont {Yan}}, \bibinfo {author}
		{\bibfnamefont {Y.-Z.}\ \bibnamefont {Wang}}, \bibinfo {author}
		{\bibfnamefont {C.}~\bibnamefont {Gong}},\ and\ \bibinfo {author}
		{\bibfnamefont {A.}~\bibnamefont {Wang}},\ }\href
	{https://doi.org/10.1103/PhysRevD.110.064044} {\bibfield  {journal} {\bibinfo
			{journal} {Phys. Rev. D}\ }\textbf {\bibinfo {volume} {110}},\ \bibinfo
		{pages} {064044} (\bibinfo {year} {2024})},\ \Eprint
	{https://arxiv.org/abs/2304.09025} {arXiv:2304.09025 [gr-qc]} \BibitemShut
	{NoStop}%
	\bibitem [{\citenamefont {Sopuerta}\ and\ \citenamefont
		{Yunes}(2009)}]{Sopuerta:2009cs}%
	\BibitemOpen
	\bibfield  {author} {\bibinfo {author} {\bibfnamefont {C.~F.}\ \bibnamefont
			{Sopuerta}}\ and\ \bibinfo {author} {\bibfnamefont {N.}~\bibnamefont
			{Yunes}},\ }\href {https://doi.org/10.1103/PhysRevD.80.064006} {\bibfield
		{journal} {\bibinfo  {journal} {Phys. Rev. D}\ }\textbf {\bibinfo {volume}
			{80}},\ \bibinfo {pages} {064006} (\bibinfo {year} {2009})},\ \Eprint
	{https://arxiv.org/abs/0904.4501} {arXiv:0904.4501 [gr-qc]} \BibitemShut
	{NoStop}%
	\bibitem [{\citenamefont {Canizares}\ \emph {et~al.}(2012)\citenamefont
		{Canizares}, \citenamefont {Gair},\ and\ \citenamefont
		{Sopuerta}}]{Canizares:2012is}%
	\BibitemOpen
	\bibfield  {author} {\bibinfo {author} {\bibfnamefont {P.}~\bibnamefont
			{Canizares}}, \bibinfo {author} {\bibfnamefont {J.~R.}\ \bibnamefont
			{Gair}},\ and\ \bibinfo {author} {\bibfnamefont {C.~F.}\ \bibnamefont
			{Sopuerta}},\ }\href {https://doi.org/10.1103/PhysRevD.86.044010} {\bibfield
		{journal} {\bibinfo  {journal} {Phys. Rev. D}\ }\textbf {\bibinfo {volume}
			{86}},\ \bibinfo {pages} {044010} (\bibinfo {year} {2012})},\ \Eprint
	{https://arxiv.org/abs/1205.1253} {arXiv:1205.1253 [gr-qc]} \BibitemShut
	{NoStop}%
	\bibitem [{\citenamefont {Li}\ \emph {et~al.}(2023)\citenamefont {Li},
		\citenamefont {Qiao}, \citenamefont {Liu}, \citenamefont {Zhu},\ and\
		\citenamefont {Zhao}}]{Li:2022dcswaveform}%
	\BibitemOpen
	\bibfield  {author} {\bibinfo {author} {\bibfnamefont {Z.}~\bibnamefont
			{Li}}, \bibinfo {author} {\bibfnamefont {J.}~\bibnamefont {Qiao}}, \bibinfo
		{author} {\bibfnamefont {T.}~\bibnamefont {Liu}}, \bibinfo {author}
		{\bibfnamefont {T.}~\bibnamefont {Zhu}},\ and\ \bibinfo {author}
		{\bibfnamefont {W.}~\bibnamefont {Zhao}},\ }\href
	{https://doi.org/10.1088/1475-7516/2023/04/006} {\bibfield  {journal}
		{\bibinfo  {journal} {JCAP}\ }\textbf {\bibinfo {volume} {2023}}\bibfield
		{number} {\bibinfo  {number} { (04)},\ \bibinfo {pages} {006}},\ }\Eprint
	{https://arxiv.org/abs/2211.12188} {arXiv:2211.12188 [gr-qc]} \BibitemShut
	{NoStop}%
	\bibitem [{\citenamefont {Li}\ \emph {et~al.}(2024)\citenamefont {Li},
		\citenamefont {Qiao}, \citenamefont {Liu}, \citenamefont {Niu}, \citenamefont
		{Hou}, \citenamefont {Zhu},\ and\ \citenamefont
		{Zhao}}]{Li:2023dcseccentric}%
	\BibitemOpen
	\bibfield  {author} {\bibinfo {author} {\bibfnamefont {Z.}~\bibnamefont
			{Li}}, \bibinfo {author} {\bibfnamefont {J.}~\bibnamefont {Qiao}}, \bibinfo
		{author} {\bibfnamefont {T.}~\bibnamefont {Liu}}, \bibinfo {author}
		{\bibfnamefont {R.}~\bibnamefont {Niu}}, \bibinfo {author} {\bibfnamefont
			{S.}~\bibnamefont {Hou}}, \bibinfo {author} {\bibfnamefont {T.}~\bibnamefont
			{Zhu}},\ and\ \bibinfo {author} {\bibfnamefont {W.}~\bibnamefont {Zhao}},\
	}\href {https://doi.org/10.1088/1475-7516/2024/05/073} {\bibfield  {journal}
		{\bibinfo  {journal} {JCAP}\ }\textbf {\bibinfo {volume} {2024}}\bibfield
		{number} {\bibinfo  {number} { (05)},\ \bibinfo {pages} {073}},\ }\Eprint
	{https://arxiv.org/abs/2309.05991} {arXiv:2309.05991 [gr-qc]} \BibitemShut
	{NoStop}%
	\bibitem [{\citenamefont {Fransen}\ and\ \citenamefont
		{Mayerson}(2022)}]{Fransen:2022jtw}%
	\BibitemOpen
	\bibfield  {author} {\bibinfo {author} {\bibfnamefont {K.}~\bibnamefont
			{Fransen}}\ and\ \bibinfo {author} {\bibfnamefont {D.~R.}\ \bibnamefont
			{Mayerson}},\ }\href {https://doi.org/10.1103/PhysRevD.106.064035} {\bibfield
		{journal} {\bibinfo  {journal} {Phys. Rev. D}\ }\textbf {\bibinfo {volume}
			{106}},\ \bibinfo {pages} {064035} (\bibinfo {year} {2022})},\ \Eprint
	{https://arxiv.org/abs/2201.03569} {arXiv:2201.03569 [gr-qc]} \BibitemShut
	{NoStop}%
	\bibitem [{\citenamefont {Muguruza}\ and\ \citenamefont
		{Sopuerta}(2026)}]{Muguruza:2026hqn}%
	\BibitemOpen
	\bibfield  {author} {\bibinfo {author} {\bibfnamefont {P.~F.}\ \bibnamefont
			{Muguruza}}\ and\ \bibinfo {author} {\bibfnamefont {C.~F.}\ \bibnamefont
			{Sopuerta}},\ }\href {https://doi.org/10.1103/l12t-9c9m} {\bibfield
		{journal} {\bibinfo  {journal} {Phys. Rev. D}\ }\textbf {\bibinfo {volume}
			{114}},\ \bibinfo {pages} {064010} (\bibinfo {year} {2026})},\ \Eprint
	{https://arxiv.org/abs/2604.06053} {arXiv:2604.06053 [gr-qc]} \BibitemShut
	{NoStop}%
	\bibitem [{\citenamefont {Tahara}\ \emph {et~al.}(2024)\citenamefont {Tahara},
		\citenamefont {Takahashi}, \citenamefont {Minamitsuji},\ and\ \citenamefont
		{Motohashi}}]{Tahara:2023pyg}%
	\BibitemOpen
	\bibfield  {author} {\bibinfo {author} {\bibfnamefont {H.~W.~H.}\
			\bibnamefont {Tahara}}, \bibinfo {author} {\bibfnamefont {K.}~\bibnamefont
			{Takahashi}}, \bibinfo {author} {\bibfnamefont {M.}~\bibnamefont
			{Minamitsuji}},\ and\ \bibinfo {author} {\bibfnamefont {H.}~\bibnamefont
			{Motohashi}},\ }\href {https://doi.org/10.1093/ptep/ptae046} {\bibfield
		{journal} {\bibinfo  {journal} {PTEP}\ }\textbf {\bibinfo {volume} {2024}},\
		\bibinfo {pages} {053E02} (\bibinfo {year} {2024})},\ \Eprint
	{https://arxiv.org/abs/2312.11899} {arXiv:2312.11899 [gr-qc]} \BibitemShut
	{NoStop}%
	\bibitem [{\citenamefont {Glampedakis}\ \emph {et~al.}(2002)\citenamefont
		{Glampedakis}, \citenamefont {Hughes},\ and\ \citenamefont
		{Kennefick}}]{Glampedakis:2002cb}%
	\BibitemOpen
	\bibfield  {author} {\bibinfo {author} {\bibfnamefont {K.}~\bibnamefont
			{Glampedakis}}, \bibinfo {author} {\bibfnamefont {S.~A.}\ \bibnamefont
			{Hughes}},\ and\ \bibinfo {author} {\bibfnamefont {D.}~\bibnamefont
			{Kennefick}},\ }\href {https://doi.org/10.1103/PhysRevD.66.064005} {\bibfield
		{journal} {\bibinfo  {journal} {Phys. Rev. D}\ }\textbf {\bibinfo {volume}
			{66}},\ \bibinfo {pages} {064005} (\bibinfo {year} {2002})},\ \Eprint
	{https://arxiv.org/abs/gr-qc/0205033} {arXiv:gr-qc/0205033} \BibitemShut
	{NoStop}%
	\bibitem [{\citenamefont {Babak}\ \emph {et~al.}(2007)\citenamefont {Babak},
		\citenamefont {Fang}, \citenamefont {Gair}, \citenamefont {Glampedakis},\
		and\ \citenamefont {Hughes}}]{Babak:2006uv}%
	\BibitemOpen
	\bibfield  {author} {\bibinfo {author} {\bibfnamefont {S.}~\bibnamefont
			{Babak}}, \bibinfo {author} {\bibfnamefont {H.}~\bibnamefont {Fang}},
		\bibinfo {author} {\bibfnamefont {J.~R.}\ \bibnamefont {Gair}}, \bibinfo
		{author} {\bibfnamefont {K.}~\bibnamefont {Glampedakis}},\ and\ \bibinfo
		{author} {\bibfnamefont {S.~A.}\ \bibnamefont {Hughes}},\ }\href
	{https://doi.org/10.1103/PhysRevD.75.024005} {\bibfield  {journal} {\bibinfo
			{journal} {Phys. Rev. D}\ }\textbf {\bibinfo {volume} {75}},\ \bibinfo
		{pages} {024005} (\bibinfo {year} {2007})},\ \bibinfo {note} {[Erratum:
		Phys.Rev.D 77, 04990 (2008)]},\ \Eprint {https://arxiv.org/abs/gr-qc/0607007}
	{arXiv:gr-qc/0607007} \BibitemShut {NoStop}%
	\bibitem [{\citenamefont {Takahashi}\ \emph {et~al.}(2023)\citenamefont
		{Takahashi}, \citenamefont {Minamitsuji},\ and\ \citenamefont
		{Motohashi}}]{Takahashi:2022mew}%
	\BibitemOpen
	\bibfield  {author} {\bibinfo {author} {\bibfnamefont {K.}~\bibnamefont
			{Takahashi}}, \bibinfo {author} {\bibfnamefont {M.}~\bibnamefont
			{Minamitsuji}},\ and\ \bibinfo {author} {\bibfnamefont {H.}~\bibnamefont
			{Motohashi}},\ }\href {https://doi.org/10.1093/ptep/ptac161} {\bibfield
		{journal} {\bibinfo  {journal} {PTEP}\ }\textbf {\bibinfo {volume} {2023}},\
		\bibinfo {pages} {013E01} (\bibinfo {year} {2023})},\ \Eprint
	{https://arxiv.org/abs/2209.02176} {arXiv:2209.02176 [gr-qc]} \BibitemShut
	{NoStop}%
	\bibitem [{\citenamefont {Bardeen}\ \emph {et~al.}(1972)\citenamefont
		{Bardeen}, \citenamefont {Press},\ and\ \citenamefont
		{Teukolsky}}]{Bardeen:1972isco}%
	\BibitemOpen
	\bibfield  {author} {\bibinfo {author} {\bibfnamefont {J.~M.}\ \bibnamefont
			{Bardeen}}, \bibinfo {author} {\bibfnamefont {W.~H.}\ \bibnamefont {Press}},\
		and\ \bibinfo {author} {\bibfnamefont {S.~A.}\ \bibnamefont {Teukolsky}},\
	}\href {https://doi.org/10.1086/151796} {\bibfield  {journal} {\bibinfo
			{journal} {Astrophys. J.}\ }\textbf {\bibinfo {volume} {178}},\ \bibinfo
		{pages} {347} (\bibinfo {year} {1972})}\BibitemShut {NoStop}%
	\bibitem [{\citenamefont {Einstein}(1918)}]{Einstein:1918btx}%
	\BibitemOpen
	\bibfield  {author} {\bibinfo {author} {\bibfnamefont {A.}~\bibnamefont
			{Einstein}},\ }\href@noop {} {\bibfield  {journal} {\bibinfo  {journal}
			{Sitzungsber. Preuss. Akad. Wiss. Berlin (Math. Phys. )}\ }\textbf {\bibinfo
			{volume} {1918}},\ \bibinfo {pages} {154} (\bibinfo {year}
		{1918})}\BibitemShut {NoStop}%
	\bibitem [{\citenamefont {Thorne}(1980)}]{Thorne:1980ru}%
	\BibitemOpen
	\bibfield  {author} {\bibinfo {author} {\bibfnamefont {K.~S.}\ \bibnamefont
			{Thorne}},\ }\href {https://doi.org/10.1103/RevModPhys.52.299} {\bibfield
		{journal} {\bibinfo  {journal} {Rev. Mod. Phys.}\ }\textbf {\bibinfo {volume}
			{52}},\ \bibinfo {pages} {299} (\bibinfo {year} {1980})}\BibitemShut
	{NoStop}%
	\bibitem [{\citenamefont {Peters}\ and\ \citenamefont
		{Mathews}(1963)}]{Peters:1963ux}%
	\BibitemOpen
	\bibfield  {author} {\bibinfo {author} {\bibfnamefont {P.~C.}\ \bibnamefont
			{Peters}}\ and\ \bibinfo {author} {\bibfnamefont {J.}~\bibnamefont
			{Mathews}},\ }\href {https://doi.org/10.1103/PhysRev.131.435} {\bibfield
		{journal} {\bibinfo  {journal} {Phys. Rev.}\ }\textbf {\bibinfo {volume}
			{131}},\ \bibinfo {pages} {435} (\bibinfo {year} {1963})}\BibitemShut
	{NoStop}%
	\bibitem [{\citenamefont {Ori}\ and\ \citenamefont
		{Thorne}(2000)}]{Ori:2000transition}%
	\BibitemOpen
	\bibfield  {author} {\bibinfo {author} {\bibfnamefont {A.}~\bibnamefont
			{Ori}}\ and\ \bibinfo {author} {\bibfnamefont {K.~S.}\ \bibnamefont
			{Thorne}},\ }\href {https://doi.org/10.1103/PhysRevD.62.124022} {\bibfield
		{journal} {\bibinfo  {journal} {Phys. Rev. D}\ }\textbf {\bibinfo {volume}
			{62}},\ \bibinfo {pages} {124022} (\bibinfo {year} {2000})},\ \Eprint
	{https://arxiv.org/abs/gr-qc/0003032} {arXiv:gr-qc/0003032} \BibitemShut
	{NoStop}%
	\bibitem [{\citenamefont {Flanagan}\ and\ \citenamefont
		{Hughes}(2005)}]{Flanagan:2005yc}%
	\BibitemOpen
	\bibfield  {author} {\bibinfo {author} {\bibfnamefont {E.~E.}\ \bibnamefont
			{Flanagan}}\ and\ \bibinfo {author} {\bibfnamefont {S.~A.}\ \bibnamefont
			{Hughes}},\ }\href {https://doi.org/10.1088/1367-2630/7/1/204} {\bibfield
		{journal} {\bibinfo  {journal} {New J. Phys.}\ }\textbf {\bibinfo {volume}
			{7}},\ \bibinfo {pages} {204} (\bibinfo {year} {2005})},\ \Eprint
	{https://arxiv.org/abs/gr-qc/0501041} {arXiv:gr-qc/0501041} \BibitemShut
	{NoStop}%
	\bibitem [{\citenamefont {Blanchet}\ \emph {et~al.}(2008)\citenamefont
		{Blanchet}, \citenamefont {Faye}, \citenamefont {Iyer},\ and\ \citenamefont
		{Sinha}}]{Blanchet:2008modes}%
	\BibitemOpen
	\bibfield  {author} {\bibinfo {author} {\bibfnamefont {L.}~\bibnamefont
			{Blanchet}}, \bibinfo {author} {\bibfnamefont {G.}~\bibnamefont {Faye}},
		\bibinfo {author} {\bibfnamefont {B.~R.}\ \bibnamefont {Iyer}},\ and\
		\bibinfo {author} {\bibfnamefont {S.}~\bibnamefont {Sinha}},\ }\href
	{https://doi.org/10.1088/0264-9381/25/16/165003} {\bibfield  {journal}
		{\bibinfo  {journal} {Class. Quant. Grav.}\ }\textbf {\bibinfo {volume}
			{25}},\ \bibinfo {pages} {165003} (\bibinfo {year} {2008})},\ \Eprint
	{https://arxiv.org/abs/0802.1249} {arXiv:0802.1249 [gr-qc]} \BibitemShut
	{NoStop}%
	\bibitem [{\citenamefont {Barack}\ and\ \citenamefont
		{Cutler}(2004)}]{Barack:2003fp}%
	\BibitemOpen
	\bibfield  {author} {\bibinfo {author} {\bibfnamefont {L.}~\bibnamefont
			{Barack}}\ and\ \bibinfo {author} {\bibfnamefont {C.}~\bibnamefont
			{Cutler}},\ }\href {https://doi.org/10.1103/PhysRevD.69.082005} {\bibfield
		{journal} {\bibinfo  {journal} {Phys. Rev. D}\ }\textbf {\bibinfo {volume}
			{69}},\ \bibinfo {pages} {082005} (\bibinfo {year} {2004})},\ \Eprint
	{https://arxiv.org/abs/gr-qc/0310125} {arXiv:gr-qc/0310125} \BibitemShut
	{NoStop}%
	\bibitem [{\citenamefont {Cutler}(1998)}]{Cutler:1997ta}%
	\BibitemOpen
	\bibfield  {author} {\bibinfo {author} {\bibfnamefont {C.}~\bibnamefont
			{Cutler}},\ }\href {https://doi.org/10.1103/PhysRevD.57.7089} {\bibfield
		{journal} {\bibinfo  {journal} {Phys. Rev. D}\ }\textbf {\bibinfo {volume}
			{57}},\ \bibinfo {pages} {7089} (\bibinfo {year} {1998})},\ \Eprint
	{https://arxiv.org/abs/gr-qc/9703068} {arXiv:gr-qc/9703068} \BibitemShut
	{NoStop}%
	\bibitem [{\citenamefont {Flanagan}\ and\ \citenamefont
		{Hughes}(1998)}]{Flanagan:1997kp}%
	\BibitemOpen
	\bibfield  {author} {\bibinfo {author} {\bibfnamefont {E.~E.}\ \bibnamefont
			{Flanagan}}\ and\ \bibinfo {author} {\bibfnamefont {S.~A.}\ \bibnamefont
			{Hughes}},\ }\href {https://doi.org/10.1103/PhysRevD.57.4566} {\bibfield
		{journal} {\bibinfo  {journal} {Phys. Rev. D}\ }\textbf {\bibinfo {volume}
			{57}},\ \bibinfo {pages} {4566} (\bibinfo {year} {1998})},\ \Eprint
	{https://arxiv.org/abs/gr-qc/9710129} {arXiv:gr-qc/9710129} \BibitemShut
	{NoStop}%
	\bibitem [{\citenamefont {Lindblom}\ \emph {et~al.}(2008)\citenamefont
		{Lindblom}, \citenamefont {Owen},\ and\ \citenamefont
		{Brown}}]{Lindblom:2008cm}%
	\BibitemOpen
	\bibfield  {author} {\bibinfo {author} {\bibfnamefont {L.}~\bibnamefont
			{Lindblom}}, \bibinfo {author} {\bibfnamefont {B.~J.}\ \bibnamefont {Owen}},\
		and\ \bibinfo {author} {\bibfnamefont {D.~A.}\ \bibnamefont {Brown}},\ }\href
	{https://doi.org/10.1103/PhysRevD.78.124020} {\bibfield  {journal} {\bibinfo
			{journal} {Phys. Rev. D}\ }\textbf {\bibinfo {volume} {78}},\ \bibinfo
		{pages} {124020} (\bibinfo {year} {2008})},\ \Eprint
	{https://arxiv.org/abs/0809.3844} {arXiv:0809.3844 [gr-qc]} \BibitemShut
	{NoStop}%
	\bibitem [{\citenamefont {Baird}\ \emph {et~al.}(2013)\citenamefont {Baird},
		\citenamefont {Fairhurst}, \citenamefont {Hannam},\ and\ \citenamefont
		{Murphy}}]{Baird:2012confidence}%
	\BibitemOpen
	\bibfield  {author} {\bibinfo {author} {\bibfnamefont {E.}~\bibnamefont
			{Baird}}, \bibinfo {author} {\bibfnamefont {S.}~\bibnamefont {Fairhurst}},
		\bibinfo {author} {\bibfnamefont {M.}~\bibnamefont {Hannam}},\ and\ \bibinfo
		{author} {\bibfnamefont {P.}~\bibnamefont {Murphy}},\ }\href
	{https://doi.org/10.1103/PhysRevD.87.024035} {\bibfield  {journal} {\bibinfo
			{journal} {Phys. Rev. D}\ }\textbf {\bibinfo {volume} {87}},\ \bibinfo
		{pages} {024035} (\bibinfo {year} {2013})},\ \Eprint
	{https://arxiv.org/abs/1211.0546} {arXiv:1211.0546 [gr-qc]} \BibitemShut
	{NoStop}%
	\bibitem [{\citenamefont {Chatziioannou}\ \emph {et~al.}(2017)\citenamefont
		{Chatziioannou}, \citenamefont {Klein}, \citenamefont {Yunes},\ and\
		\citenamefont {Cornish}}]{Chatziioannou:2017faithfulness}%
	\BibitemOpen
	\bibfield  {author} {\bibinfo {author} {\bibfnamefont {K.}~\bibnamefont
			{Chatziioannou}}, \bibinfo {author} {\bibfnamefont {A.}~\bibnamefont
			{Klein}}, \bibinfo {author} {\bibfnamefont {N.}~\bibnamefont {Yunes}},\ and\
		\bibinfo {author} {\bibfnamefont {N.}~\bibnamefont {Cornish}},\ }\href
	{https://doi.org/10.1103/PhysRevD.95.104004} {\bibfield  {journal} {\bibinfo
			{journal} {Phys. Rev. D}\ }\textbf {\bibinfo {volume} {95}},\ \bibinfo
		{pages} {104004} (\bibinfo {year} {2017})},\ \Eprint
	{https://arxiv.org/abs/1703.03967} {arXiv:1703.03967 [gr-qc]} \BibitemShut
	{NoStop}%
	\bibitem [{\citenamefont {Vallisneri}(2008)}]{Vallisneri:2007ev}%
	\BibitemOpen
	\bibfield  {author} {\bibinfo {author} {\bibfnamefont {M.}~\bibnamefont
			{Vallisneri}},\ }\href {https://doi.org/10.1103/PhysRevD.77.042001}
	{\bibfield  {journal} {\bibinfo  {journal} {Phys. Rev. D}\ }\textbf {\bibinfo
			{volume} {77}},\ \bibinfo {pages} {042001} (\bibinfo {year} {2008})},\
	\Eprint {https://arxiv.org/abs/gr-qc/0703086} {arXiv:gr-qc/0703086}
	\BibitemShut {NoStop}%
\end{thebibliography}
%

\end{document}